\documentclass[twocolumn,aps,prd,nofootinbib,superscriptaddress,showpacs]{revtex4-1}
\usepackage{amsmath,graphicx,bm,amsbsy,aas_macros}
\pdfoutput=1

\begin{document} 

\title{A Fast Method for Power Spectrum and Foreground Analysis for 21\,cm Cosmology}

\author{Joshua S. Dillon}
\email{jsdillon@mit.edu}
\affiliation{Dept. of Physics and MIT Kavli Institute, Massachusetts Institute of Technology, Cambridge, MA 02139, USA}
\author{Adrian Liu}
\affiliation{Dept. of Physics and MIT Kavli Institute, Massachusetts Institute of Technology, Cambridge, MA 02139, USA}
\affiliation{Dept. of Astronomy and Berkeley Center for Cosmological Physics, Berkeley, CA 94702, USA}
\author{Max Tegmark}
\affiliation{Dept. of Physics and MIT Kavli Institute, Massachusetts Institute of Technology, Cambridge, MA 0213
9, USA}

\newcommand{\x}{\mathbf{x}}
\newcommand{\y}{\mathbf{y}}
\newcommand{\pHat}{\widehat{\mathbf{p}}}
\newcommand{\C}{\mathbf{C}}
\newcommand{\R}{\mathbf{R}}
\newcommand{\U}{\mathbf{U}}
\newcommand{\G}{\mathbf{G}}
\newcommand{\N}{\mathbf{N}}
\newcommand{\F}{\mathbf{F}}
\newcommand{\Q}{\mathbf{Q}}
\newcommand{\I}{\mathbf{I}}
\newcommand{\V}{\mathbf{V}}
\newcommand{\ev}{\mathbf{v}}
\newcommand{\Cinv}{\mathbf{C}^{-1}}
\newcommand{\BigO}{\mathcal{O}}
\newcommand{\Pre}{\mathbf{P}}
\newcommand{\trans}{\mathsf{T}}
\newcommand{\Proj}{\mathbf{\Pi}}
\newcommand{\rv}{\mathbf{r}}
\newcommand{\kv}{\mathbf{k}}
\newcommand{\D}{\mathbf{D}}
\newcommand{\Lam}{\mathbf{\Lambda}}
\newcommand{\Oof}{\mathcal{O}}
\newcommand{\Ubar}{\overline{\U}}
\newcommand{\Gbar}{\overline{\G}}
\newcommand{\lbar}{\overline{\lambda}}
\newcommand{\Gam}{\mathbf{\Gamma}}
\newcommand{\GammaBar}{\overline{\Gam}}
\newcommand{\beq}{\begin{equation}}
\newcommand{\eeq}{\end{equation}}
\newcommand{\T}{\mathbf{T}}
\newcommand{\perpInt}{\int_{k_{\perp}^\alpha -\Delta k_{\perp}/2}^{k_{\perp}^\alpha +\Delta k_{\perp}/2}}
\newcommand{\paraIntPos}{\int_{k_{\|}^\alpha -\Delta k_{\|}/2}^{k_{\|}^\alpha +\Delta k_{\|}/2}}
\newcommand{\paraIntNeg}{\int_{-k_{\|}^\alpha +\Delta k_{\|}/2}^{-k_{\|}^\alpha -\Delta k_{\|}/2}}

\date{November 9, 2012; Accepted: January 24, 2013}

 
\begin{abstract}
We develop and demonstrate an acceleration of the Liu \& Tegmark quadratic estimator formalism for inverse variance foreground subtraction and power spectrum estimation in 21 cm tomography from $\BigO(N^3)$ to $\BigO(N\log N)$, where $N$ is the number of voxels of data.  This technique makes feasible the megavoxel scale analysis necessary for current and upcoming radio interferometers by making only moderately restrictive assumptions about foreground models and survey geometry.  We exploit iterative and Monte Carlo techniques and the symmetries of the foreground covariance matrices to quickly estimate the 21 cm brightness temperature power spectrum, $P(k_\|,k_\perp)$, the Fisher information matrix, the error bars, the window functions, and the bias.  We also extend the Liu \& Tegmark foreground model to include bright point sources with known positions in a way that scales as $\BigO[(N\log N) \times (\text{N point sources})] \leq \BigO(N^{5/3})$. As a first application of our method, we forecast error bars and window functions for the upcoming 128-tile deployment of the Murchinson Widefield Array, showing that 1000 hours of observation should prove sufficiently sensitive to detect the power spectrum signal from the Epoch of Reionization.
\end{abstract} 
 
 
\maketitle
   

\section{Introduction}
Neutral hydrogen tomography with the 21 cm line promises to shed light on vast and unexplored epoch of the early universe.  As a cosmological probe, it offers the opportunity to directly learn about the evolution of structure in our universe during the cosmological dark ages and the subsequent Epoch of Reionization (EoR) \cite{BLreview,FurlanettoReview,miguelreview,PritchardLoebReview}.  More importantly, the huge volume of space and wide range of cosmological scales probed makes 21 cm tomography uniquely suited for precise statistical determination of the parameters that govern modern cosmological and astrophysical models for how our universe transitioned from hot and smooth to cool and clumpy \citep{Matt3,Santos2, juddjackiemiguel1,Whitepaper2,wyithe2008,ChangDE,Rees, Tozzi2, Tozzi, Iliev,furlanetto1,Loeb1,furlanetto2,Barkana1,Mack,Yi,ClesseBackgroundReionizationOmniscopes}.  It has the potential to surpass even the Cosmic Microwave Background (CMB) in its sensitivity as a cosmological probe \citep{Yi}. 

The central idea behind 21 cm tomography is that images produced by low frequency radio interferometers at different frequencies can create a series of images at different redshifts, forming a three dimensional map of the 21 cm brightness temperature.  Yet we expect that our images will be dominated by synchrotron emission from our galaxies and others.  In fact, we expect those foreground signals to dominate over the elusive cosmological signal by about four orders of magnitude \citep{Angelica,bernardi}.  

One major challenge for 21 cm cosmology is the extraction of the brightness temperature power spectrum, a key prediction of theoretical models of the dark ages and the EoR, out from underneath a mountain of foregrounds and instrumental noise.  Liu \& Tegmark (\citep{LT11}, hereafter ``LT") presented a method for power spectrum estimation that has many advantages over previous approaches (on which we will elaborate in Section \ref{bruteForce}).  It has, however, one unfortunate drawback: it is very slow.  The LT method relies on multiplying and inverting very large matrices, operations that scale as $\BigO(N^3)$, where $N$ is the number of voxels of data to analyze.  

The goal of the present paper is to develop and demonstrate a way of achieving the results of the LT method that scales only as $\BigO(N\log N)$.  Along the way, we will also show how LT can be extended to take advantage of additional information about the brightest point sources in the map while maintaining a reasonable algorithmic scaling with $N$.  Current generation interferometers, including the Low Frequency Array (LOFAR, \citep{LOFARinstrument}), the Giant Metrewave Radio Telescope (GMRT, \citep{GMRT}), the Murchinson Widefield Array (MWA, \citep{MWAstatus}), and the Precision Array for Probing the Epoch of Reionization (PAPER, \citep{PAPER}) are already producing massive data sets at or near the megavoxel scale (e.g. \citep{ChrisMWA}).  These data sets are simply too large to be tackled by the LT method.  We expect next generation observational efforts, like the Hydrogen Epoch of Reionization Array \cite{HERA}, a massive Omniscope \citep{FFTT2}, or the Square Kilometer Array \cite{SKAspecifications}, to produce even larger volumes of data.  Moreover, as computer processing speed continues to grow exponentially, the ability to observe with increasingly fine frequency resolution will enable the investigation of the higher Fourier modes of the power spectrum at the cost of yet larger data sets.  The need for an acceleration of the LT method is pressing and becoming more urgent.

Our paper has a similar objective to \citep{UeLiFast}, which also seeks to speed up algorithms for power spectrum estimation with iterative and Monte Carlo techniques.  The major differences between the paper arise from the our specialization to the problem of 21 cm cosmology and the added complications presented by foregrounds, especially with regard to the basis in which various covariance matrices are easiest to manipulate.   Our paper also shares similarities to \citep{GibbsPSE}.  Like \citep{UeLiFast}, \citep{GibbsPSE} does not extend its analysis to include foregrounds.  It differs also from this paper in spirit because that it seeks to go from interferometric visibilities to a power spectrum within a Bayesian framework rather than from a map to a power spectrum and because it considers one frequency channel at a time.  In this paper, we take advantage of many frequency channels simultaneously in order to address the problem of foregrounds.

This paper is organized as follows.  We begin with Section \ref{bruteForce} wherein we review the motivation for and details of the LT method.  In Section \ref{Fast} we present the novel aspects of our technique for measuring the 21 cm brightness temperature power spectrum.  We discuss the extension of the method to bright point sources and the assumptions we must make to accelerate our analysis.  In Section \ref{results} we demonstrate end-to-end tests of the algorithm and show some of its first predictions for the ability of the upcoming 128-tile deployment of the MWA to detect the statistical signal of the Epoch of Reionization.
 

\section{The Brute Force Method}\label{bruteForce} 

The solution to the problem of power spectrum estimation in the presence of foregrounds put forward by LT offers a number of improvements over previous proposals that rely primarily on line of sight foreground information \citep{xiaomin,nusserforegrounds,Judd08,paper1,LOFAR,Harker,paper2,LOFAR2,ChoForegrounds}.  The problem of 21 cm power spectrum estimation shares essential qualities with both CMB and galaxy survey power spectrum estimation efforts.  Like with galaxy surveys, we are interested in measuring a three dimensional power spectrum.  On the other hand, our noise and foreground contaminants bear more similarity to the problems faced by CMB studies---though the foregrounds we face are orders of magnitude larger.

The LT method therefore builds on the literature of both CMB and galaxy surveys, providing a unified framework for the treatment of geometric and foreground effects by employing the quadratic estimator formalism for inverse variance foreground subtraction and power spectrum estimation.  In Section \ref{21cmStatistics} we will review precisely how it is implemented.  

The LT formalism has a number of important advantages over its predecessors.  By treating foregrounds as a form of correlated noise, both foregrounds and noise can be downweighted in a way that is unbiased and lossless in the sense that it maintains all the cosmological information in the data.  Furthermore, the method allows for the simultaneous estimation of both the errors on power spectrum estimates and the window functions or ``horizontal" error bars.  

Unfortunately, the LT method suffers from computational difficulties.  Because it involves inverting and multiplying very large matrices, it cannot be accomplished faster than in $\BigO(N^3)$ steps, where $N$ is the number of voxels in the data to be analyzed.  This makes analyzing large data sets with this method infeasible.  The primary goal of this paper is to demonstrate an adaptation of the method that can be run much faster.  But first, we need to review the essential elements of the method to put our adaptations and improvements into proper context.  In Sections \ref{dataOrganization} and \ref{PSdiscretization}, we describe our conventions and notation and explain the relationship between the measured quantities and those we seek to estimate.  In Section \ref{21cmStatistics}, we review the LT statistical estimators and how the Fisher information matrix is used to calculate statistical errors on power spectrum measurements.  Then in \ref{AdrianModels} we explain the LT model of noise and foregrounds in order to motivate and justify our refinements that will greatly speed up the algorithm in Section \ref{Fast}.

\subsection{Data Organization and Conventions} \label{dataOrganization}
We begin with a grid of data that represents the brightness temperatures at different positions on the sky as a function of frequency from which we wish to estimate the 21 cm brightness temperature power spectrum.  We summarize that information using a data vector $\mathbf{x}$ which can be thought of as a one dimensional object of length $n_{x} n_{y} n_{z}\equiv N$,\footnote{While it is helpful to think of $\mathbf{x}$ as a vector in the matrix operations below, it is important to remember that the index $i$ in $x_i$, which refers to the different components of $\mathbf{x}$, actually runs over different values of the spatial coordinates $x$, $y$, and $z$.} the number of voxels in the data cube.

\begin{figure}
  \centering 
    \includegraphics[width=.45\textwidth]{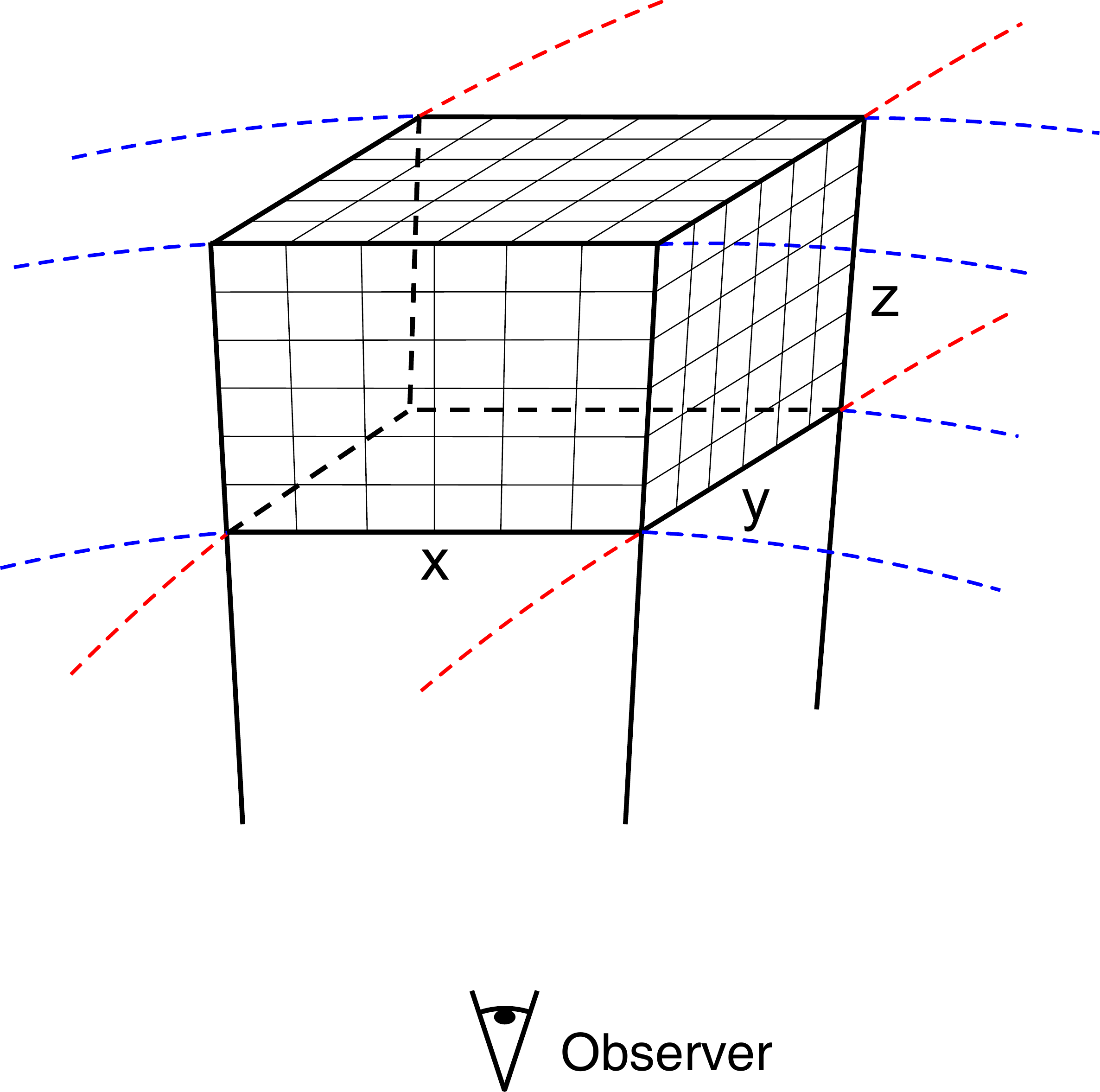}
  \caption{This exaggerated schematic illustrates the flat sky approximation.  It shows great circles (colored and dashed) approximated linearly in the region considered, with lines tracing back to the observer treated as if they were parallel.  Our data cube contains the measured brightness temperatures for every small voxel.}
  \label{flatsky}
\end{figure}

Although the LT technique works for arbitrary survey geometries, we restrict ourselves to the simpler case of a data ``cube" that corresponds to a relatively small rectilinear section of our universe of size $\ell_x \times \ell_y \times \ell_z$ in comoving coordinates.\footnote{This restriction and its attendant approximations lie at the heart of our strategy for speeding up these calculations, as we explain in Section \ref{Fast}.}  We pick our box to be a subset of the total 21 cm brightness temperature 3D map that a large interferometric observatory would produce.  Unlike the LT method, our fast method requires that the range of positions on the sky must be small enough for the flat sky approximation to hold (Figure \ref{flatsky}).
Similarly, our range of frequencies (and thus redshifts) in the data cube must correspond to an epoch short enough so that $P(k,z)$ might be approximated as constant in time. Following simulations by \cite{McQuinnLyman}, \cite{Yi} argued that we can conservatively extend the redshift ranges of our data cubes to about $\Delta z \lesssim 0.5$.  At typical EoR redshifts, such a small range in $\Delta z$ allows a very nearly linear mapping between the frequencies measured by an interferometer to a regularly spaced range of comoving distances, $d_C(z)$, although in general $d_C(z)$ is not a linear function of $z$ or $\nu$.  This also justifies the approximation that our data cube corresponds to an evenly partitioned volume of our universe.

If the measured brightness temperatures, $x_i$, were only the result of redshifted 21 cm radiation, then each measurement would represent the average value in some small box of volume $\Delta x \Delta y \Delta z$ centered on $\rv_i$ of a continuous brightness temperature field $x(\rv)$ \citep{Maxgalaxysurvey1}: 
\beq
x_i \equiv \int \psi_i(\rv)x(\rv) d^3r, \label{voxAvg}
\eeq
where our discretization function $\psi_i$ is defined as $\psi_i(\rv) \equiv \psi_0(\rv -\rv_i)$, where
\beq
\psi_0(\rv) \equiv \frac{\Pi(\frac{x}{\Delta x})\Pi(\frac{y}{\Delta y})\Pi(\frac{z}{\Delta z})}{\Delta x \Delta y \Delta z} \label{tophat}
\eeq
and where $\Pi(x)$ is the normalized, symmetric boxcar function ($\Pi(x)= 1$ if $|x|< \frac{1}{2}$ and $0$ otherwise).  This choice of pixelization encapsulates the idea that each measured brightness temperature is the average over a continuous temperature field inside each voxel.  In this paper, we improve on the LT method by including the effect of finite pixelation.  This will manifest itself as an extra $\Phi(\kv)$ term that will we define in Equation \ref{FTpsi} and that will reappear throughout this paper.

\subsection{The Discretized 21 cm Power Spectrum} \label{PSdiscretization}
Ultimately, the goal of this paper is to estimate the 21 cm power spectrum $P(\kv)$, defined via
\beq
\langle \widetilde{x}^*(\kv) \widetilde{x}(\kv') \rangle \equiv (2\pi)^3 \delta(\kv - \kv') P(\kv),
\eeq
where $\widetilde{x}(\kv)$ is the Fourier transformed brightness temperature field and where angle brackets denote the ensemble average of all possible universes obeying the same statistics.  

Our choice of pixelization determines the relationship between the continuous power spectrum, $P(\kv)$, and the 21 cm signal covariance matrix, which we call $\mathbf{S}$ for \textbf{S}ignal.  It is fairly straightforward to show, given Equation \ref{voxAvg} and the definition of the power spectrum \citep{Maxgalaxysurvey1}, that:
\beq
S_{ij} \equiv \langle x_i x_j\rangle - \langle x_i\rangle\langle x_j\rangle = \int \widetilde{\psi}_{i}(\kv)\widetilde{\psi}_{j}^{*}(\kv)P(\kv)\frac{d^3k}{(2\pi)^{3}}, \label{CDef}
\eeq
where $\widetilde{\psi}_{i}(\kv)$ is the Fourier transform of $\psi_i(\rv)$:
\beq
\widetilde{\psi}_{i}(\kv) \equiv  \int e^{-i\kv\cdot\rv}\psi_{i}(\rv)d^{3}\rv.
\eeq
Separating this integral into each of the three Cartesian coordinates and integrating yields
\begin{align}
\widetilde{\psi}_{i}(\kv) &= e^{i \kv \cdot \rv_i} \Phi(\kv), \text{ where} \nonumber \\
\Phi(\kv) &\equiv j_0\left(\frac{k_x \Delta x}{2}\right) j_0\left(\frac{k_y \Delta y}{2}\right) j_0\left(\frac{k_z \Delta z}{2}\right), \label{FTpsi}
\end{align}
where $j_0(x) = \sin x / x$ is the zeroth spherical Bessel function.  Because we can only make a finite number of measurements of the power spectrum, we parametrize and discretize $P(\kv)$ by approximating it as a piecewise constant function:
\beq
P(\kv) \approx \sum_\alpha p^\alpha \chi^\alpha(\kv), \label{QSpectral}
\eeq
where the ``band power" $p^\alpha$ gives the power in region $\alpha$ of Fourier space,\footnote{In contrast to lowered Latin indices, which we use to pick out voxels in a real space or Fourier space data cube, we will use raised Greek indices to pick out power spectrum bins, which will generally each run a range in $k_\|$ and in $k_\perp$.} specified by the characteristic function $\chi^\alpha(\kv)$ which equals 1 inside the region and vanishes elsewhere.

Combining Equations \ref{CDef} and \ref{QSpectral} we can write down $S_{ij}$:
\begin{align}
S_{ij} &= \sum_\alpha p^\alpha Q^\alpha_{ij}, \text{ where} \nonumber \\
Q^\alpha_{ij} &\equiv \int \widetilde{\psi}_{i}(\kv)\widetilde{\psi}_{j}^{*}(\kv)\chi^\alpha(\kv)\frac{d^3k}{(2\pi)^{3}}.
\end{align}
We choose these $\chi^\alpha(\kv)$ to produce band powers that reflect the symmetries of the observation.  Our universe is isotropic in three dimensions, but due to redshift space distortions, foregrounds, and other effects, our measurements will be isotropic only perpendicular to the line of sight \citep{Yi,Barkana2,AliAP, NusserAP,BarkanaAP}.  This suggests cylindrical binning of the power spectrum; in the directions perpendicular to the line of sight, we bin $k_x$ and $k_y$ together radially to get a region in k-space extending from from $k_\perp^\alpha -\Delta k_\perp /2$ to $k_\perp^\alpha +\Delta k_\perp /2$ where $k_\perp^2 \equiv k_x^2 + k_y^2$.  Likewise, in the direction parallel to the line of sight, we integrate over a region of k-space both from $k_\|^\alpha -\Delta k_\| /2$ to $k_\|^\alpha +\Delta k_\| /2$ and, because the power spectrum only depends on $k_\| \equiv |k_z|$, from $-k_\|^\alpha +\Delta k_\| /2$ to $-k_\|^\alpha -\Delta k_\| /2$.  Therefore, we have
\begin{align}
Q_{ij}^{\alpha} &= \frac{1}{(2\pi)^{3}} \left[ \paraIntPos - \paraIntNeg \right] \nonumber \\ &  \perpInt |\Phi(\kv)|^2 e^{i\mathbf{k}\cdot(\mathbf{r}_{i}-\mathbf{r}_{j})} k_{\bot}d\theta dk_\perp dk_{\|}. \label{Qderivation}
\end{align}
Without the factor of $|\Phi(\kv)|^2$, the LT method was able to evaluate this integral analytically.  With it, the integral must be evaluated numerically if it is to be evaluated at all.  This is of no consequence; we will return to this formula in Section \ref{fastPSE} to show how the matrix $\Q^\alpha$ naturally lends itself to approximate multiplication by vectors using fast Fourier techniques.

\subsection{21 cm Power Spectrum Statistics} \label{21cmStatistics}
In order to interpret the data from any experiment, we need to be able to estimate both the 21 cm brightness temperature power spectrum and the correlated errors induced by the survey parameters, the instrument, and the foregrounds.  The LT method does both at the same time; with it, the calculation of the error bars immediately enables power spectrum estimation. 
 
\subsubsection{Inverse Variance Weighted Power Spectrum Estimation}
The LT method adapts the inverse variance weighted quadratic estimator formalism \citep{Maxpowerspeclossless, BJK} for calculating 21 cm power spectrum statistics.  The first step towards constructing the estimator $\widehat{p}^\alpha$ for $p^\alpha$ is to compute a quadratic quantity, called $\widehat{q}^\alpha$ whose relationship\footnote{Unlike the notation in LT, we do not include the bias term in $\widehat{q}^\alpha$ but will later include it in our power spectrum estimator.  The result is the same.} to $\widehat{p}^\alpha$ we will explain shortly:
\begin{equation}
\widehat{q}^\alpha \equiv \frac{1}{2}(\mathbf{x} - \langle \x \rangle)^{\trans}\mathbf{C}^{-1}\mathbf{Q}^{\alpha}\mathbf{C}^{-1}(\mathbf{x} - \langle \x \rangle). \label{estimator}
\end{equation}
Here $\mathbf{C}$ is the covariance matrix of $\mathbf{x}$, so
\begin{equation}
\mathbf{C} \equiv \langle\mathbf{x}\mathbf{x}^{\trans}\rangle - \langle\mathbf{x}\rangle\langle\mathbf{x}\rangle^{\trans}.
\end{equation}
For any given value of $\alpha$, the right-hand side of Equation (\ref{estimator}) yields a scalar. Were both our signal and foregrounds Gaussian, this estimator would be optimal in the sense that it preserves all the cosmological information contained in the data.  Of course, with a non-Gaussian signal, the power spectrum cannot contain all of the information, though it still can be very useful \citep{Maxpowerspeclossless}.
 
Our interest in the quadratic estimators $\widehat{q}^\alpha$ lies in their simple relationship to the underlying band powers.  In \cite{Maxpowerspeclossless}, it is shown that:
\beq
\left< \widehat{\mathbf{q}} \right> = \mathbf{F}\mathbf{p} + \mathbf{b} \label{expectqhat}
\eeq
where each $b^\alpha$ is the bias in the estimator and $\F$ is the Fisher information matrix, which is related to the probability of having measured our data given a particular set of band powers, $f(\mathbf{x}|p^{\alpha})$.  The matrix is defined \citep{fisher} as:
\begin{equation}
F^{\alpha\beta} \equiv -\left<\frac{\partial^{2}\mbox{ln}f(\mathbf{x}|p^{\alpha})}{\partial p^{\alpha}\partial p^{\beta}}\right>.
\end{equation}
The LT method employs the estimators by calculating both $\F$ and $\mathbf{b}$ using relationships derived in \citep{Maxpowerspeclossless}:
\begin{align}
F^{\alpha\beta} &= \frac{1}{2}\mbox{tr}[\mathbf{C}^{-1}\mathbf{Q}^{\alpha}\mathbf{C}^{-1}\mathbf{Q}^{\beta}] \text{ and} \label{fisherTrace} \\
b^\alpha &= \frac{1}{2}\mbox{tr}[(\C - \mathbf{S})\mathbf{C}^{-1}\mathbf{Q}^{\alpha}\mathbf{C}^{-1}]. \label{biasTrace}
\end{align}

We want our $\widehat{p}^\alpha$ to be unbiased estimators of the true underlying band powers, which means that we will have to take care to remove the biases for each band power, $b^\alpha$.  We construct our estimators\footnote{Here were differ slightly from the LT method in the normalization, which does not have the property from Equation \ref{weightedAverage}.  We instead follow \citep{THX2df}. } as linear combinations of the quadratic estimators $\widehat{q}^\alpha$ that have been corrected for bias:
\beq
\widehat{\mathbf{p}} = \mathbf{M}(\widehat{\mathbf{q}} - \mathbf{b}), \label{phatdef}
\eeq
where $\mathbf{M}$ is a matrix which correctly normalizes the power spectrum estimates; the form of $\mathbf{M}$ represents a choice in the trade-off between small error bars and narrow window functions, as we will explain shortly.

How do we expect this estimator to behave statistically?  The only random variable on the right hand side of Equation \ref{phatdef} is $\mathbf{\widehat{q}}$, so we can combine Equations \ref{expectqhat} and \ref{phatdef} to see that our choice of $\widehat{\mathbf{p}}$ indeed removes the bias term:
\beq
\left< \widehat{\mathbf{p}} \right> = \mathbf{M}\F \mathbf{p} +\mathbf{Mb} - \mathbf{Mb} = \mathbf{M}\F \mathbf{p} = \mathbf{W}\mathbf{p}. \label{expectphat} 
\eeq
We have defined the matrix of ``window functions" $\mathbf{W} \equiv \mathbf{M}\F$ because Equation \ref{expectphat} tells us that we can expect our band power spectrum estimator, $\widehat{\mathbf{p}}$, be be a weighted average of the true, underlying band powers, $\mathbf{p}$.  That definition imposes the condition on $\mathbf{W}$ that
\beq
\sum_\beta W^{\alpha\beta} = 1 \label{weightedAverage}
\eeq
which is equivalent to the statement that the weights in a weighted average must add up to one.  The condition on $\mathbf{W}$ constrains our choice of $\mathbf{M}$, though as long as $\mathbf{M}$ is an invertible matrix,\footnote{None of the choices of $\mathbf{M}$ involve anything more computationally intensive than inverting $\F$. This is fine, since $\F$ is a much smaller matrix than $\C$.} the choice of $\mathbf{M}$ does not change the information content of our power spectrum estimate, only the way we choose to represent our result.

\subsubsection{Window Functions and Error Bars}

In this paper, we choose a form of $\widehat{\mathbf{p}}$ where $\mathbf{M} \propto \F^{-1/2}$.  Two other choices for $\mathbf{M}$ are presented in \cite{THX2df}: one where $\mathbf{M} \propto \I$ and another where $\mathbf{M} \propto \F^{-1}$.  The former produces the smallest possible error bars, but at the cost of wide window functions and correlated measurement errors.  The latter produces $\delta$-function windows, but large and anticorrelated measurement errors.  This choice of $\mathbf{M} \propto \F^{-1/2}$ has proven to be a happy medium between those other two choices for $\mathbf{M}$.  It produces reasonably narrow window functions and reasonably small error bars which have the added advantage of being completely uncorrelated, so that each measurement contains a statistically independent piece of information.  Because $\mathbf{W} \equiv \mathbf{M} \F$ and because of the condition on $\mathbf{W}$ in Equation \ref{weightedAverage}, there is only one such $\mathbf{M}$:
\beq
M^{\alpha\beta} \equiv \frac{\left(\mathbf{F}^{-1/2}\right)^{\alpha\beta}}{\sum_\gamma (\F^{1/2})^{\alpha\gamma}}.
\eeq
With this choice of $\mathbf{M}$ we get window functions of the form
\beq
W^{\alpha\beta} = \frac{(\F^{1/2})^{\alpha\beta}}{\sum_\gamma (\F^{1/2})^{\alpha\gamma}}
\eeq
which we can use to put ``horizontal error bars" on our power spectrum estimates.

Using Equation \ref{phatdef} and the fact derived in \citep{Maxpowerspeclossless} that an equivalent formula for $\F$ is given by
\beq
\mathbf{F} = \langle\widehat{\mathbf{q}}\widehat{\mathbf{q}}^{\trans}\rangle - \langle\widehat{\mathbf{q}}\rangle\langle\widehat{\mathbf{q}}\rangle^{\trans}, \label{covq}
\eeq
we can see that the covariance of $\widehat{\mathbf{p}}$ takes on a simple form:
\beq
\langle\widehat{\mathbf{p}}\widehat{\mathbf{p}}^{\trans}\rangle - \langle\widehat{\mathbf{p}}\rangle\langle\widehat{\mathbf{p}}\rangle^{\trans} = \mathbf{M}\F \mathbf{M}^\trans.
\eeq
This allows us to write down the ``vertical error bars" on our individual power spectrum estimates:
\beq
\Delta \widehat{p}^\alpha = \left[\left(\mathbf{M}\F \mathbf{M}^\trans\right)^{\alpha\alpha}\right]^{1/2} = \frac{1}{\sum_\gamma (\F^{1/2})^{\alpha\gamma}}.
\eeq
As in LT, we can transform our power spectrum estimates and our vertical error bars into temperature units:
\beq
\widehat{T}^\alpha \equiv \left[\frac{(k^\alpha_\perp)^2 k^\alpha_\|}{2\pi^2} p^\alpha \right]^{1/2}
\eeq
and likewise,
\beq
\Delta \widehat{T}^\alpha = \left[\frac{(k^\alpha_\perp)^2 k^\alpha_\|}{2\pi^2 \left(\sum_\gamma (\F^{1/2})^{\alpha\gamma}\right)} \right]^{1/2}. \label{TemperatureUnits}
\eeq
This makes it easier to compare to theoretical predictions, which are often quoted in units of K or mK.

\subsection{Foreground and Noise Models}\label{AdrianModels}
The structure of the matrix $\C$ that goes into our inverse variance weighted estimator depends on the way we model our foregrounds, noise, and signal.  We assume that those contributions are the sum of five uncorrelated components:
\begin{align}
\mathbf{C} &= \sum_{c \text{ } \in \text{ components}}\langle\mathbf{x}_c\mathbf{x}_c^{\trans}\rangle - \langle\mathbf{x}_c\rangle\langle\mathbf{x}_c\rangle^{\trans} \nonumber \\ &\equiv \mathbf{S} +  \mathbf{R} + \mathbf{U} + \mathbf{G} + \mathbf{N}.
\end{align}
These are the covariance matrices due to 21 cm \textbf{S}ignal, bright point sources \textbf{R}esolved from one another, \textbf{U}nresolved point sources, the \textbf{G}alactic synchrotron, and detector \textbf{N}oise, respectively.  This deconstruction of $\mathbf{C}$ is both physically motivated and will ultimately let us approximate $\mathbf{C}^{-1}(\mathbf{x}-\langle \x \rangle)$ much more quickly than by just inverting the matrix.   

Following LT, we neglect the small cosmological $\mathbf{S}$ because it is only important for taking cosmic variance into account.  It is straightforward to include the $\mathbf{S}$ matrix in our method, especially because we expect it to have a very simple form, but this will only be necessary once the experimental field moves from upper limits to detection and characterization of the 21 cm brightness temperature power spectrum.  

In this paper, we will develop an accelerated version of the LT method using the models delineated in LT.  That speed-up relies on the fact that all of these covariance matrices can be multiplied by vectors $\mathcal{O}(N\mbox{log}N)$ time.  However, our techniques for acceleration will work on a large class of models for $\C$ as long as certain assumptions about translation invariance and spectral structure are respected.  In this section, we review the three contaminant matrices from LT: $\U$, $\G$, and $\N$.  When we discuss methods to incorporate these matrices into a faster technique in Section \ref{FastCov}, we will also expand the discussion of foregrounds to include $\R$, which is a natural extension of $\U$.
\begin{figure*} 	
	\centering 
	\includegraphics[width=.8\textwidth]{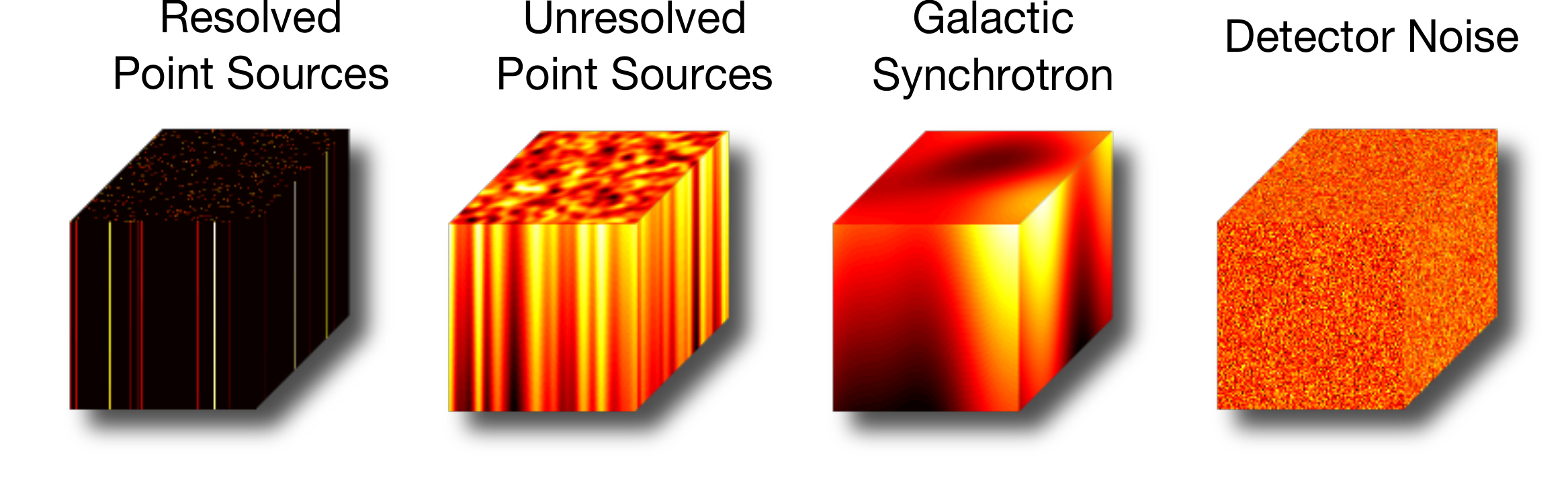}
	\caption{These example data cubes (with the line of sight drawn vertically) illustrate the strong or weak correlations between different voxels in the same cube.  In Section \ref{random} we explain how these simulated data cubes are generated quickly. The addition of resolved point sources, which is not included in LT, is discussed in Section \ref{RFast}.  To best exemplify the detailed structure of the models, the color scales are different for each of the cubes.}
	\label{DataCubes}
\end{figure*}

\subsubsection{Unresolved Point Sources} \label{AdrianU}
For a typical current generation or near future experiment, the pixels perpendicular to the line of sight are so large that every one is virtually guaranteed to have a point source in it bright enough to be an important foreground to our 21 cm signal.  These confusion limited point sources are taken into account using their strong correlations parallel to the line of sight and weaker correlations perpendicular to the line of sight, both of which are easily discerned in Figure \ref{DataCubes}.

Following LT we split $\U$ into the tensor product of two parts, one representing correlations perpendicular to the line of sight and the other parallel to the line of sight:
\beq
\U \equiv  \U_\perp \otimes \U_\|
\eeq
Covariance perpendicular to the line of sight is modeled as an unnormalized Gaussian:
\beq
(U_\perp)_{ij} \equiv \exp\left[\ \frac{\left((\mathbf{r}_\perp)_{i} - (\mathbf{r}_\perp)_{j}\right)^2}{2 \sigma^2_\perp} \right] \label{Uperp}
\eeq
where $\sigma_\perp$ represents the correlation length perpendicular to the line of sight.  Following LT, we take this to be a comoving distance corresponding to 7 arcminutes, representing the weak clustering of point sources. 

The covariance along the line of sight assumes a Poisson distributed number of point sources below some flux cut, $S_{\text{cut}}$, which we take to be 0.1 Jy, each with a spectral index drawn from a Gaussian distribution with mean $\bar{\kappa}$ and standard deviation $\sigma_\kappa$.  Given a differential source count \citep{dimatteo1} of
\begin{align}
\frac{dn}{dS} =& (4000 \text{ Jy}^{-1} \text{sr}^{-1}) \times \nonumber \\ &\begin{cases} \left(\frac{S}{0.880 \text{ Jy}}\right)^{-2.51} & \text{for S $>$ 0.880 Jy} \\ \left(\frac{S}{0.880 \text{ Jy}}\right)^{-1.75} & \text{for S $\leq$ 0.880 Jy,} \end{cases} \label{sourceCounts}
\end{align}
we get a covariance parallel to the line of sight of 
\begin{align}
(U_\|)_{ij} = & (1.4\times10^{-3}\text{ K})^2\ (\eta_i \eta_j)^{-2-\overline{\kappa}}\left(\frac{\Omega_{pix}}{\text{1 sr}}\right)^{-1} \times \nonumber \\ 
& \text{exp}\left[\frac{\sigma^2_\kappa}{2}(\text{ln}(\eta_i \eta_j))^2\right] I_2(S_{cut}). \label{Uexact}
\end{align}
where we have assumed a power law spectrum for the point sources where $\eta_i \equiv \nu_i / \nu_*$, $\nu_* = $ 150 MHz, and $\overline{\kappa}$ and $\sigma_\kappa$ are the average value and standard deviation of the distribution of spectral indices of the point sources.  We define $I_2(S_{cut})$ as
\beq
I_2(S_{cut}) \equiv \int_0^{S_{\text{cut}}}S^2\frac{dn}{dS}dS \label{I2fluxintegral}
\eeq
Following LT, we take $\overline{\kappa}=0.5$ and $\sigma_\kappa = 0.5$, both of which are consistent with the results of \citep{ChrisMWA}. In Section \ref{Ufast}, we will return to Equation \ref{Uexact} and show how it can be put into an approximate form that can be quickly multiplied by a vector.

\subsubsection{Galactic Synchrotron Radiation} \label{AdrianG}

Following LT, we model Galactic synchrotron emission in the same way that we model unresolved point sources.  Fundamentally, both are spatially correlated synchrotron signals contributing to the brightness temperature of every pixel in our data cube.  However the galactic synchrotron is much more highly correlated spatially, which can be clearly seen in the sample data cube in Figure \ref{DataCubes}. This leads to our adoption of a much larger value of $\sigma_\perp$; we take $\sigma_\perp$ to be a comoving distance corresponding to $30^\circ$ on the sky.  Following LT, we take $\overline{\kappa}=0.8$ and $\sigma_\kappa = 0.4$.

This is an admittedly crude model for the galactic synchrotron, in part because it fails to take into account the roughly planar spatial distribution of the Galactic synchrotron.  A more sophisticated model for $\G$ that incorporates a more informative map of the Galactic synchrotron can only produce smaller error bars and narrower window functions.  However, such a model might involve breaking the assumption of the translational invariance of correlations, which could be problematic for the technique we use in Section \ref{Fast} to speed up this algorithm.  In practice, we expect very little benefit from an improved spatial model of the Galactic synchrotron due to the restriction imposed by the flat sky approximation that our map encompass a relatively small solid angle.

\subsubsection{Instrumental Noise} \label{AdrianN}

Here we diverge from LT to adopt a form of the noise power spectrum from \citep{FFTT} that is more readily adaptable to the pixelization scheme we will introduce:
\beq
P^N(\kv,\lambda) = \frac{\lambda^2 T_\text{sys}^2 y d_M^2}{A f^\text{cover} \tau} B^{-2}(\kv,\lambda). \label{NPowerSpectrum}
\eeq
Here $T_\text{sys}$ is the system temperature (which is sky noise dominated in our case), $A$ is the total effective collecting area of the array, and $\tau$ is the is the total observing time.  $B(\kv,\lambda)$ is a function representing the $uv$-coverage, normalized to peak at unity, which changes with wavelength. Lastly,  $y$ is the conversion from bandwidth to the comoving length of the box parallel to the line of sight and $d_M$ is the transverse comoving distance\footnote{The transverse comoving distance, $d_M(z)$, is the ratio of an object's comoving size the angle it subtends, as opposed to the angular diameter distance, $d_A(z)$, which is the ratio of its physical size to the angle it subtends.  It is sometimes called the ``comoving angular diameter distance" and it is even sometimes written as $d_A(z)$.  See \cite{HoggDistance} for a helpful summary of these often confusingly named quantities.}, so $y d_M^2 \Omega_\text{pix} \Delta \nu = \Delta x \Delta y \Delta z$ with $\Omega_\text{pix}$ being the angular size of our pixels and $\Delta \nu$ being the frequency channel width. This form of the noise power spectrum assumes that the entire map is observed for the same time $\tau$, which is why the ratio of the angular size of the map to the field of view does not appear.

We use Equation \ref{CDef} to discretize the power spectrum and get $\N$:
\beq
N_{ij} = \int e^{i \kv \cdot \rv_i} e^{-i \kv \cdot \rv_j} |\Phi(\kv)|^2 P^N(\kv,\lambda) \frac{d^3k}{(2\pi)^{3}} \label{Ndef}
\eeq
Instead of evaluating this integral, we will show in Section \ref{Nfast} that it can be approximated using the discrete Fourier transform.

\subsection{Computational Challenges to the Brute Force Method}
For a large data cube, the LT method requires the application of large matrices that are memory-intensive to store and computationally infeasible to invert.  However, we need to be able to multiply by and often invert these large matrices to calculate our quadratic estimators (Equations \ref{Qderivation} and \ref{estimator}), the Fisher matrix (Equation \ref{fisherTrace}), and the bias (Equation \ref{biasTrace}).  A $10^6$ voxel data cube, for example, would take $\BigO(10^{18})$ computational steps to analyze.  This is simply infeasible for next-generation radio interferometers and we have therefore endeavored to find a faster way to compute 21 power spectrum statistics.


\section{Our Fast Method} \label{Fast}
To avoid the computational challenges of the LT method, we seek to exploit symmetries and simpler forms in certain bases of the various matrices out of which we construct our estimate of the 21 cm power spectrum and its attendant errors and window functions.  In this section, we describe the mathematical and computational techniques we employ to create a fast and scalable algorithm.  

Our fast method combines the following six separate ideas:
\begin{enumerate}
\item A Monte Carlo technique for computing the Fisher information matrix and the bias (Section \ref{fisherMC}).
\item An FFT-based technique for computing band powers using the $\mathbf{Q}^\alpha$ matrices (Section \ref{fastPSE}).
\item An application of the conjugate gradient method that eliminates the need to invert $\C$ (Section \ref{CGPSE}).
\item A Toeplitz matrix technique for multiplying vectors quickly by the constituent matrices of $\C$ (Section \ref{FastCov}).  
\item A combined FFT and spectral technique for preconditioning $\C$ to improve converge of the conjugate gradient method (Section \ref{FastPrecon})
\item A technique using spectral decomposition and Toeplitz matrices for rapid simulation of data cubes for our Monte Carlo (Section \ref{random}).
\end{enumerate}
In this Section, we explain how all six are realized and how they fit into our fast method for power spectrum estimation.  Finally, in Section \ref{fisherConverge}, we verify the algorithm in an end-to-end test.

\subsection{Monte Carlo Calculation of the Fisher Information Matrix} \label{fisherMC}
In order to turn the results of our quadratic estimator into estimates of the power spectrum with proper vertical and horizontal error bars, we need to be able to calculate the Fisher information matrix and the bias term.  Instead of using the form of $\F$ in Equation \ref{fisherTrace} that the LT method employs, we take advantage of the relationship between $\F$ and $\widehat{\mathbf{q}}$ in Equation \ref{covq} that $\F = \langle\widehat{\mathbf{q}}\widehat{\mathbf{q}}^{\trans}\rangle - \langle\widehat{\mathbf{q}}\rangle\langle\widehat{\mathbf{q}}\rangle^{\trans}$.  If we can generate a large number of simulated data sets $\mathbf{x}$ drawn from the same covariance $\C$ and then compute $\widehat{\mathbf{q}}$ from each one, then we can iteratively approximate $\F$ with a Monte Carlo. In other words, a solution to the problem of quickly calculating $\widehat{\mathbf{q}}$ also provides us with a way to estimate $\F$.  What's more, the solution is trivially parallelizable; creating artificial data cubes and analyzing them can be done by many CPUs simultaneously.

In calculating $\F$, we can get $b^\alpha$ out essentially for free.  If we take the average of all our $\widehat{\mathbf{q}}$ vectors, we expect to that 
\begin{align}
\langle \widehat{q}^\alpha \rangle &= \left< \frac{1}{2}(\mathbf{x} - \langle \x \rangle)^{\trans}\mathbf{C}^{-1}\mathbf{Q}^{\alpha}\mathbf{C}^{-1}(\mathbf{x} - \langle \x \rangle) \right> \nonumber \\ &= \mbox{tr}\left[ \left< (\mathbf{x} - \langle \x \rangle)(\mathbf{x} - \langle \x \rangle)^{\trans} \right> \mathbf{C}^{-1}\mathbf{Q}^{\alpha}\mathbf{C}^{-1}\right] \nonumber \\ &= \mbox{tr}\left[ \mathbf{Q}^\alpha \C^{-1} \right] = b^\alpha
\end{align}
in the limit where $\mathbf{S}$ is negligibly small.  This implies that $\widehat{\mathbf{p}}$ can be written in an even simpler way:
\beq
\widehat{p}^\alpha = \frac{1}{\sum_\gamma F^{\alpha\gamma}}\left(\widehat{q}^\alpha - \left< \widehat{q}^\alpha \right>\right)
\eeq
where, recall, $\mathbf{F}$ is calculated as the sample covariance of our $\widehat{\mathbf{q}}$ vectors.  We therefore can calculate all the components of our power spectrum estimate and its error bars using a Monte Carlo.

In Section \ref{fisherConverge} we will return to assess how well the Monte Carlo technique works and its convergence properties.  But first, we need to tackle the three impediments to computing $\widehat{q}^\alpha$ in Equation \ref{estimator} quickly: generating a random $\x$ drawn from $\C$, computing $\C^{-1}(\x-\langle \x \rangle)$, and applying $\mathbf{Q}^\alpha$.

\subsection{Fast Power Spectrum Estimation Without Noise or Foregrounds}\label{fastPSE}
If we make the definition that
\beq
\mathbf{y} \equiv \mathbf{C}^{-1}(\mathbf{x} - \langle \x \rangle) 
\eeq
to simplify Equation \ref{estimator} to
\beq
q^\alpha = \mathbf{y}^\trans \mathbf{Q}^\alpha \mathbf{y} \label{yQy},
\eeq
we can see that even if we have managed to calculate $\mathbf{y}$ quickly, we still need to multiply it by a $N\times N$ element $\mathbf{Q}^\alpha$ matrix for each band power $\alpha$.   Though each $\Q^\alpha$ respects translation invariance that could make multiplying by vectors faster, there exists an even faster technique that can calculate every entry of $\widehat{\mathbf{p}}$ simultaneously using fast Fourier transforms.

To see that this is the case, we substitute Equation \ref{Qderivation} into Equation \ref{yQy}, reversing the order of summation and integration and factoring the integrand:
\begin{align}
\widehat{q}^{\alpha} = \text{ }& \frac{1}{2} \left[ \paraIntPos - \paraIntNeg \right] \nonumber \\ &  \perpInt \bigg(\sum_i y_i e^{i\mathbf{k}\cdot\mathbf{r}_i}\bigg) \bigg(\sum_j y_j e^{-i\mathbf{k}\cdot\mathbf{r}_j}\bigg) \times \nonumber \\ &  |\Phi(\kv)|^2 \frac{k_{\perp} d\theta dk_\perp dk_{\|}}{(2\pi)^{3}}. \label{integralEstimator}
\end{align}
The two sums inside the integral are very nearly discrete, 3D Fourier transforms.  All that remains is to discretize the Fourier space conjugate variable $\kv$ as we have already discretized the real space variable $\rv$.

In order to evaluate the outer integrals, we approximate them as a sum over grid points in Fourier space. The most natural choice for discretization in $\mathbf{k}$ is one that follows naturally from the FFT of $\mathbf{y}$ in real space.  If our box is of size $\ell_x \ell_y  \ell_z$ and broken into $n_x \times n_y \times n_z$ voxels\footnote{For simplicity and consistency we assume that $n_x$, $n_y$, and $n_z$ are all even and we take the origin the to be the second of the two center bins.} we have that 
\begin{align}
\mathbf{r}_j =& \left( \frac{j_x\ell_x}{n_x}, \frac{j_y\ell_y}{n_y},\frac{j_z\ell_z}{n_z} \right)  \label{rvector} \\ &\mbox{where } j_x,j_y,j_z \in \left\{-\frac{n_{x,y,z}}{2},...,0,...,\frac{n_{x,y,z}}{2}-1\right\}.\nonumber 
\end{align}
The natural 3D Fourier space discretization is
\begin{align}
\mathbf{k}_m =& \left( \frac{2 \pi m_x }{\ell_x}, \frac{2 \pi m_y }{\ell_y},\frac{2 \pi m_z }{\ell_z} \right) \\ & \mbox{where } m_x,m_y,m_z \in \left\{-\frac{n_{x,y,z}}{2},...,0,...,\frac{n_{x,y,z}}{2}-1\right\} \nonumber 
\end{align}
with a Fourier space voxel volume 
\beq
(\Delta k)^3 = \frac{2 \pi}{\ell_x} \times \frac{2 \pi}{\ell_y} \times \frac{2 \pi}{\ell_z}.
\eeq

With this choice of discretization, we will simplify our integrals by sampling Fourier space with delta functions, applying the approximation in the integrand of Equation \ref{integralEstimator} that 
\beq
1 \approx \sum_m \frac{(2 \pi)^3 \delta^3(\mathbf{k} - \mathbf{k_m})}{\ell_x \ell_y \ell_z}. \label{FourierApprox}
\eeq
This simplifies Equation \ref{integralEstimator} considerably:
\begin{align}
\widehat{q}^{\alpha} = \text{ }& \frac{1}{2} \sum_m \bigg(\sum_i y_i e^{i\mathbf{k}_m\cdot\mathbf{r}_i}\bigg) \bigg(\sum_j y_j e^{-i\mathbf{k}_m\cdot\mathbf{r}_j}\bigg)  \times \nonumber \\ &\chi^\alpha(\mathbf{k}_m) |\Phi(\kv_m)|^2 / (\ell_x \ell_y \ell_z). 
\end{align}
If we define $\widetilde{y}_m \equiv \sum_j y_i  e^{-i \mathbf{k}_m \cdot \mathbf{r}_j}$, then we can write $\widehat{\mathbf{q}}$ as:
\begin{align}
\widehat{q}^\alpha &\approx \frac{1}{2 \ell_x \ell_y \ell_z} \sum_m \widetilde{y}_m^* \widetilde{y}_m \chi^\alpha(\mathbf{k}_m) |\Phi(\kv_m)|^2 \nonumber \\&= \frac{1}{2 \ell_x \ell_y \ell_z} \sum_m |\widetilde{y}_m|^2 \chi^\alpha(\mathbf{k}_m) |\Phi(\kv_m)|^2. \label{FFTPSE}
\end{align}
This result makes a lot of sense: after all, the power spectrum is---very roughly speaking---the data Fourier transformed, squared, and binned with an appropriate convolution kernel.  

This is a very quick way to calculate $\widehat{\mathbf{q}}$ because we can compute $\widetilde{\y}$ in $\mathcal{O}(N\log N)$ time (if we already have $\y$) and then we simply need to add $|\widetilde{y}_m|^2$ for every $m$, weighted by the value of the analytic function $|\Phi(\kv_m)|^2$ to the appropriate band power $\alpha$.\footnote{For simplicty, we choose band power spectrum bins with the same width as our Fourier space bins (before zero padding).  This linear binning scheme makes plotting, which is typically logarithmic in the literature, more challenging.  On the other hand, it better spreads out the number of Fourier space data cube bins assigned to each band power.}  Each value of $|\widetilde{y}_m|^2$ gets mapped uniquely to one value of $\alpha$, so there are only $N$ steps involved in performing the binning.  Unlike in the LT method, we perform the calculation of $\widehat{q}^\alpha$ for all values of $\alpha$ simultaneously.

\begin{figure} 
	\centering 
	\includegraphics[width=.45\textwidth]{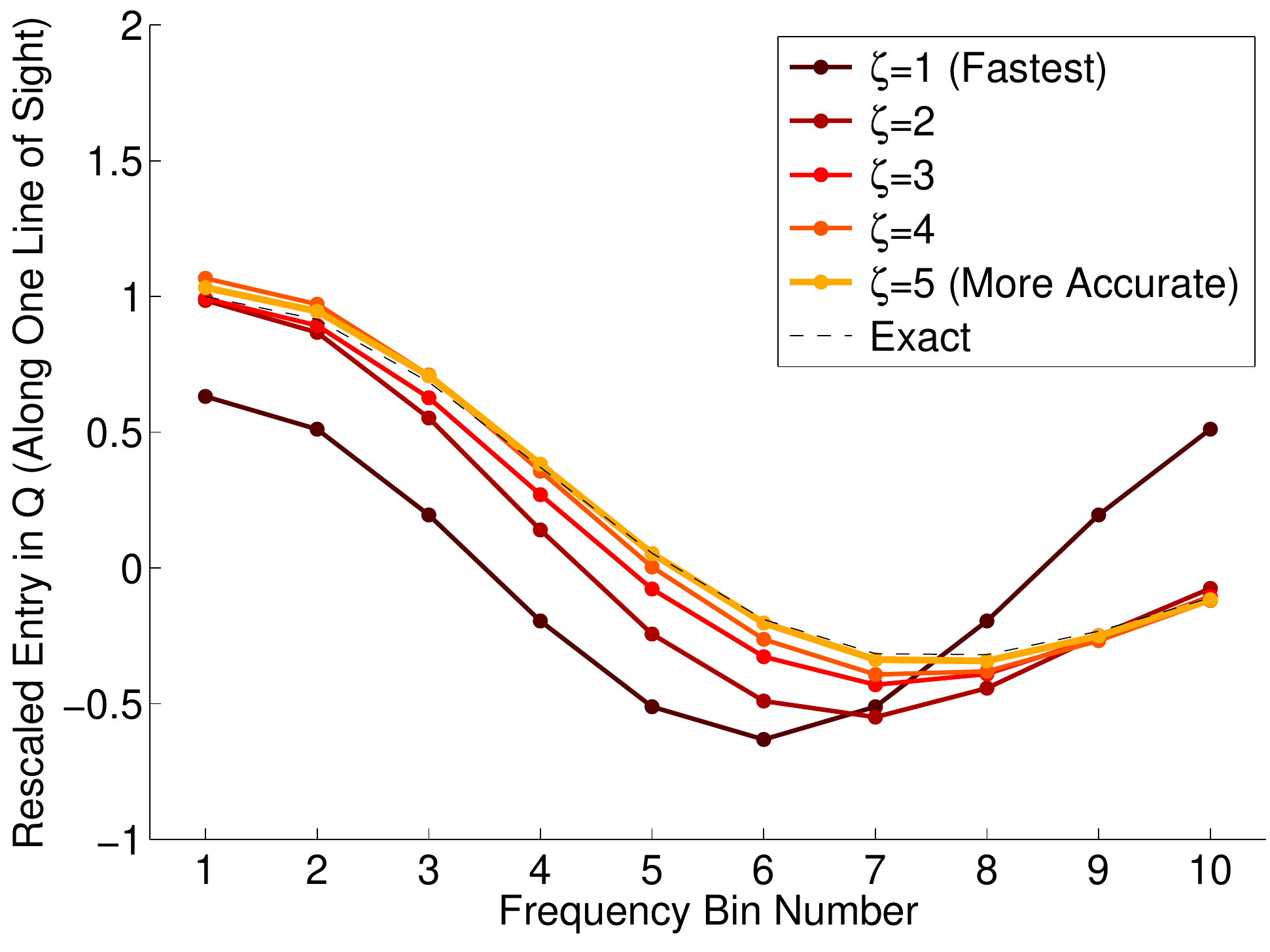}
	\caption{We use an FFT-based technique to approximate the action of the matrix $\Q^\alpha$ that encodes the Fourier transforming, binning, and pixelization factors.  In this Figure, we show how the approximation improves with different factors of the zero padding parameter, $\zeta$, while varying a single coordinate of one of the $\mathbf{Q}^\alpha$ matrices.  For a fairly small value of $\zeta$, the approximation is quite good, meaning that the binning and Fourier transforming step contributes subdomninantly to the complexity of the overall algorithm.}
	\label{FFTapproximationToQ}
\end{figure}

However, the FFT approximation to $Q_{ij}^\alpha$ from Equation \ref{FourierApprox} does not work very well at large values of $(\mathbf{r}_i - \mathbf{r}_j)$ because the discrete version of $\mathbf{Q}^\alpha$ does not sample the continuous version of $\mathbf{Q}^\alpha$ very finely.  This can be improved by zero padding the input vector $y_i$, embedding it inside of a data cube of all zeros a factor of $\zeta^3$ larger.  For simplicity, we restrict $\zeta$ to integer values where $\zeta = 1$ represents no zero padding.  By increasing our box size, we decrease the step size in Fourier space and thus the distance between each grid point in Fourier space where we sample $k$ with delta functions.  Repeating the derivation from Equations \ref{rvector} through \ref{FFTPSE} yields: 
\beq
\widehat{q}^\alpha \approx \frac{1}{2\ell_x \ell_y \ell_z \zeta^3} \sum_m |\widetilde{y}_m|^2 \chi^\alpha(\mathbf{k}_m) |\Phi(\kv_m)|^2,
\eeq
where $\widetilde{\y}$ has been zero padded and then Fourier transformed.  This technique of power spectrum estimation scales as $\mathcal{O}(\zeta^3 N\log N)$, which is fine as long as $\zeta$ is small,\footnote{Though not the computational bottleneck, this step is the most memory intensive; it involves writing down an array of $\zeta^3 N$ double-precision complex numbers.  This can reach into the gigabytes for very large data cubes.}  In Figure \ref{FFTapproximationToQ} we see how increasing $\zeta$ from 1 to 5 greatly improves accuracy.

\subsection{Inverse Variance Weighting with the Conjugate Gradient Method} \label{CGPSE}
We now know how to calculate $\widehat{q}^\alpha$ quickly provided that we can also calculate $\y \equiv \C^{-1}(\x - \langle \x \rangle)$ quickly.  The latter turns out to be the most challenging part of the problem; we will address the various difficulties that it presents in this Section through Section \ref{FastPrecon}.  We take our inspiration for a solution from a similar problem that the WMAP team faced in making their maps.  They employed the preconditioned conjugate gradient method to great success \citep{OhCMBConjGrad,WMAPconjugategrad}.  

The conjugate gradient method \cite{ConjGradOriginal} is an iterative technique for solving a system of linear equations such as $\C\y = (\x - \langle \x \rangle) $.  Although directly solving this system involves inverting the matrix $\C$, the conjugate gradient method can approximate the solution to arbitrary precision with only a limited number of multiplications of vectors by $\C$.  If we can figure out a way to quickly multiply vectors by $\C$ by investigating the structure of its constituent matrices, then we can fairly quickly approximate $\y$.  We will not spell out the entire algorithm here but rather refer the reader to the helpful and comprehensive description of it in \citep{ConjGrad}.

Whenever iterative algorithms are employed, it is important to understand how quickly they converge and what their rates of convergence depend upon.  If we are trying to achieve an error $\varepsilon$ on our approximation $\y_\text{CGM}$ to $\y$ where
\beq
\varepsilon \equiv \frac{|\C\y_\text{CGM} - (\x - \langle \x \rangle)|}{|\x - \langle \x \rangle|}.
\eeq
and where $|\x| \equiv \left(\sum_i x_i^2\right)^{1/2}$ is the length of the vector $\x$, then the number of iterations required to converge (ignoring the accumulation of round-off error) is bounded by \citep{ConjGrad}:
\beq
n \le \frac{1}{2}\sqrt{\kappa}\ln\left(\frac{2}{\varepsilon}\right)
\eeq
where $\kappa$ is the condition number of the matrix (not to be confused with $\kappa$ used elsewhere as a spectral index), defined as the ratio of its largest eigenvalue to its smallest:
\beq
\kappa(\C) \equiv \frac{\lambda_{\max}(\C)}{\lambda_{\min}(\C)}
\eeq
Because $n$ only depends logarithmically on $\varepsilon$, the convergence of the conjugate gradient method is exponential.  In order to make the algorithm converge in only a few iterations, it is necessary to ensure that $\kappa$ is not too large. This turns out to be a major hurdle that we must overcome, because we will routinely need to deal with covariance matrices with $\kappa(\C)\approx 10^{8}$ or worse.  This dynamic range problem is unavoidable; it comes directly from the ratio of the brightest foregrounds, typically hundreds of kelvin, to the noise and signal, typically tens of millikelvin.  That factor, about $10^4$, enters squared into the covariance matrices, yielding condition numbers of roughly $10^{8}$.  In Section \ref{FastPrecon} we will explain the efforts we undertake to mitigate this problem.

\subsection{Foreground and Noise Covariance Matrices} \label{FastCov}

Before we can go about ensuring that the conjugate gradient method converges quickly, we must understand the detailed structure of the constituent matrices of $\C$.  In particular, we will show that these matrices can all be multiplied by vectors in $\BigO(N\log N)$ time.  We will first examine the new kind of foreground we want to include, resolved point sources, which will also provide a useful example for how the foreground covariances can be quickly multiplied by vectors. 

\subsubsection{Resolved Point Sources} \label{RFast}

Unlike LT, we do not assume that bright point sources have already been cleaned out of our map.  Rather we wish to unify the framework for accounting for both resolved and unresolved foregrounds by inverse covariance weighting. This will allow us to directly calculate how our uncertainties about the fluxes and spectral indices of these point sources affect our ability to measure the 21 cm power spectrum.

In contrast to the unresolved point sources modeled by $\U$, we model $N_R$ bright resolved point sources as having known positions\footnote{If the data cube is not overresolved, this assumption should be pretty good.  If a point source appears to fall in two or more neighboring pixels, it could be modeled as two independent point sources in this framework.  An even better choice would be to include the correlations between the two pixels, which would be quite strong.  Modeling those correlations could only improve the results, since it would represent including additional information about the foregrounds, though it might slow down the method slightly.  Not accounting for position uncertainty will cause the method to underestimate the ``wedge" feature \citep{Dattapowerspec,juddjackiemiguel1,VedanthamWedge,MoralesPSShapes,Trottwedge}.} with different fluxes $S_n$ (at reference frequency $\nu_*$) and spectral indices $\kappa_n$, neither of which is known perfectly.  We assume that resolved point source contributions to $\x$ are uncorrelated with each other, so we can define an individual covariance matrix $\mathbf{R}_n$ for each point source.  This means that our complete model for $\R$ is:
\beq
\mathbf{R} \equiv \sum_n \mathbf{R}_n.
\eeq
 
Following LT, we can express the expected brightness temperature in a given voxel along the line of sight of the $n^\text{th}$ point source by a probability distribution for flux, $p_{S_n}(S')$, and spectral index, $p_{\kappa_n}(\kappa')$, that are both Gaussians with means $S_n$ and $\kappa_n$ and standard deviations $\sigma_{S_n}$ and $\sigma_{\kappa_n}$, respectively.  Following the derivation in LT, this yields:
\begin{eqnarray} \label{spectralIndexAveraging}
\left<x_i\right>_n & = & \delta_{ii_n} (1.4\times 10^{-3}\mbox{K}) \left(\frac{\Omega_{pix}}{1 \mbox{ sr}}\right)^{-1} \eta_i^{-2} \nonumber \times
\\ \nonumber && \int_{-\infty}^{\infty} \left(\frac{S'}{1 \mbox{ Jy}}\right) p_{S_n}(S') dS' \int_{-\infty}^{\infty} \eta_i^{-\kappa'} p_{\kappa_n}(\kappa') d\kappa'
\\ & = & \delta_{ii_n}(1.4\times 10^{-3}\mbox{K}) \left(\frac{S_n}{1 \mbox{ Jy}}\right) \left(\frac{\Omega_{pix}}{1 \mbox{ sr}}\right)^{-1} \times \nonumber
\\ && \eta_i^{-2-\kappa_n} \mbox{exp}\left[\frac{\sigma_{\kappa_n}^2}{2}(\mbox{ln}\eta_i)^2\right],
\end{eqnarray}
where again $\eta_i \equiv \nu_i / \nu_*$.  Here $\delta_{ii_n}$ is a Kronecker delta that forces $\left<x_i\right>$ to be zero anywhere other than the line of sight corresponding to the $n^\text{th}$ resolved point source.  Likewise, we can write down the second moment:
\begin{align}
\left< x_i x_j \right>_n =&\text{ } \delta_{ii_n}\delta_{jj_n}(1.4\times 10^{-3}\text{K})^2 (\eta_i\eta_j)^{-2} \times \nonumber \\
& \left(\int_{-\infty}^{\infty} \left(\frac{S'}{1 \mbox{ Jy}}\right)^2 p_{S_n}(S') dS' \right)\times \nonumber \\ &\left(\frac{\Omega_{pix}}{1 \mbox{ sr}}\right)^{-2} \left(\int_{-\infty}^{\infty} (\eta_i\eta_j)^{-\kappa'} p_{\kappa_n}(\kappa') d\kappa' \right) \nonumber \\
=& \text{ } \delta_{ii_n}\delta_{jj_n}(1.4\times 10^{-3}\text{K})^2 (\eta_i\eta_j)^{-2-\kappa_n} \times \nonumber \\
& \left(\frac{S_n^2 + \sigma^2_{S_n}}{(1\text{ Jy})^2}\right) \left(\frac{\Omega_{pix}}{1 \mbox{ sr}}\right)^{-2} \nonumber \times \\ &\exp\left[ \frac{\sigma^2_{\kappa_n}}{2}(\text{ln} \eta_i \eta_j )^2 \right] 
\end{align}
where we assume $\sigma_{S_n} \approx 5\%$ of $S_n$ and $\sigma_{\kappa_n} \approx 0.2$. 

We know that $\left<x_i\right>_n\left<x_j\right>_n$ can be quickly multiplied by a vector because it is a rank 1 matrix.  Therefore, in order to show that all of $\R$ can be quickly multiplied, we recast $\left< x_i x_j \right>_n$ as the product of matrices that can be multiplied by a vector in $\BigO(N\log N)$ or faster.  If we then ignore the constants and just look at the parts of this matrix that depend on coordinates, we have that:
\begin{align}
\left< x_i x_j \right>_n \propto & \text{ }\delta_{ii_n}\delta_{jj_n} (\eta_i \eta_j)^{-2-\kappa_n} \exp\left[ \frac{\sigma^2_{\kappa_n}}{2}(\text{ln} \eta_i \eta_j )^2 \right] 
 \nonumber \\
= & \text{ } \delta_{ii_n} (\eta_i )^{-2-\kappa_n} \exp\left[ \sigma^2_{\kappa_n}(\text{ln}\eta_i)^2 \right] \times \nonumber \\
& \exp\left[-\frac{\sigma^2_{\kappa_n}}{2}\left(\text{ln}\frac{\eta_i}{\eta_j}\right)^2\right]  \times \nonumber \\
&\delta_{jj_n} (\eta_j )^{-2-\kappa_n}  \exp\left[ \sigma^2_{\kappa_n}(\text{ln}\eta_j)^2 \right] \label{2ndmoment}.
\end{align}
This matrix can be separated into the product of three matrices: one diagonal matrix that only depends on $\eta_i$, an inner matrix that includes the logarithm of a quotient of $\eta_i$ and $\eta_j$ in the exponent, and another diagonal matrix that only depends on $\eta_j$.  The diagonal matrices can be multiplied by a vector in $\BigO(n_z)$.  Moreover, because our cubes have redshift ranges $\Delta z < 0.5$ the frequencies at $i$ and $j$ are never very far apart, we can make the approximation that:
\begin{align}
\text{ln}\left(\frac{\eta_i}{\eta_j}\right) &= \text{ln}\left(\frac{\nu_i}{\nu_j}\right) \nonumber \\ 
&\approx \frac{\nu_0 + \Delta \nu_i}{\nu_0 + \Delta \nu_j} -1 = \frac{1}{\nu_0}(\Delta \nu_i - \Delta \nu_j) \label{freqApprox}
\end{align}
where $\Delta\nu_i \equiv \nu_i - \nu_0$ and $\nu_0$ is a constant reference frequency close to both $\nu_i$ and $\nu_j$.  We choose the center frequency of the data cube to be $\nu_0$. We can see now by combining Equations \ref{2ndmoment} and \ref{freqApprox} that the inner matrix in our decomposition of the second moment depends only on the magnitude of the difference between $\nu_i$ and $\nu_j$.  In the approximation that the physical size of the data cube is small enough that frequencies map linearly to distances, this shows that $\R_n$ respects translational invariance along the line of sight.
 
Because the entries in this inner part of $\R_n$ only depend on differences in frequencies, the inner matrix is a diagonal-constant or ``Toeplitz" matrix.  Toeplitz matrices have the fortuitous property that they can be multiplied by vectors in $\mathcal{O}(N\mbox{log}N)$, as we explain in Appendix \ref{ToeplitzAppendix}.  Therefore, we can multiply $\R_n$ by a vector in $\BigO(n_z \log n_z)$ and we can multiply $\R$ by a vector faster than $\BigO(N \log N)$.

We can understand this result intuitively as a consequence of the fact that the inner part of $\R_n$ is translationally invariant along the line of sight.  Matrices that are translationally invariant in real space are diagonal in Fourier space.  That we need to utilize this trick involving circulant and Toeplitz matrices is a consequence of the fact that our data cube is neither infinite nor periodic.

\subsubsection{Unresolved Point Sources and the Galactic Synchrotron Radiation}\label{Ufast}
Let us now take what we learned in Section \ref{RFast} (and Appendix \ref{ToeplitzAppendix}) to see if $\U$ can also be quickly multiplied by a vector.  Looking back at Equation \ref{Uperp}, we can see that our job is already half finished; $(U_\perp)_{ij}$ only depends on the absolute differences between $(r_\perp)_i$ and $(r_\perp)_j$.  Likewise, we can perform  the exact same trick we employed in Equations \ref{2ndmoment} and \ref{freqApprox} to write down the relevant parts of $(U_\|)_{ij}$ from Equation \ref{Uexact} with the approximation that $\Delta \nu_i$ is always small relative to $\nu_0$:
\beq
(U_\|)_{ij} \propto  \exp\left[-\frac{\sigma^{2}_{\kappa_n}}{2\nu_0^{2}}(\Delta\nu_i - \Delta\nu_j)^{2}\right].
\eeq

In fact, we can decompose $\U$ as a tensor product of three matrices sandwiched between two diagonal matrices:
\beq
\U = \D_\U [\U_x \otimes \U_y \otimes \U_z] \D_\U. \label{formOfU}
\eeq
where all three inner matrices are Toeplitz matrices.  When we wish to multiply $\U$ by a vector, we simply pick out one dimension at a time and multiply every segment of the data by the appropriate Toeplitz matrix (e.g. every line of sight for $\U_z$). All together, the three sets of multiplications can be done in $\BigO(n_x n_y n_f \log n_f) + \BigO(n_x n_f n_y \log n_y) + \BigO(n_y n_f n_x \log n_x) = \BigO(N \log N)$ time.

Moreover, since $\G$ has exactly the same form as $\U$, albeit with different parameters, $\G$ too can be multiplied by a vector in $\BigO(N \log N)$ time by making the same approximation that we made in Equation \ref{freqApprox}.

\subsubsection{Instrumental Noise}\label{Nfast}
Lastly, we return now to the form of $\N$ we introduced in Section \ref{AdrianN}.  To derive a form we combine Equations \ref{NPowerSpectrum} and \ref{Ndef}.  The details are presented in Appendix \ref{NoiseAppendix}, so here we simply state the result:
\beq
\N = \F_\perp^\dagger \widetilde{\N} \F_\perp,
\eeq
where $\F_\perp$ and $\F_\perp^\dagger$ are the unitary discrete 2D Fourier and inverse Fourier transforms and where:
\beq
\widetilde{N}_{lm} =  \frac{\lambda^4 T_\text{sys}^2 j_0^2(k_{x,l} \Delta x /2) j_0^2(k_{y,l} \Delta y /2) }{A^2_\text{ant} (\Omega_\text{pix})^2 n_x n_y\Delta\nu}   \frac{\delta_{lm}}{t_l}.  \label{Nfourier}
\eeq
Here, $A_\text{ant}$ is the effective area of a single antenna, $\Delta\nu$ is the frequency channel width, $l$ and $m$ are indices that index over both $uv$-cells and frequencies, and $t_l$ is the total observation time in a particular $uv$-cell at a particular frequency. 

Because this matrix is diagonal, we have therefore shown that $\N$, along with $\R$, $\U$, and $\G$, can be multiplied by a vector in $\BigO(N \log N)$.  We have summarized the results for all four matrices in Table \ref{fastCovBasis}.
\begin{table*}
	\begin{center}   
	\begin{tabular}{| l | l | l |}
	\hline
   \textbf{Covariance Matrix} & \textbf{Parallel to the Line of Sight } & \textbf{Perpendicular to the Line of Sight} \\ \hline
	$\R$: Resolved Point Sources & Toeplitz symmetry & Diagonal in real space  \\ \hline
	$\U$: Unresolved Point Sources & Toeplitz symmetry & Toeplitz symmetry \\ \hline
	$\G$: Galactic Synchrotron Radiation & Toeplitz symmetry & Toeplitz symmetry \\ \hline
   $\N$: Instrumental Noise & Diagonal in real space & Diagonal in Fourier space \\ \hline
   \end{tabular}
	\caption{Due to the symmetries or approximate symmetries our models for the foreground and noise covariance matrices, they can all be multiplied by a vector in $\BigO(N\log N)$ time or faster. Summarized above are the reasons why each matrix can be quickly multiplied by.  Either the matrices respect translation invariance and thus Toeplitz symmetry, their components are uncorrelated between lines of sight or  frequencies, making them diagonal in real space, or they are uncorrelated and thus diagonal in Fourier space.  In each case, the symmetries rely on the separabiltiy of the modeled covariance matrices into the tensor product of parts parallel or perpendicular to the line of sight. \label{fastCovBasis}}
  	\end{center}
\end{table*}

\subsubsection{Eliminating Unobserved Modes with the Psuedo-Inverse}\label{psuedoinverse}
In our expression for the noise covariance in Equation \ref{Nfourier}, we are faced with the possibility that $t_l$ could be zero for some values of $l$, leading to infinite values of $N_{ij}$.  Fourier modes with $t_l = 0$ correspond to parts of the $uv$-plane that are not observed by the instrument, i.e. to modes containing no cosmological information. We can completely remove these modes by means of the ``psuedo-inverse" \citep{Maxgalaxysurvey1}, which replaces $\C^{-1}$ in the expression $\C^{-1}(\x - \langle \x \rangle)$ and optimally weights all observed modes (this removal can itself be thought of as an optimal weighting---the optimal weight being zero).  The psuedo-inverse involves $\Proj$, a projection matrix ($\Proj^\dagger = \Proj$ and $\Proj^2 = \Proj$) whose eigenvalues are 0 for modes that we want to eliminate and 1 for all other modes. It can be shown \citep{Maxgalaxysurvey1} that the quantity we want to calculate for inverse variance weighting is not $\C^{-1}(\x - \langle \x \rangle)$ but rather the quantity where:
\beq
\C^{-1} \longrightarrow \Proj \left[\Proj \C \Proj + \gamma (\I - \Proj) \right]^{-1} \Proj.
\eeq
In this equation, $\gamma$ can actually be any number other than 0.  The term in brackets in the above equation replaces the eigenvalues of the contaminated modes of $\C$ with $\gamma$.  The outer $\Proj$ matrices then project those modes out after inversion.  In this paper, we take $\gamma = 1$ as the convenient choice for the preconditioner we will develop in Section \ref{FastPrecon}.

The ability to remove unobserved modes is also essential for analyzing real data cubes produced by an interferometer.  Interferometers usually produce so-called ``dirty maps," which are corrected for the effects of the primary beam but have been convolved by the synthesized beam, represented by the matrix $\mathbf{B}$:
\beq
\x_{\text{dirty map}} = \mathbf{B} \x.
\eeq
To compute $\x$ for our quadratic estimator, we need to invert $\mathbf{B}$.  Since the synthesized beam matrix is diagonal in Fourier space, this would be trivial were it not for unobserved baselines that make $\mathbf{B}$ uninvertable.  This can be accomplished with the psuedoinverse as well, since the modes that would have been divided by 0 when inverting $\mathbf{B}$ are precisely the modes that we will project out via the psuedoinverse.  We can therefore comfortably take 
\beq
\x = \F^\dagger_\perp \Proj [\Proj \widetilde{\mathbf{B}} \Proj + \gamma(\I - \Proj)]^{-1} \Proj \F_\perp \x_{\text{dirty map}},
\eeq
where $\mathbf{B} \equiv \F^\dagger_\perp \widetilde{\mathbf{B}} \F_\perp$ and $\widetilde{\mathbf{B}}$ is diagonal.

The psuedo-inverse formalism can be usefully extended to any kind of mode we want to eliminate.  One especially useful application would be to eliminate frequency channels contaminated by radio frequency interference or adversely affected by aliasing or other instrumental issues.

\subsection{Preconditioning for Fast Conjugate Gradient Convergence} \label{FastPrecon}
\begin{figure*} 
	\centering 
	\includegraphics[width=1\textwidth]{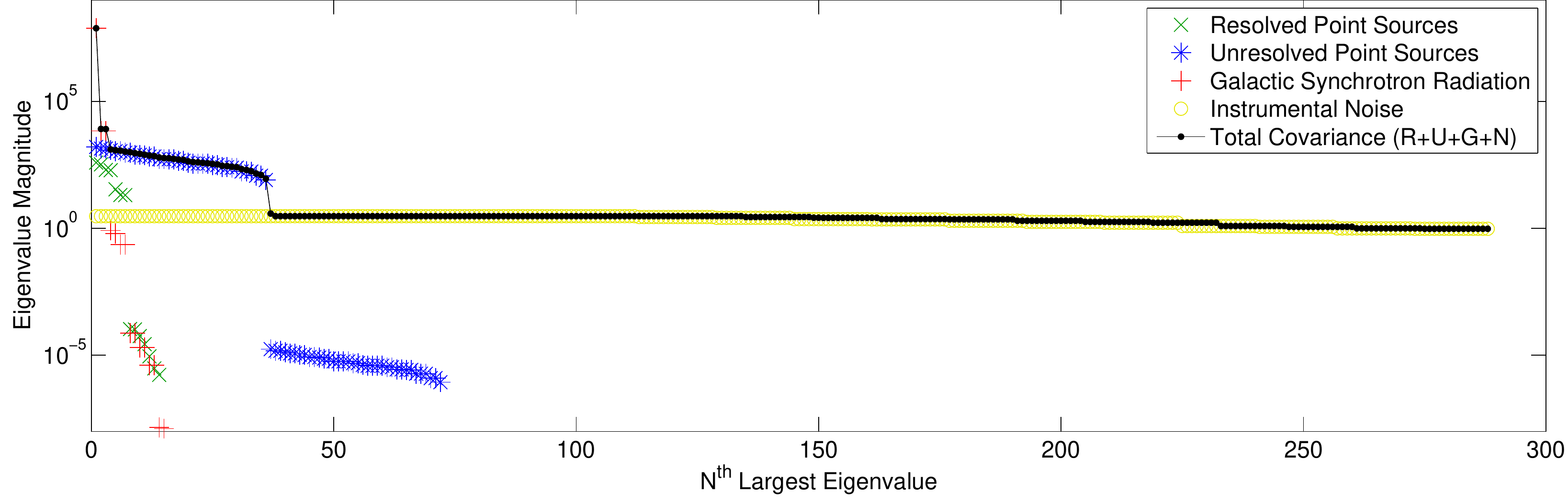}
	\caption{The distinct patterns in the eigenvalue spectrum of our covariance matrix provide an angle of attack for making the calculation of $\C^{-1} (\x - \langle \x \rangle)$ numerically feasible via preconditioning.  The plotted eigenvalue spectra of the covariance for a very small data cube exemplifies many of the important characteristics of the constituent matrices.  First, notice that the noise eigenvalue spectrum, while flatter than any of the others, is not perfectly flat.  The condition number of $N$ is related the ratio of the observing times in the most and least observed cell in the $uv$-plane.   Sometimes this factor can be $10^3$ or $10^4$.  Another important pattern to notice are the fundamental differences between the eigenvalue spectra of $\U$, $\G$, and $\R$.  First off, $\R$ has mostly zero eigenvalues, because $\R$ is a block diagonal matrix with most of its blocks equal to zero.  Second, despite the fact that $\U$ and $\G$ have nearly identical mathematical forms, $\U$ has stair-stepping eigenvalue spectrum while that of $\G$ is a much clearer exponential falloff.  This is due to the much stronger correlations perpendicular to the line of sight in $\G$. }
	\label{noPrecon}
\end{figure*}

We have asserted that the quantity $\mathbf{y} \equiv \C^{-1} (\x - \langle \x \rangle)$ can be estimated quickly using the conjugate gradient method as long as the condition number $\kappa(\C)$ is reasonably small.  Unfortunately, this is never the case for any realistic data cube we might analyze.  In Figure \ref{noPrecon} we plot the eigenvalues of $\C$ and its constituent matrices for a small data cube (only $6\times 6\times 8$ voxels) taken from a larger, more representative volume.  
In this example, $\kappa(\C) \approx 10^8$, which would cause the conjugate gradient method to require tens of thousands of iterations to converge.  This is typical; as we discussed in Section \ref{CGPSE}, values of around $10^{8}$ are to be expected.  We need to do better.

\subsubsection{The Form of the Preconditioner}

The core idea behind ``preconditioning" is to avoid the large value of $\kappa(\C)$ by introducing a pair of preconditioning matrices $\Pre$ and $\Pre^\dagger$.  Instead of solving the linear system $\C\y=(\x - \langle \x \rangle)$, we solve the mathematically equivalent system: 
\beq
\C' \y' = \Pre (\x - \langle \x \rangle),
\eeq
where $\C' \equiv \Pre \C \Pre^\dagger$ and $\y' \equiv (\Pre^\dagger)^{-1}\y$.  If we can compute $\Pre(\x - \langle \x \rangle)$ and, using the conjugate gradient method on $\C'$, we can solve for $\y'$ and thus finally find $\y = \Pre^\dagger \y'$.  If $\Pre$ and $\Pre^\dagger$ are matrices that can be multiplied by quickly and if $\kappa(\C') \ll \kappa(\C)$, then we can greatly speed up our computation of $\y = \C^{-1}(\x - \langle \x \rangle)$.  Our goal is to build up preconditioning matrices specialized to the forms of the constituent matrices of $\C$.  We construct preconditioners for $\C = \N$, generalize them to $\C = \U + \N$, and then finally incorporate $\R$ and $\G$ to build the full preconditioner.  

The result is the following:
\beq
\C' = \F_\perp^\dagger \Pre_\U \Pre_\Gam \Pre_\N (\C) \Pre_\N^\dagger \Pre_\Gam^\dagger \Pre_\U^\dagger \F_\perp.
\eeq
Where $\Pre_\U$, $\Pre_\Gam$ and $\Pre_\N$ and preconditioners for $\U$, $\Gam \equiv \R + \G$, and $\N$ respectively.  A complete and pedagogical explanation of this preconditioner and the motivation for its construction and complex form can be found in Appendix \ref{PreconAppendix}.  The definitions of the matrices can be found in Equations \ref{PUmulti}, \ref{PreGammaMulti}, and \ref{PNdef} respectively.

\begin{figure} 
	\centering 
	\includegraphics[width=.45\textwidth]{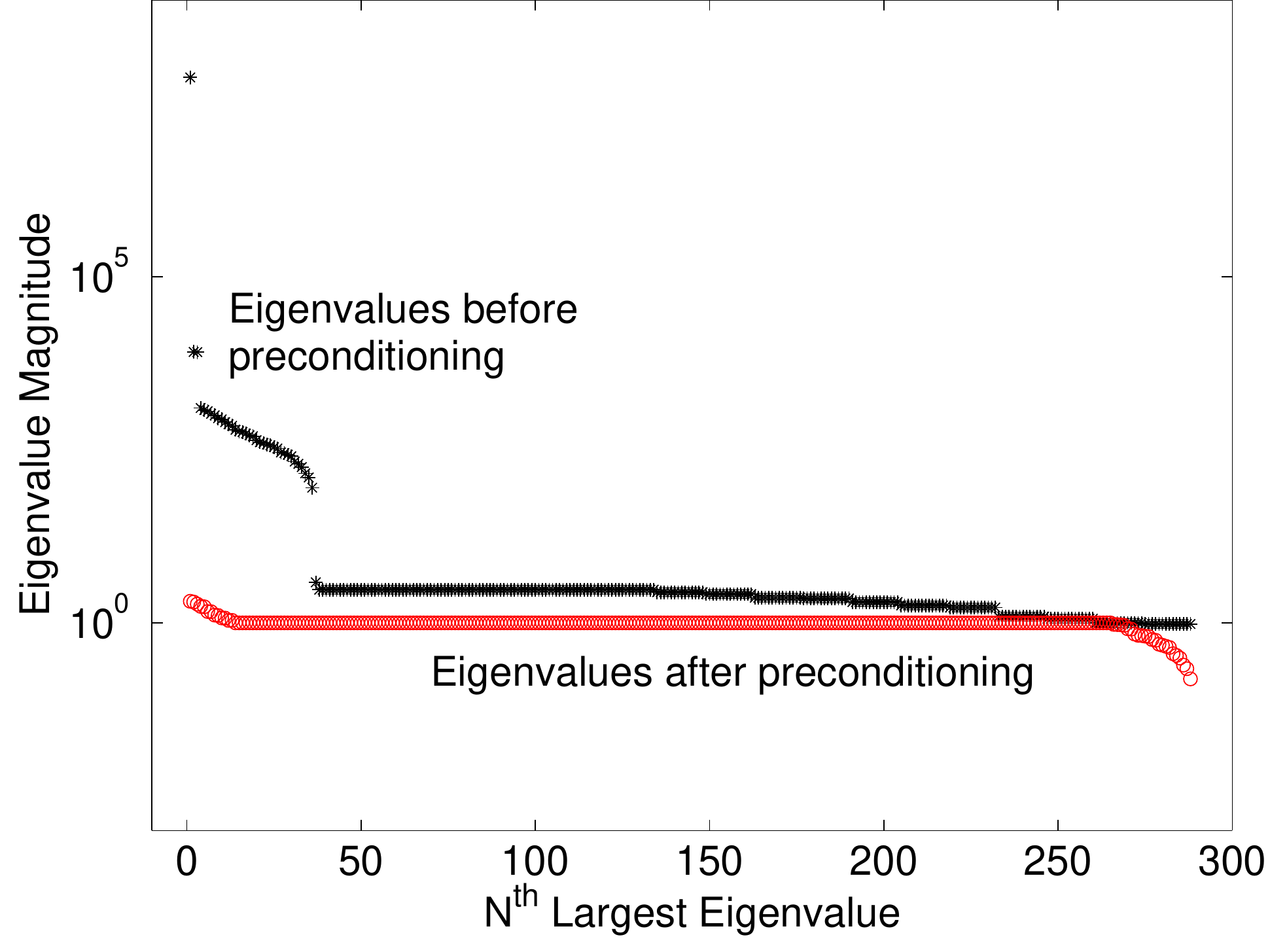}
	\caption{The preconditioner for the conjugate gradient method that we have devised significantly decreases the range of eigenvalues of $\C$.  Our preconditioner attempts to whiten the eigenvalue spectra of the constituent matrices of $\C$ sequentially, first $\N$, then $\R$ and $\G$ together, and finally $\U.$   By preconditioning, the condition number $\kappa(\C)$, the ratio of the largest to smallest eigenvalues, is reduced from over $10^8$ to about $10^1$.}
	\label{preconditionedEigs}
\end{figure}

Despite its complex form and construction, the procedure reduces $\kappa(\C)$ by many orders of magnitude.  In Figure \ref{preconditionedEigs}, in explicit contrast to Figure \ref{noPrecon}, we see a demonstration of that effect.

\subsubsection{Computational Complexity of the Preconditioner} \label{PreconComplexity}
In Appendix \ref{PreconAppendix}, we briefly discuss how the different steps in computing and applying this preconditioner scale with the problem size.  If any of them scale too rapidly with $N$, we can quickly lose the computational advantage of our method over that of LT.\footnote{This section may be difficult to follow without first reading Appendix \ref{PreconAppendix}.  However, the key results can be found in Table \ref{preconComplexity}.}  

First, let us enumerate the complexity of setting up the preconditioner for each matrix.  $\Pre_\N$ requires no setup since it only involves computing powers of the diagonal matrix $\widetilde{\N}$ (see Appendix \ref{PreconNSection}). $\Pre_\U$ requires the eigenvalue decomposition of $\U_z$, the component of $\U$ along the line of sight, which takes $\BigO (n_z^3)$ time (see Appendix \ref{PreconForU}).

We need the eigensystems of $\R$ and $\G$ to compute the eigensystem of $\Gam$ for $\Pre_\Gam$ (see Appendix \ref{PreconForGamma}). $\R$ requires performing one eigenvalue decomposition of an $n_z \times n_z$ matrix for every resolved point source; that takes $\BigO( N_R n_z^3)$ time. $\G$ simply requires three eigenvalue decompositions: one for each matrix like those that appear for $\U$ in Equation \ref{Uouterproduct} whose total outer product is $\G$.  Thus, the complexity is $\BigO(n_x^3) + \BigO(n_y^3) + \BigO(n_z^3)$.  

Next, we need to compute the eigenvalues of $\Gam_{\perp,k}$, the components of $\Gam$ perpendicular to the line of sight corresponding to each of the ``relevant" (i.e. much bigger than the noise floor) eigenvalues of $\Gam$ along the line of sight (see Appendix \ref{PreconForGamma} for a more rigorous definition).  Using the notation we develop in Appendix \ref{PreconForU}, we denote the number of relevant eigenvalues of a matrix $\mathbf{M}$ as $m(\mathbf{M})$.  The number of times we need to decompose an $n_x n_y \times n_x n_y$ matrix is generally equal to the number of relevant eigenvalues of $\G_z$, since the number of relevant eigenvectors is almost always the same for $\G$ and $\R$.  So we have then a computational complexity of $\BigO(m(\G_z)(n_x n_y)^3)$.  Given the limited angular resolution of the experiment and the flat sky approximation, we generally expect $n_x$ and $n_y$ to be a good deal smaller than $n_f$, making this scaling more tolerable.  All these scalings are summarized in Table \ref{preconComplexity}.
\begin{table}
	\begin{center}   
	\begin{tabular}{| l | l |}
	\hline
   \textbf{Operation} & \textbf{Complexity} \\ \hline
	Compute $\U$ eigensystem & $\BigO(n_z^3)$ \\ \hline
	Compute $\G$ eigensystems & $\BigO(n_x^3) + \BigO(n_y^3) + \BigO(n_z^3)$ \\ \hline
	Compute $\R$ eigensystems & $\BigO(N_R n_z^3)$ \\ \hline
	Compute $\Gam$ eigensystems & $\BigO(m(\G_z)(n_x n_y)^3)$ \\ \hline \hline
	Apply $\Pre_\N$ & $\BigO(N\log N)$ \\ \hline
	Apply $\Pre_\U$ & $\BigO(N m(\U_z))$ \\ \hline
	Apply $\Pre_\Gam$ & $\BigO(N m(\G)) + \BigO(N N_R m(\R_n))$ \\ \hline
   \end{tabular}
	\caption{The computational complexity of setting up the preconditioner is, at worst, roughly $\BigO(N^2)$, though this operation only needs to be performed once. Even for large data cubes, this is not the rate-limiting step in power spectrum estimation.    The computational complexity of applying the preconditioner ranges from $\BigO(N \log N)$ to $\BigO(NN_R)$.  For large data cubes with hundreds of bright point sources, the preconditioning time is dominated by $\Pre_\Gam$, which is in turn dominated by preconditioning associated with individual point sources.  The computational complexity of the preconditioner therefore depends on the number of point sources considered ``resolved," which scales with both field of view and with the flux cut.  Here $N_R$ is the number of resolved point sources in our field of view, $n_d$ is the size of the box in voxels along the $d^{\text{th}}$ dimension, and $m$ is the number of relevant eigenvalues of a matrix above the noise floor that need preconditioning.\label{preconComplexity}}
  	\end{center}
\end{table}
     
\begin{figure} 
	\centering 
	\includegraphics[width=.45\textwidth]{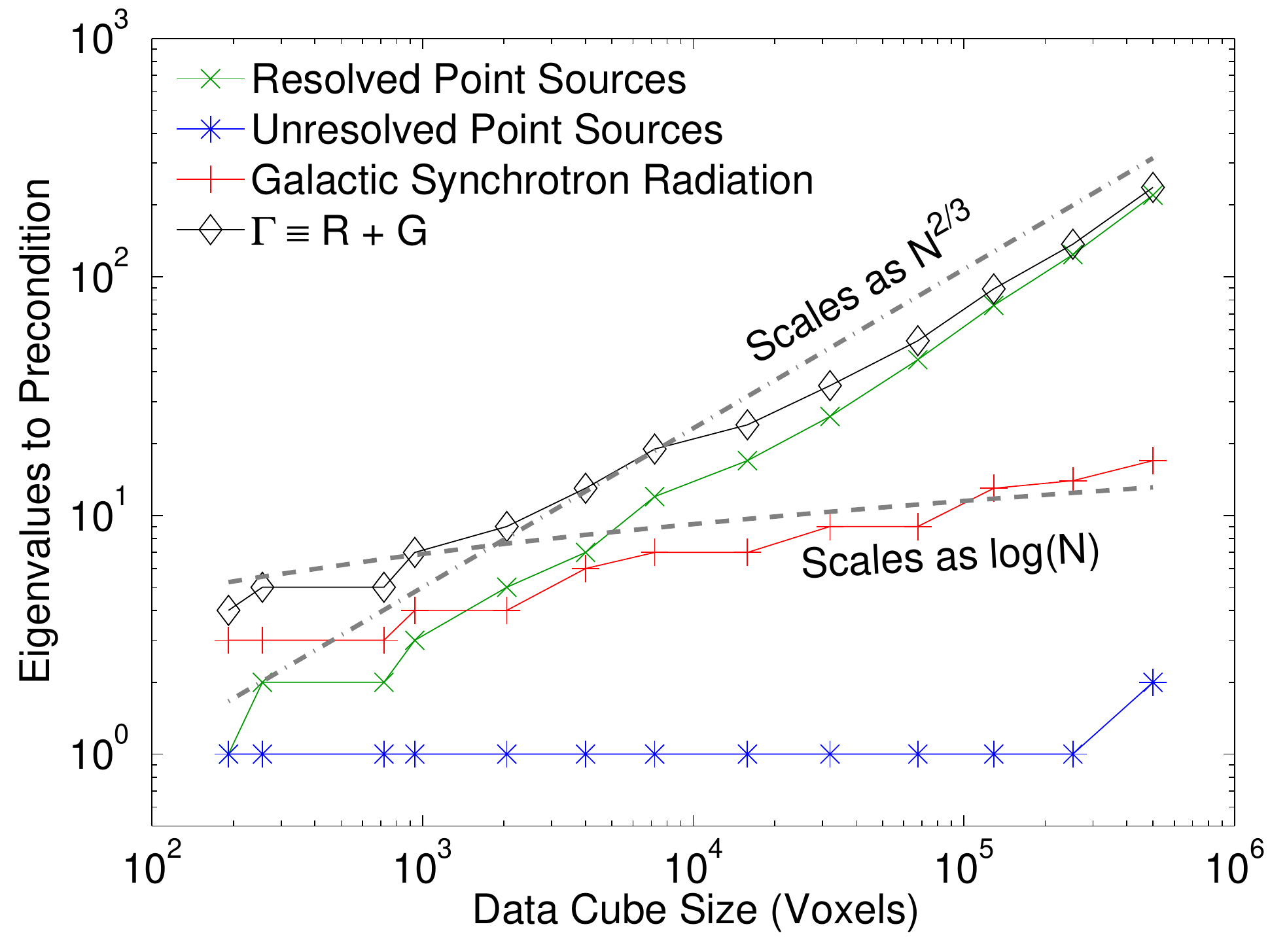}
	\caption{For large data cubes and a fixed definition of what constitutes a ``bright" point source, the complexity of preconditioning is dominated by the number of resolved point sources.  Specifically, the complexity of preconditioning for $\Gamma$ scales as $N^{2/3}$ because the number of resolved point sources is simply proportional to the solid angle of sky surveyed, which scales with the survey volume (and thus number of voxels, assuming fixed angular and frequency resolution) to the $\frac{2}{3}$ power. This also confirms our assertion that the number of important eigenvalues of $\G$ and $\U$ should scale logarithmically with data cube size (albeit with a different prefactor).  Each of the data cubes is taken from the same survey with the same ratio of width to depth.  The number of eigenvalues to precondition is computed assuming an eigenvalue threshold of $\theta = 1$.}
	\label{PreconScaling}
\end{figure}

Until now, all of our complexities have been $\BigO(N \log N)$ or smaller.  Because these small incursions into bigger complexity classes are only part of the set-up cost, they are not intolerably slow as long as $m(\G_z)$ is small. This turns out to be true because the eigenvalue spectra of $\R_n$ and $\G_z$ fall off exponentially, meaning that we expect the number of relevant eigenvalues to grow only logarithmically.  This is borne out in Figure \ref{PreconScaling} where we see exactly how the number of eigenvalues that need to be preconditioned scales with the problem size.

Let us now turn to a far more important scaling: that of multiplying the preconditioner by a vector.  The set-up needs to be done only once per Fisher matrix calculation; the preconditioning needs to happen for every iteration of the conjugate gradient method.  $\Pre_\N$ is the easiest; we only ever need to perform a Fourier transform or multiply by a diagonal matrix.  The complexity is merely $\BigO (N \log N)$.  $\Pre_\U$ only involves multiplying by vectors for each relevant eigenvalue of $\U_z$, so the total complexity is $\BigO(N m(\U_z))$.

Finally, we need to assess the complexity of applying $\Pre_\Gam$.  When performing the eigenvalue decomposition of $\Gam_{\perp,k}$, we expect roughly the same number of eigenvalues to be important that would have been important from $\R$ and $\G$ separately for that $k$ index.  Each of those eigenvectors takes $\BigO(N)$ time to multiply by a vector.  So we expect to deal with $m(\G)$ eigenvalues from $\G$ and one eigenvalue from each resolved point source for each relevant value of $k$, or about $N_R m(R_n)$.  Applying $\Pre_\Gam$ therefore is $\BigO(N m(\G)) + \BigO(N N_R m(\R_n))$.  If we keep the same minimum flux for the definition of a resolved point source and if we scale our cube uniformly in all three spatial directions, then $N_R \propto N^{2/3}$.  

This turns out to be the rate-limiting step in the entire algorithm.  If we decide instead to only consider the brightest $N_R$ to be resolved, regardless of box size, then applying $\Pre_\Gam$ reduces to $\BigO(N\log N)$.  Likewise, if we are only interested in expanding the frequency range of our data cube, the scaling also reduces to $\BigO(N\log N).$  We can comfortably say then that the inclusion of a model for resolved point sources introduces a complexity bounded by $\BigO(N\log N)$ and $\BigO(N^{5/3})$.  We can see the precise computational effect of the preconditioner when we return in Section \ref{CompScaling} to assess the overall scaling of the entire algorithm. These results are also summarized in Table \ref{preconComplexity}.

\subsubsection{Preconditioner Results} \label{PreconResults}
\begin{figure} 
	\centering 
	\includegraphics[width=.45\textwidth]{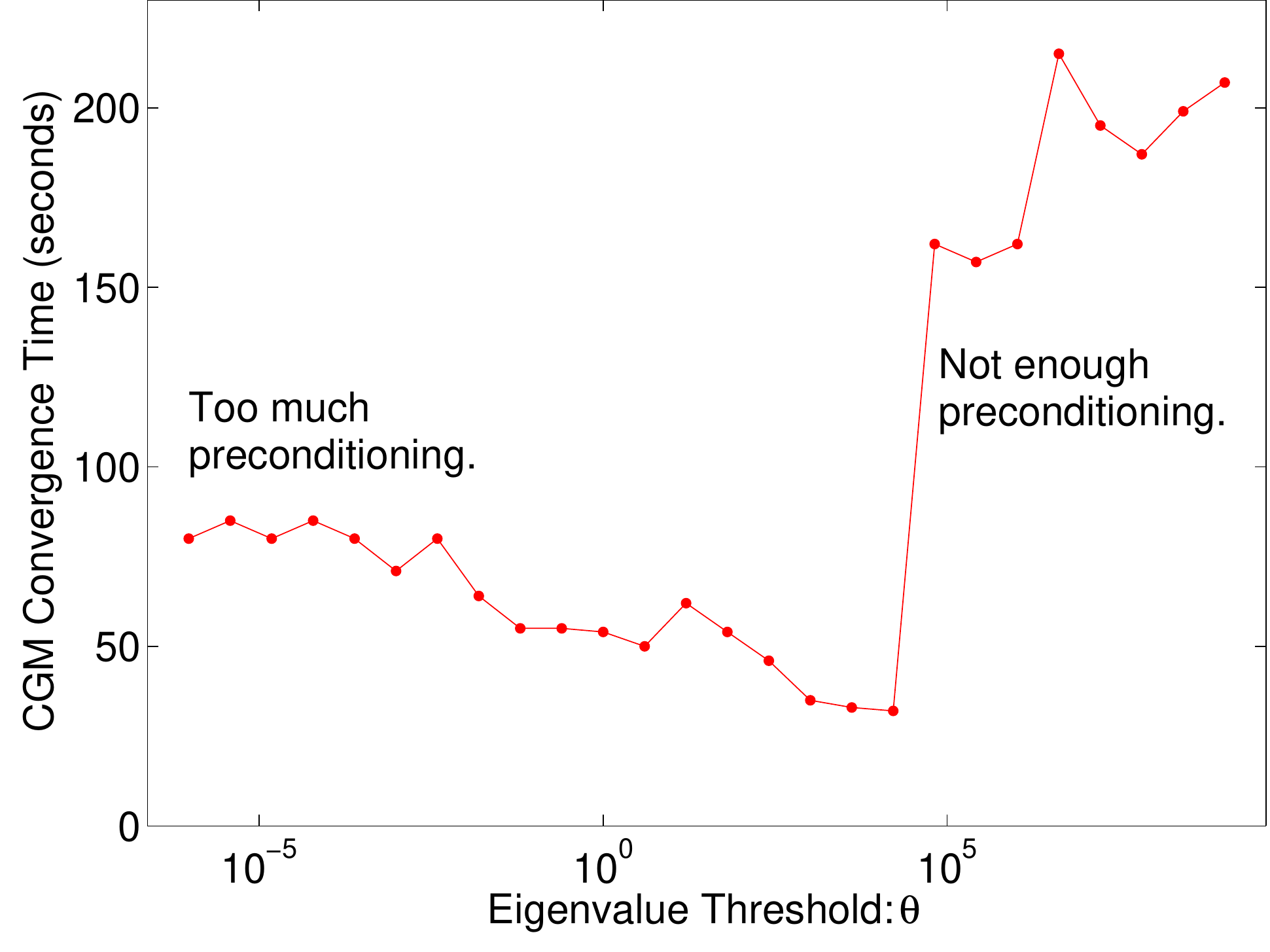}
	\caption{This plot shows how computational time scales with $\theta$, the threshold for preconditioning, for the conjugate gradient method performed on an $N \approx 10^4$ voxel data cube.  It appears that for this particular covariance matrix, a minimum exists near $\theta = 10^4$.  At the minimum, the greater number of conjugate gradient iterations are balanced by quicker individual iterations (since each iteration involves less preconditioning).  We can see from this plot that there exists a critical value of $\theta$ around $5\times 10^4$ where the preconditioning of a small number of additional eigenvalues yields a large effect on the condition number of the resultant matrix.  Without preconditioning, sufficiently large values of $\kappa(\C)$ could also lead to the accumulation of roundoff error that prevents convergence of the conjugate gradient method.}
	\label{PreconPerformance}
\end{figure}

Choosing which eigenvalues are ``relevant" in the constituent matrices of $\C$ and therefore need preconditioning depends on how these eigenvalues compare to the noise floor.  In Appendix \ref{PreconForU}, we define a threshold $\theta$ which distinguishes relevant from irrelevant eigenvalues by comparing them to $\theta$ times the noise floor.  Properly choosing a value for $\theta$, the threshold below which we do not precondition eigenvalues of $\Ubar$ and $\GammaBar$, presents a tradeoff.  We expect that that too low of a value of $\theta$ will precondition inconsequential eigenvalues, thus increasing the conjugate gradient convergence time.  We also expect that too large of a value of $\theta$ will leave some of the most important eigenvalues without any preconditioning, vastly increasing convergence time.  Both of these expectations are borne out by our numerical experiments, which we present in Figure \ref{PreconPerformance}.

In this work, we choose $\theta = 1$ (all foreground eigenvalues above the noise floor are preconditioned) for simplicity and to be sure that we are not skipping the preconditioning of any important foreground eigenvalues.  One might also worry that more iterations of the algorithm provides more opportunity for round-off error to accumulate and prevent convergence, as has sometimes proven the case in our numerical experiments.  For lengthy or repeated calculations of the Fisher matrix, it is wise to explore the performance of several levels of preconditioning, especially if it can garner us a another factor of 2 in speed.

\subsection{Fast Simulation of Foregrounds and Noise} \label{random}
We concluded Section \ref{fisherMC} with the fact that a Monte Carlo calculation of the Fisher matrix required the ability to compute $\widehat{\mathbf{q}}$ from many different realizations of the foregrounds and noise modeled by $\C$.  In Sections \ref{fastPSE} through \ref{FastPrecon}, we have shown how to quickly calculate $\widehat{\mathbf{q}}$ from a data vector $\x$ using Equation \ref{estimator}.

But where does $\x$ come from?  When we want to estimate the 21 cm temperature power spectrum of our universe, $\x$ will come from data cubes generated from real observations.  But in order to calculate $\F$, which is essential both to measuring $\widehat{\mathbf{p}}$ and estimating the error on that measurement, we must first be able to create many realizations of $\x$ drawn from our models for noise and foregrounds that we presented in Section \ref{FastCov}.

A mathematically simple way to draw $\x$ from the right covariance matrix is to create a vector $\mathbf{n}$ of independent and identically distributed random numbers drawn from a normal distribution with mean 0 and standard deviation 1.  Then, it is easy to see that 
\beq
\x \equiv \C^{1/2} \mathbf{n} \label{CovHalfPower}
\eeq
is a random vector with mean $\mathbf{0}$ and covariance $\C$.  Unfortunately, computing $\C^{1/2}$ is just as computationally difficult as computing $\C^{-1}$.

In this last section of our presentation of our fast method for power spectrum estimation and statistics, we will explain how a vector can be created randomly with covariance $\C$.  We do so by creating vectors randomly from each constituent matrix of $\C$, since each contribution to the measured signal is uncorrelated.  In Section \ref{randomResults}, we will demonstrate numerically that these simulations can be performed quickly while still being accurately described by the underlying statistics.

\subsubsection{Resolved Point Sources} \label{Rrandom}
The simplest model to reproduce in a simulation is the one for resolved point sources, because the covariance was created from a supposed probability distribution over their true fluxes and spectral indices.  We start with a list of point sources with positions and with a specified but uncertain fluxes and spectral indices.  These fluxes can either come from a simulation, in which case we draw them from our source count distribution (Equation \ref{sourceCounts}) and spectral indices from a Gaussian distribution, or from a real catalog of sources with its attendant error bars.  The list of sources does not change over the course of calculating the Fisher matrix. 

In either case, calculating a random $\x_\R$ requires only picking two numbers, a flux and a spectral index, for each point source and then calculating a temperature in each voxel along that particular line of sight.  The latter is easy, since we assume it is drawn from a Gaussian.  The former can be quickly accomplished by numerically calculating the cumulative probability distribution from Equation \ref{sourceCounts} and inverting it.  Each random $\x_\R$ is therefore calculable in $\BigO(N_R n_z) < \BigO(N)$ time.

\subsubsection{Unresolved Point Sources} \label{Urandom}
We next focus on $\U$, which is more difficult.  Our goal is to quickly produce a vector with specified mean and covariance.  LT has already established what value we want for $\left<\x_\U\right>$ and $\left<\x_\G\right>$ with a calculation very similar to Equation \ref{spectralIndexAveraging}.  We need to figure out how to produce a vector with zero mean and the correct covariance.

One way around the problem of calculating $\C^{1/2}$ is to take advantage of the eigenvalue decomposition of the covariance matrix.  That is because if $\C = \mathbf{Q\Lambda Q}^\trans$, where $\mathbf{Q}$ is the matrix that transforms into the eigenbasis and $\mathbf{\Lambda}$ is a diagonal matrix made up of the eigenvalues, then $\C^{1/2} = \mathbf{Q\Lambda}^{1/2}\mathbf{Q}^\trans$.    We already found the few important eigenvalues of $\U$ for our preconditioner (see Section \ref{PreconForU}), so does this technique solve our problem?

Yes and no.  In the direction parallel to the line of sight, this technique works exceedingly well because only a small number of eigenvectors correspond to non-negligible eigenvalues.  We can, to very good approximation, ignore all but the largest eigenvalues (which correspond to the first few ``steps" in Figure \ref{noPrecon}.)  We can therefore generate random unresolved point source lines of sight in $\BigO(n_z m(\U_z))$ with the right covariance.

A problem arises, however, when we want to generate $\x_\U$ with the proper correlations perpendicular to the line of sight.  Unlike the extremely strong correlations parallel to the line of sight, these correlations are quite weak.  Weak correlations entail many comparable eigenvalues; in the limit that point sources were uncorrelated, $\U_x \otimes \U_y \rightarrow \I_{\perp}$ and all the eigenvalues would be 1 (though the eigenvectors would of course be much simpler too).  Utilizing the same technique as above would require a total complexity of $\BigO(N n_x n_y)$ time, which is slower than we would like.
 
However, the fact that both $\U_x$ and $\U_y$ are Toeplitz matrices allows us to use the same sort of trick we employed to multiply our Toeplitz matrices by vectors in Section \ref{RFast} to draw random vectors from $\U_x \otimes \U_y$ \citep{ToeplitzSimulation}.  It turns out that the circulant matrix in which we embed our covariance matrix must be positive-semidefinite for this technique to work.  Although there exists such an embedding for any Gaussian covariance matrix, only Gaussians with coherence lengths small compared to the box size can be embedded in a reasonably small circulant matrix---exactly the situation we find ourselves in with $\U_\perp$.  As such, we can generate random $\x_\U$ vectors in $\BigO(N m(\U_z) \log(n_x n_y)) \approx \BigO(N \log N)$. 

\subsubsection{Galactic Synchrotron Radiation} \label{Grandom}
The matrix $\G$ differs from $\U$ primarily in the coherence length perpendicular to the line of sight.  Unlike $\U$, $\G$ has only a small handful of important eigenvalues, which means that random $\x_\G$ vectors can be generated in the same way we create line of sight components for $\x_\U$ vectors, which we described above.  Since $m(\G)$ is so small (see Figure \ref{noPrecon}) and grows so slowly with data cube size (see Figure \ref{PreconScaling}), we can create random $\x_\G$ vectors in approximately $\BigO(N)$.

\subsubsection{Instrumental Noise} \label{Nrandom}
Finally, we turn to $\N$, which is also mathematically simple to simulate.  First off, $\left< \x_\N \right> = 0$.  Next, because $\N$ is diagonal in the Fourier basis, we can simply use Equation \ref{CovHalfPower}.  Because $\N = \F_\perp^\dagger \widetilde{\N} \F_\perp$,
\beq
\N^{1/2} = \F_\perp^\dagger \widetilde{\N}^{1/2} \F_\perp,
\eeq
which is computationally easy to multiply by $\mathbf{n}$ because $\widetilde{\N}$ is a diagonal matrix.  The most computationally intensive step in creating random $\x_\N$-vectors is the fast Fourier transform, which of course scales as $\BigO(N\log N)$.

\subsubsection{Data Simulation Speed and Accuracy} \label{randomResults}
\begin{figure} 
	\centering 
	\includegraphics[width=.45\textwidth]{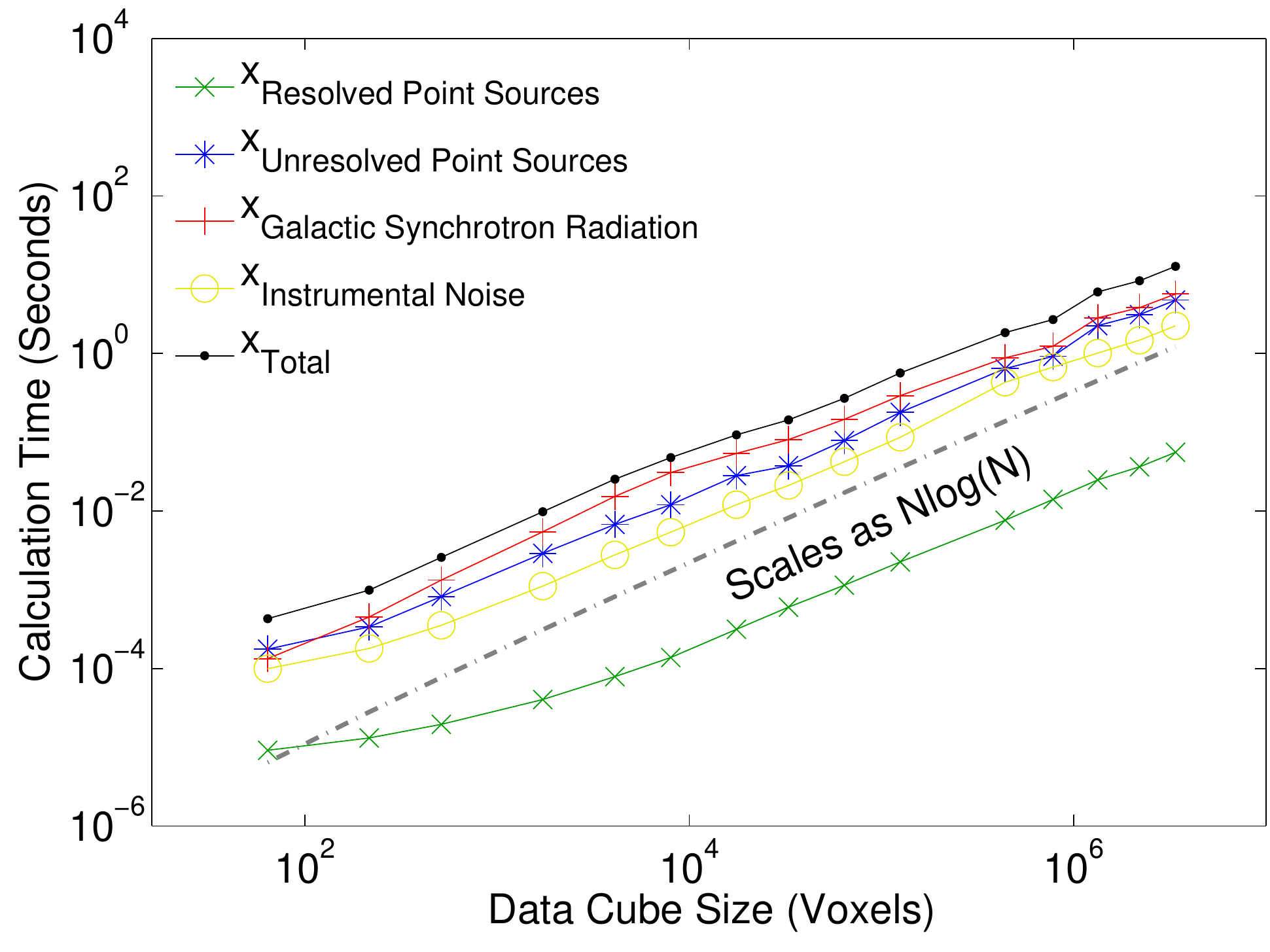}
	\caption{In order to estimate the Fisher matrix via a Monte Carlo, we need to draw random data cubes from our modeled covariance.  Here we show that we can do so in $\BigO(N \log N)$ by plotting computational time as a function of problem size for generating a random $\x$ for each of the constituent sources of $\x$.  In practice, generating random $\x$ vectors is never the rate-limiting step in calculating $\F$.}
	\label{RandomFieldScaling}
\end{figure}

Before we conclude this section and move on to the results of our method as a whole, we verify what we have claimed in the above sections: namely that we can quickly generate data cubes with the correct covariance properties.  Figure \ref{RandomFieldScaling} verifies the speed, showing that the algorithm is both fast and well-behaved for large data cubes.

\begin{figure} 
	\centering 
	\includegraphics[width=.45\textwidth]{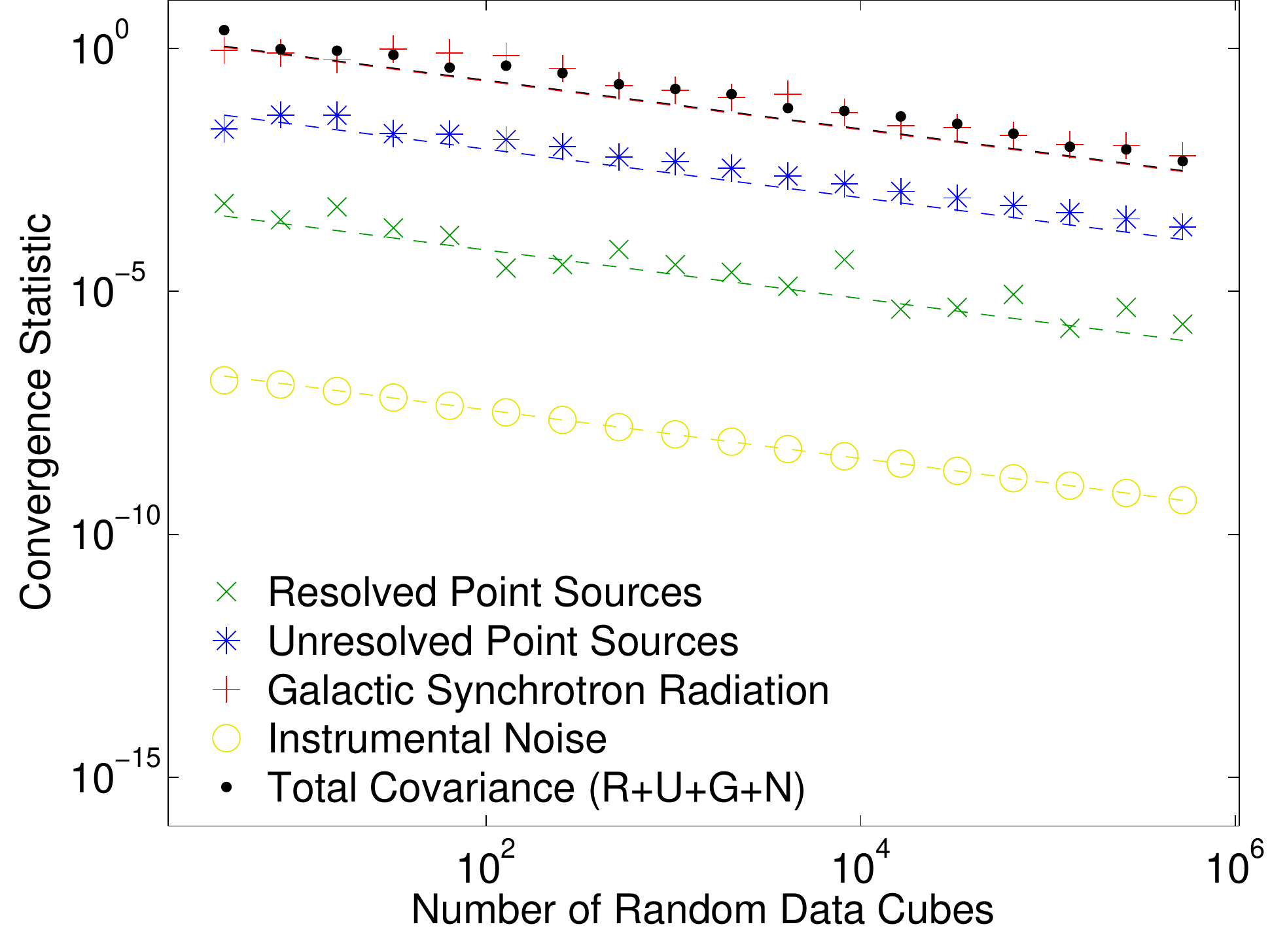}
	\caption{We verify that our technique for quickly generating random data cubes actually reproduces the correct statistics by generating a large number of such cubes and calculating their sample covariances.  Plotted here is the error statistic detailed in Equation \ref{errorStatisic}.  The color-matched dotted lines are the expected convergences for correlated Gaussians from Equation \ref{expectedConvergence}.}
	\label{RandomFieldConvergence}
\end{figure}

In order to show that the sample covariance of a large number of random $\x$ vectors converges to the appropriate covariance matrix, we must first define a convergence statistic, $\varepsilon$.  We are interested in how well the matrix converges relative to the total covariance matrix $\C$.  For example, for $\R$ we choose:
\beq
\varepsilon(\widehat{\R}) \equiv \sqrt{\frac{\sum_{ij}\left|\widehat{R}_{ij} - R_{ij}\right|^2}{\sum_{ij}\left|C_{ij}\right|^2}} \label{errorStatisic}
\eeq
where $\widehat{\R}$ is the sample covariance of $n$ random $\x_\R$ vectors drawn from $\R$.  If each $\x$ is a Gaussian random vector then the expected RMS value of $\varepsilon$ is:
\beq
\sqrt{\left< \varepsilon(\widehat{\R})^2 \right>} = \frac{1}{\sqrt{n}}\left[\frac{\sum_{i,j}R_{ij}^2 + (\text{tr} \R)^2}{\sum_{ij}C_{ij}^2}  \right]. \label{expectedConvergence}
\eeq
In Figure \ref{RandomFieldConvergence}, we see that all four constituent matrices of $\C$ converge like $n^{-1/2}$, as expected, with very nearly the prefactor predicted by Equation \ref{expectedConvergence}.  We can be confident, therefore, in both the speed and accuracy of our technique for generating random vectors.

\subsection{Method Accuracy and Convergence} \label{fisherConverge}
Before we move on to discuss some of the results of our method, it is worthwhile to check that no unwarranted approximations prevent it from  converging to the exact form of the Fisher information matrix in Equation \ref{fisherTrace}.  Since calculating $\F$ exactly can only be done in $\BigO(N^3)$ time, we perform this test in two parts.
 
\begin{figure} 
	\centering 
	\includegraphics[width=.45\textwidth]{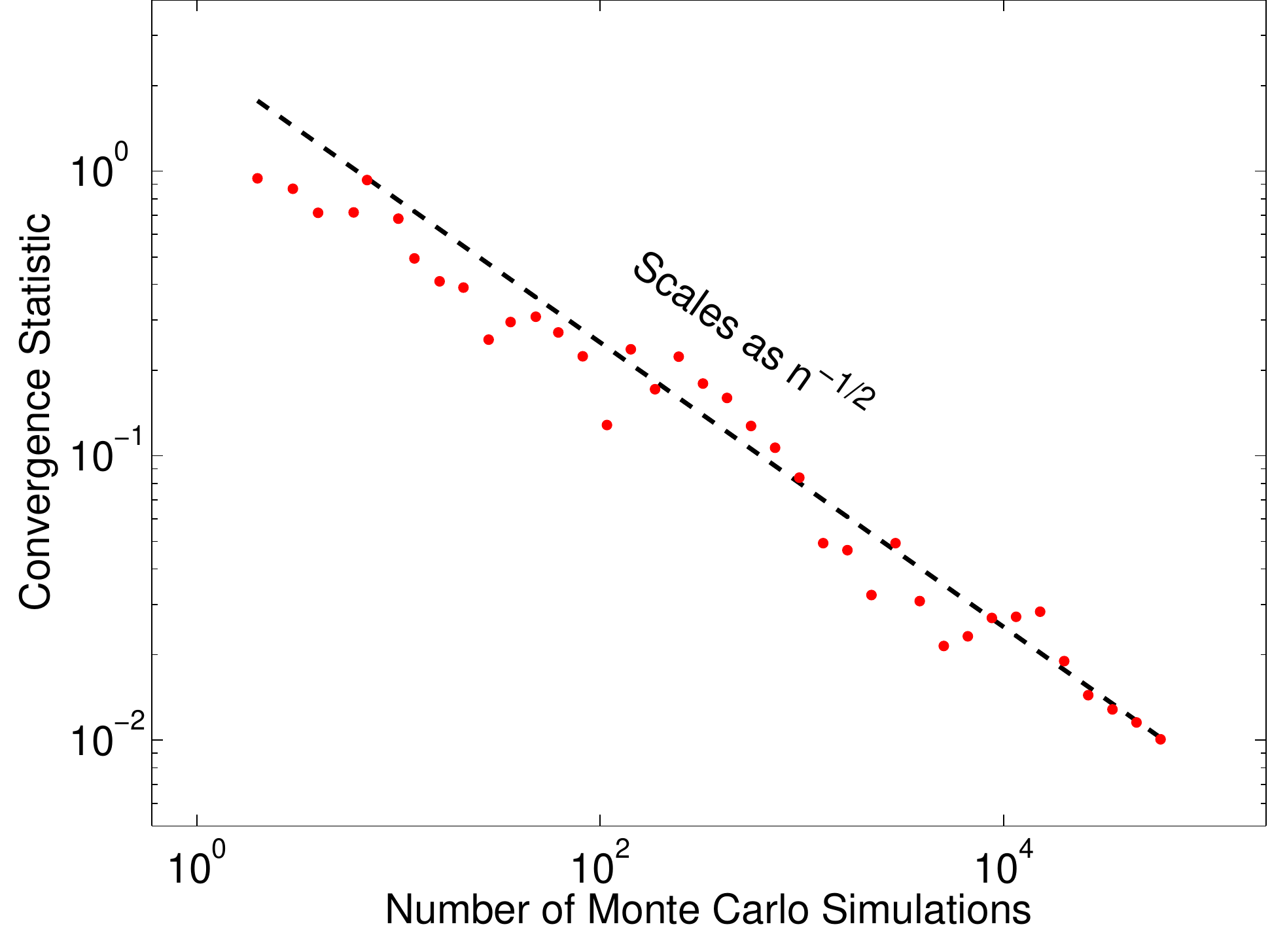}
	\caption{Our Monte Carlo converges to the correct Fisher matrix as $n^{-1/2}$, as expected.  In this plot, we compare the sample covariance of many $\widehat{\mathbf{q}}$ vectors generated from small data cubes to an exact calculation of $\F$ by calculating the relative error of their diagonals.}
	\label{MCFisherConvergence}
\end{figure}

First, we measure convergence to the exact Fisher matrix for a very small data cube with only $6\times 6 \times 8$ voxels.  Taking advantage of Equation \ref{covq}, we generate an estimate of $\F$, which we call $\widehat{\F}$, from the sample covariance of many independent $\widehat{\mathbf{q}}$ vectors.  We compare these $\widehat{\F}$, which we calculate periodically along the course of the Monte Carlo, with the $\F$ that we calculated directly using Equation \ref{fisherTrace}.  As we show in Figure \ref{MCFisherConvergence}, the sample covariance of our $\widehat{\mathbf{q}}$ vectors clearly follows the expected $n^{-1/2}$ convergence to the correct result.

\begin{figure} 
	\centering 
	\includegraphics[width=.45\textwidth]{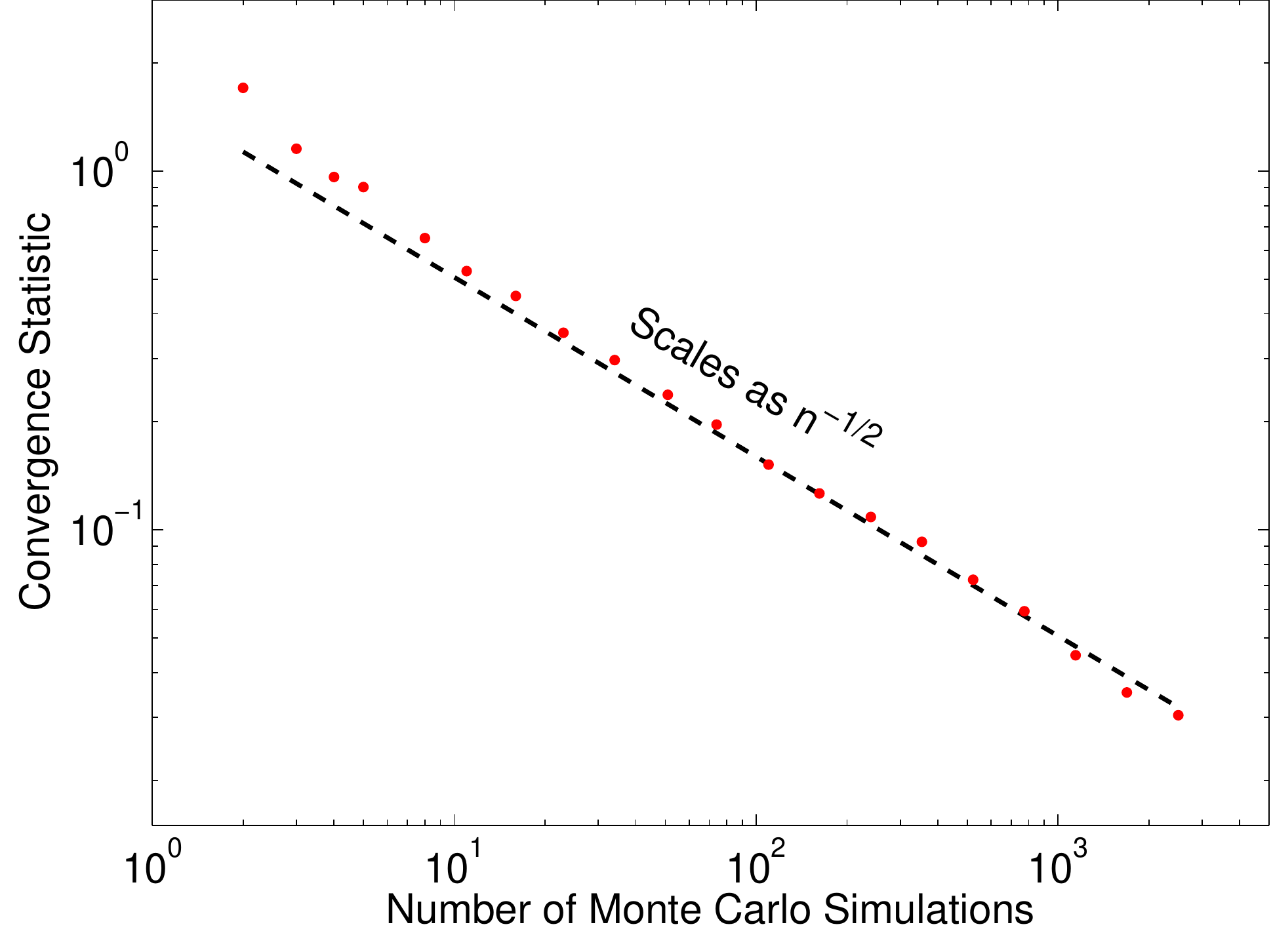}
	\caption{For large data cubes, the convergence of the sample covariance of our $\widehat{\mathbf{q}}$ vectors to $\F$ also nicely follows the expected $n^{-1/2}$ scaling.  We perform this analysis on a data cube with $1.5\times10^5$ voxels analogously to that which we performed in Figure \ref{MCFisherConvergence}, except that we use the sample covariance of double the number of Monte Carlo iterations as our ``true" Fisher matrix.  This explains the artificially fast convergence we see in the last few points of the above plot.}
	\label{MCFisherConvergence2}
\end{figure}

However, we are more concerned with the accuracy of the method for large data cubes which cannot be tackled by the LT method.  Unfortunately, for such large data cubes, we cannot directly verify our result except in the case where $\C = \I$.  In concert with other tests for agreement  with LT, we also check that the method does indeed converge as $n^{-1/2}$ by comparing the convergence of subsets of the $\widehat{q}^\alpha$ vectors up to $n/2$ Monte Carlo iterations to the reference Fisher matrix, which we take to be the sample covariance of all $n$ iterations.  As we show in Figure \ref{MCFisherConvergence2}, our expectation is borne out numerically.

\subsection{Method Summary}

We have constructed a technique that accelerates the LT technique to $\BigO(N\log N)$ and extends it to include bright point sources in, at worst, $\BigO(N^{5/3})$.  We do so by generating random data vectors with the modeled foreground and noise covariances and calculating the Fisher information matrix via Monte Carlo.  We are able to calculate individual inverse variance weighted power spectrum estimates quickly using the conjugate gradient method with a specially adapted preconditioner.  

Our method makes a number of assumptions, most of which are not shared with the LT method.  Our method can analyze larger data sets but at a slight loss of generality. Although we have mentioned these assumptions throughout this work, it is useful to summarize them in one place:

\begin{itemize}
\item Our method relies on a small enough data cube perpendicular to the line of sight that it can be approximated as rectilinear (see Figure \ref{flatsky}).  

\item We approximate the natural log of the quotient of frequencies in the exponent of our point source covariance matrix by a leading-order Taylor expansion (Equation \ref{freqApprox}).  This assumption makes the foreground covariances translationally invariant along the line of sight and thus amenable to fast multiplication using Toeplitz matrix techniques.  This is a justified assumption as long as the coherence length of the foregrounds is much longer than the size of the box along the line of sight. 

\item Our ability to precondition our covariance matrix for the conjugate gradient method depends on the approximation that the correlation length of $\U$ perpendicular to the line of sight, due to weak spatial clustering of point sources, is not much bigger than the pixel size. For the purposes of preconditioning, we approximate $\U_\perp$ to be the identity (see Section \ref{PreconForU}).  The longer the correlation length of $\U_\perp$, the longer the conjugate gradient algorithm will take to converge.

\item Likewise, the speed of the preconditioned conjugate gradient algorithm depends on the similarities of the eigenmodes of the covariances for  $\R$, $\U$, and $\G$ along the line of sight.  The more similar the eigenmodes are (though their accompanying eigenvalues can be quite different) the more the preconditioning algorithm can reduce the condition number of $\C$.  We believe that this similarity is a fairly general property of models for foregrounds, though the introduction of a radically different foreground model might require a different preconditioning scheme.

\item We assume that the number of Monte Carlo iterations needed to estimate the Fisher information matrix is not so large that that it precludes analyzing large data cubes.  Because the process of generating more artificial $\widehat{\mathbf{q}}$ vectors is trivially parallelizable, we do not expect getting down to the requisite precision on the window functions to be an insurmountable barrier. 
\end{itemize}

One common theme among these assumptions, especially the last three, is that the approximations we made to speed up the algorithm can be relaxed as long as we are willing to accept longer runtimes.  This reflects the flexibility of the method, which can trade off speed for accuracy and vice versa.


\section{Results}\label{results}
Now that we are confident that our method can accurately estimate the Fisher information matrix and can therefore calculate both power spectrum estimates from data and the attendant error bars and window functions, we turn to the first end-to-end results of the algorithm.  In this Section, we demonstrate the power our method and the improvements that it offers over that of LT.  First, in Section \ref{CompToLT11} we show that our technique reproduces the results of LT in the regions of Fourier space where they overlap.  Then in Section \ref{CovMatBuildup}, we highlight the improvements stemming from novel aspects of our method, especially the inclusion of the pixelization factor $|\Phi(\kv)|^2$ in $\Q^\alpha$ and $\N$ and the separation of point sources into resolved point sources ($\R$) and unresolved point sources ($\U$), by showing how different parts of our algorithm affect $\F$.  In Section \ref{CompScaling} we examine just how much faster our algorithm is than that of LT, and lastly, in Section \ref{MWA} we forecast the cosmological constraining power of the 128-tile deployment of the MWA.

\subsection{Comparison to Liu \& Tegmark} \label{CompToLT11}

\begin{figure*}
\centering 
	\includegraphics[width=1\textwidth]{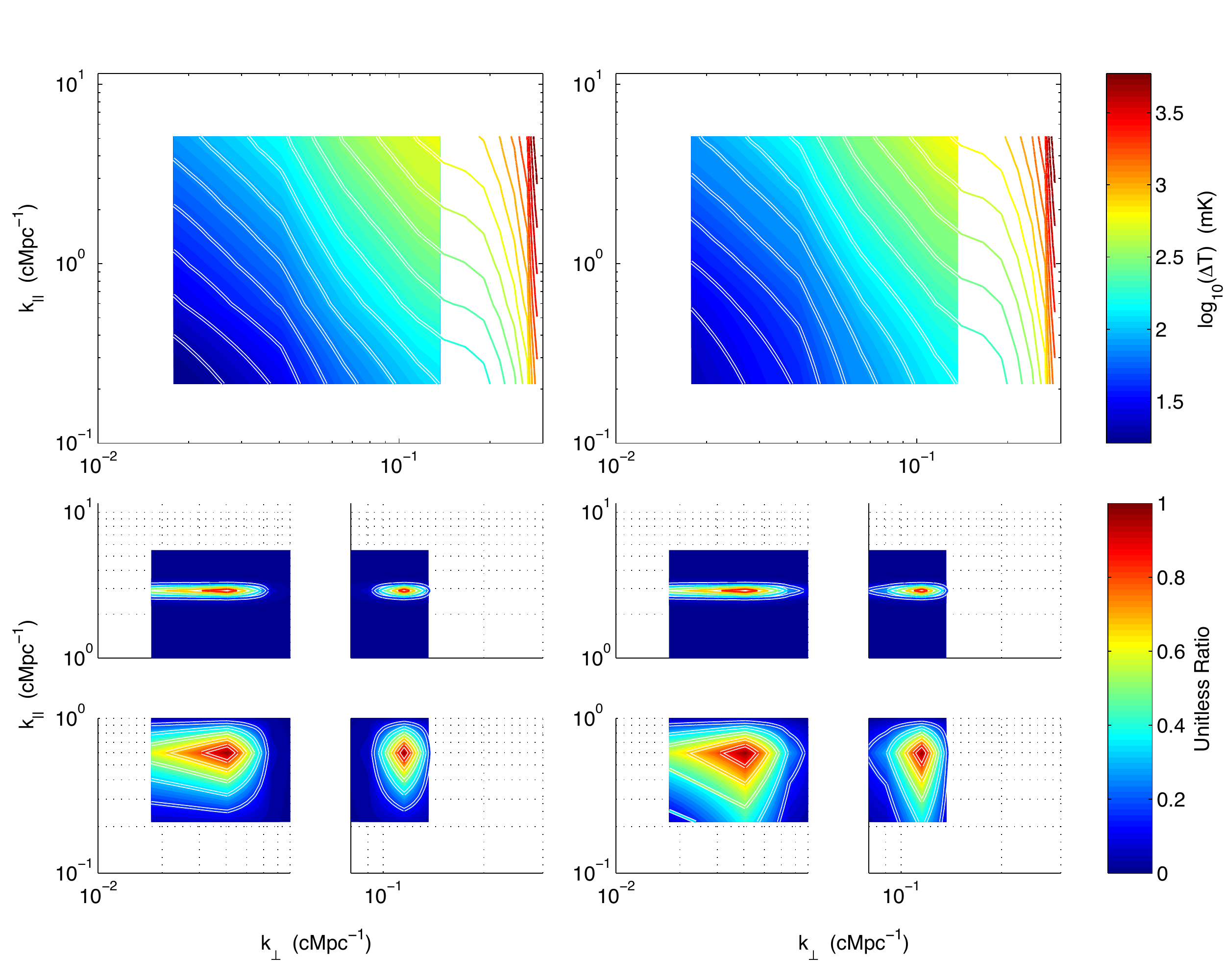}
	\caption{Our method faithfully reproduces the results of LT in the regions of Fourier space accessible to both methods.  Here we recreate both the vertical error bar contours from LT's Figure 8 (top two panels) and a few selected window functions from LT's Figure 2 (bottom two groups of panels).  The shaded regions represent the LT results; the white-outlined, colored contours are overplotted to show our results. Both are on the same color scale.  Following LT, we have plotted both the case without foregrounds ($\C = \N$, left two panels) and the case with foregrounds ($\C = \N + \U + \G$, right two panels), which allows us to get a sense for the effects of foregrounds on our power spectrum estimates, error bars, and window functions. }  
	\label{ComparisonToAdrian}
\end{figure*}

First we want to verify that our method reproduces that of LT in the regions of Fourier space accessible to both methods.  Figure \ref{ComparisonToAdrian} provides an explicit comparison to LT's Figures 2 and 8.  These plots show the shaded regions representing their method and over-plotted, white-outlined contours representing ours.  Both are on the same color scale.  These plots show error bars in temperature units in $k_\perp$-$k_\|$ space and a selection of window functions in both the case where $\C=\N$ and $\C = \U + \G + \N$.  They are generated from the same survey geometry with identical foreground and noise parameters.  In the regions where the methods overlap, we see very good agreement between the two methods.

In addition to the modes shown in the shaded regions in Figure \ref{ComparisonToAdrian}, the LT method can access Fourier modes longer than the box size, which we cannot.  This is no great loss---these modes are poorly constrained by a single data cube.  Moreover, they are generally those most contaminated by foregrounds; the low-$k_\perp$ modes will see heavy galactic synchrotron contamination while the low-$k_\|$ modes will be contaminated by types of the foregrounds.  We imagine that very low-$k_\perp$ Fourier modes, those that depend on correlations between data cubes that cannot be joined without violating the flat-sky approximation, will still be analyzed by the LT method.  Because our method can handle many more voxels, it excels in measuring both medium and high-$k$ modes that require high spectral and spatial resolution.

\subsection{Novel Effects on the Fisher Matrix} \label{CovMatBuildup}

\begin{figure*}	\centering 
	\includegraphics[width=.95\textwidth]{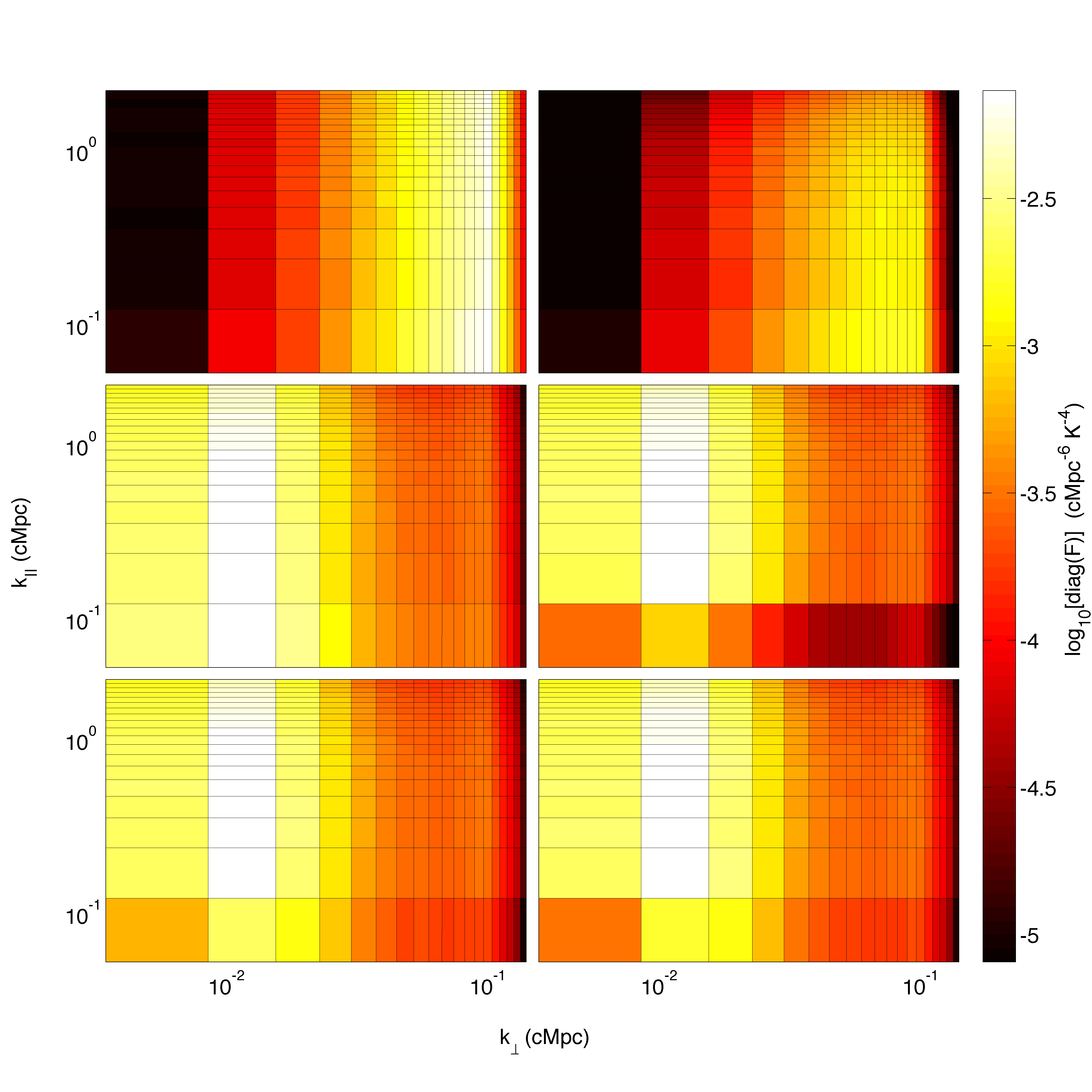}
	\caption{The Fisher information matrix provides a useful window into understanding the challenges presented to measuring the 21 cm power spectrum by the various contaminants.  In this figure, we add these effects one by one, reading from left to right and top to bottom, to see how the diagonal of the Fisher matrix (expanded along the $k_\perp$ and $k_\|$ directions), is affected.  Brighter regions represent, roughly speaking, more information and thus smaller error bars (in power spectrum units).  We comment in more detail on each panel individually in Section \ref{CovMatBuildup}, including upon the advantages of the novel aspects of our technique. \emph{Top left:} the covariance matrix is taken to be the identity and pixelization effects on $\Q^\alpha$ are ignored.  \emph{Top right:} the pixelization factor $|\Phi(\kv)|^2$ is included and not set to $1$.  \emph{Middle left:} the noise expected from 1000 hours of observation with the MWA 128-tile configuration is included. \emph{Middle right:} all point sources (up to 200 Jy) are modeled as unresolved; all information about their positions is ignored.  \emph{Bottom left:} resolved point sources are included in the model, with all point sources dimmer than 100 mJy considered unresolved. \emph{Bottom right:} in addition to bright point sources, galactic synchrotron is also included. }\label{fisher_buildup} 

\end{figure*}

A simple way to understand the different effects that our forms of $\C$ and $\Q^\alpha$ have on the Fisher information matrix, especially the novel inclusions of $\R$ and $|\Phi(\kv)|^2$, is to build up the Fisher matrix component by component.  In Figure \ref{fisher_buildup} we do precisely that by plotting the diagonal elements of $\F$.  These diagonal elements are related to the vertical error bars on our band powers.  Large values of $F^{\alpha\alpha}$ correspond to band powers about which we have more information.

In the top two panels of Figure \ref{fisher_buildup}, we show the first novel effect that our method takes into account.  In them, we can see how modeling the finite size of our voxels affects the information available in the case where $\C = \I$ (the color scale for these two panels only is arbitrary).  In the top left panel, we have set $|\Phi(\kv)|^2 = 1$, which corresponds to the delta function pixelization of LT.  We see that the amount of information depends only on $k_\perp$.  This is purely a binning effect: our bins in $k_\perp$ are concentric circles with constant increments in radius; higher values of $k_\perp$ incorporate more volume in Fourier space, except at the high-$k_\perp$ edge where the circles are large enough to only include the corners of the data cube. In the top right panel, we see that including $|\Phi(\kv)|^2\neq 1$ affects our ability to measure high-$k$ modes, which depends increasingly on our real space resolution and is limited by the finite size of our voxels.  

In the middle left panel, we now set $\C = \N$. In comparison to $\C = \I$, the new covariance matrix (and thus new vector $\x$ for the Monte Carlo calculation of $\F$), shifts the region of highest information to a much lower value of $k_\perp$.  Though there are fewer Fourier modes that sample this region, there are far more baselines in the array configuration at the corresponding baseline length. Our noise covariance is calculated according to our derivation in Section \ref{Nfast} for 1000 hours of observation with the 128-tile deployment of the MWA \citep{MWAstatus}.

We next expand to $\C = \U + \N$ for the middle right panel, where we have classified all point sources as ``unresolved."  In other words, we take $S_\text{cut}$ in Equation \ref{I2fluxintegral} to be large (we choose 200 Jy, which is representative of some of the brightest sources at our frequencies).  As we expect, smooth spectrum contamination reduces our ability to measure power spectrum modes with low values of $k_\|$.  This is because of the exponentially decaying eigenvalue spectrum of $\U_\|$, most of which is smaller than the eigenvalues of $\N$. The effect is seen across $k_\perp$ because the characteristic clustering scale of unresolved point sources is smaller than the pixel size; localized structure in real space corresponds to unlocalized power in Fourier space.

\begin{figure*}	\centering 
	\includegraphics[width=.95\textwidth]{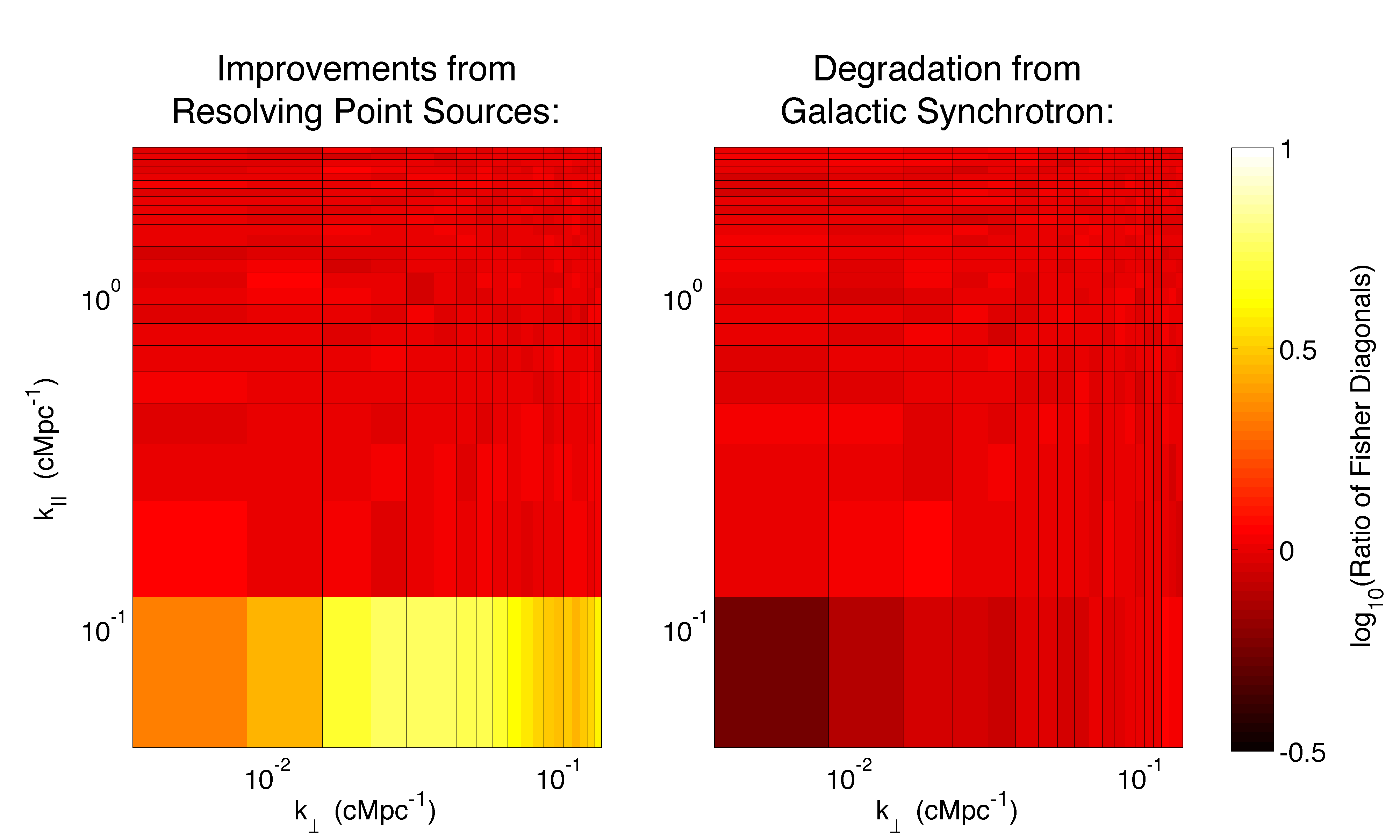} 
	\caption{By comparing the Fisher matrices arising from the various covariance models we explore in Section \ref{CovMatBuildup} and Figure \ref{fisher_buildup}, the precise improvements are brought into sharper focus.  In the left panel, we can see the information we gain by explicity including resolved point sources.  Shown here is the ratio of the bottom left panel of Figure \ref{fisher_buildup} to the middle right panel.  By taking into account precise position, flux, and spectral information and uncertainties, we improve our ability to measure the power spectrum at the longest scales parallel to the line of sight, effectively ameliorating the effects of foregrounds.  In the right panel, we see the remarkably small effect that the galactic synchrotron radiation has on our abilty  the measure the 21 cm power spectrum.  Shown here is the ratio of the bottom right panel of Figure \ref{fisher_buildup} to the botton left.  Because we take spatial information into account, the strong spatial and spectral coherence of the signal from our Galaxy is confined to the bottom left corner of the $k_\|$-$k_\perp$ plane.} \label{fisherRatios} 
\end{figure*}

In the bottom left panel, we have included information about the positions of roughly 200 resolved point sources above 100 mJy, with random fluxes drawn from our source count distribution (Equation \ref{sourceCounts}) and random spectral indices drawn from a Gaussian centered on $\kappa_n = 0.5$ with a width of $0.15$.  By doing this, we reduce $S_\text{cut}$ in our model for $\U$ down to 100 mJy.  Including all this extra information---positions, fluxes, flux uncertainties, spectral indices, and spectral index uncertainties---provides us with significantly more Fisher information at low-$k_\|$ where foregrounds dominate and thus smaller errors on those modes.  Additionally, by incorporating resolved point sources as part of our inverse covariance weighting, we no longer have to worry about forward propagating errors from any external point-source subtraction scheme. In the left panel of Figure \ref{fisherRatios} we see the ratio of this panel to the middle right panel.  

Finally, in the bottom right panel of Figure \ref{fisher_buildup} we show the effect of including Galactic synchrotron radiation.  Adding $\G$ has the expected effect; we already know that $\G$ has only a few important eigenmodes which correspond roughly to the lowest Fourier modes both parallel and perpendicular to the line of sight.  As a result, we only see a noticeable effect in the bottom left corner of the $k_\perp$-$k_\|$ plane; we include the ratio of the two figures in the right panel of Figure \ref{fisherRatios} for clarity.  Otherwise, our Galaxy has very little effect on the regions of interest.  In fact, the similarity between the this panel and the middle left panel tells us something very striking: in the regions of Fourier space that our data most readily probes, foregrounds (once properly downweighted) should not prove an insurmountable obstacle to power spectrum estimation. 

The set of plots in Figure \ref{fisher_buildup} is useful for developing a heuristic understanding of how noise and foregrounds affect the regions in which we can most accurately estimate the 21 cm power spectrum.  With it, we can more easily identify the precise regions of $k$-space that we expect to be minimally contaminated by foregrounds and can thus tailor our instruments and our observations to the task of measuring the 21 cm power spectrum.  

\subsection{Computational Scaling of the Method} \label{CompScaling}
\begin{figure}  
	\centering 
	\includegraphics[width=.5\textwidth]{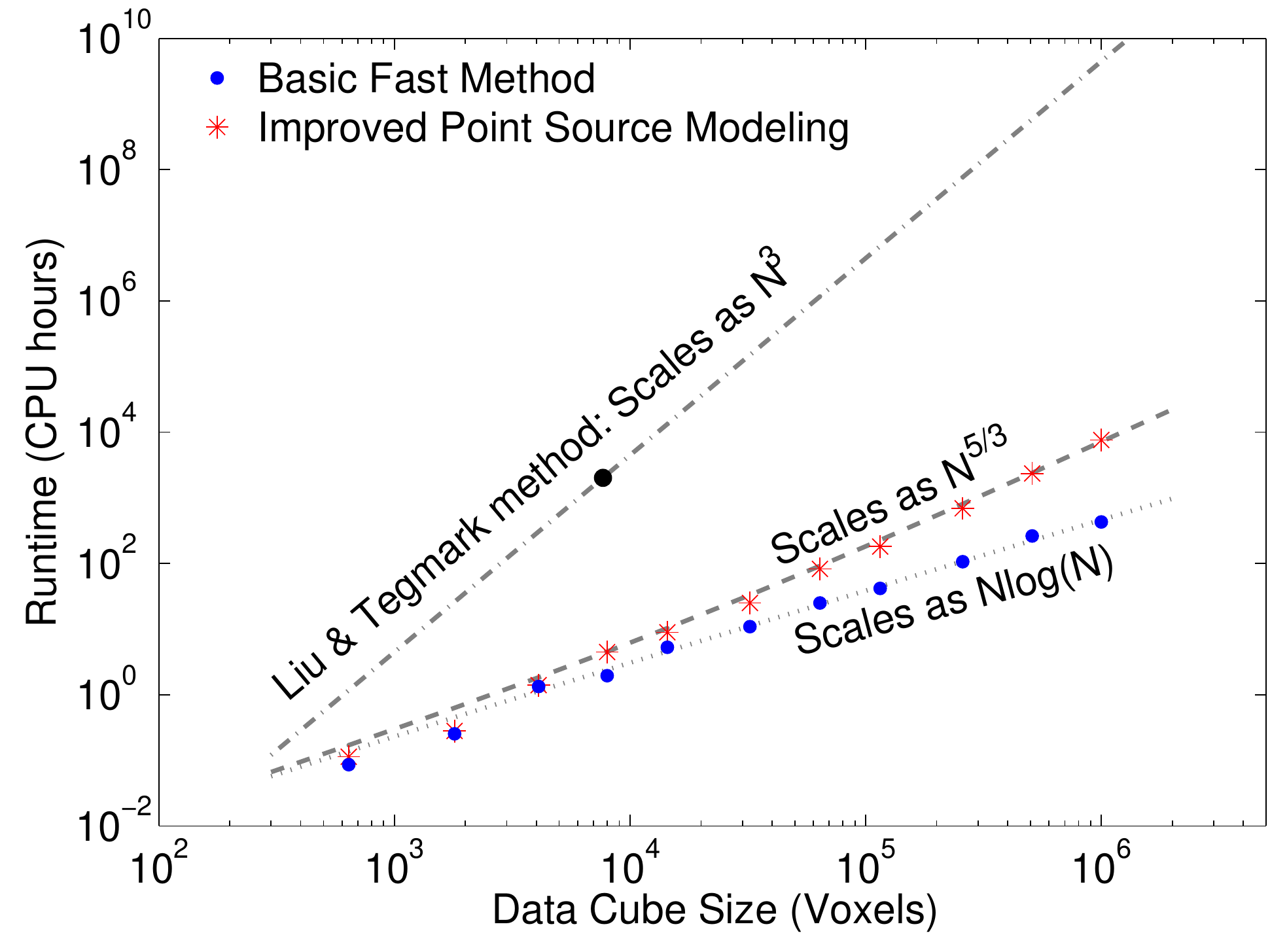}
	\caption{Our algorithm scales with the number of voxels, $N$, as $\BigO(N \log N)$ in the best case and as $\BigO(N^{5/3})$ in the worst, depending on the treatment of bright point sources.  If we choose to ignore all information about the location, brightness, and spectra of bright point sources, we can estimate the power spectrum in $\BigO(N \log N)$.  If we choose to take into account this extra information, the algorithmic complexity increases to $\BigO(N N_R)$, where $N_R$ is the number of bright, resolved point sources.  For a fixed minimum flux for ``bright" sources, this leads to $\BigO(N^{5/3})$ complexity for uniform scaling in all three dimensions. Both scenarios represent a major improvement over the LT method, which scales as $\BigO(N^3)$.}  
	\label{ComputationalTimeResults} 
\end{figure}

Now that we understand how our technique works, we want to also see that it works as quickly as promised by and achieves the desired computational speed up over the LT method.  Specifically, we want to show that we can achieve the theoretical $\BigO(N \log N)$ performance in reproducing the results of LT.  We also want to better understand the computational cost of including resolved point sources so as to compare that cost to benefits outlined in Section \ref{CovMatBuildup}. We have therefore tested the algorithm's speed for a wide range of data cube sizes; we present the results of that study in Figure \ref{ComputationalTimeResults}.

In this figure, we show the combined setup and runtime for power spectrum estimates including 1000 Monte Carlo simulations of $\widehat{\mathbf{q}}$ for estimating the Fisher matrix on a single modern CPU.  For each successive trial, we scale the box by the same ratio in all three dimensions.  Because we maintain a fixed flux cut, increasing the linear size of the box by a factor of two increases the number of resolved point sources in the box by a factor of 4 and the number of voxels by a factor of 8.  With any more than a few point sources, the computational cost becomes dominated by point sources, leading to an overall complexity of $\BigO(N N_R)$.  In this case, the largest data cubes include about 400 point sources over a field of about 50 square degrees, accounting for about 15\% of the lines of sight in the data cube.  

For any given analysis project, these exists a trade-off between including additional astrophysical information into the analysis and the computational complexity of that analysis; at some point the marginal cost of a slower algorithm exceeds marginal benefit of including more bright point sources.  It is beyond the scope of this paper to prescribe a precise rubric for where to draw the line between resolved and unresolved point sources.  However, we can confidently say that the algorithm runs no slower than $\BigO(N^{5/3})$ and can often run at or near $\BigO(N \log N)$ if only the brightest few point sources are treated individually. 

\subsection{Implications for Future Surveys}\label{MWA}
Though the primary purpose of this paper is to describe an efficient method for 21 cm power spectrum analysis, our technique enables us immediately to make predictions about the potential performance of upcoming instruments.  In this section we put all our new machinery to work in order to see just how well the upcoming 128-tile deployment of the MWA can perform.

\begin{figure} 
	\centering 
	\includegraphics[width=.48\textwidth]{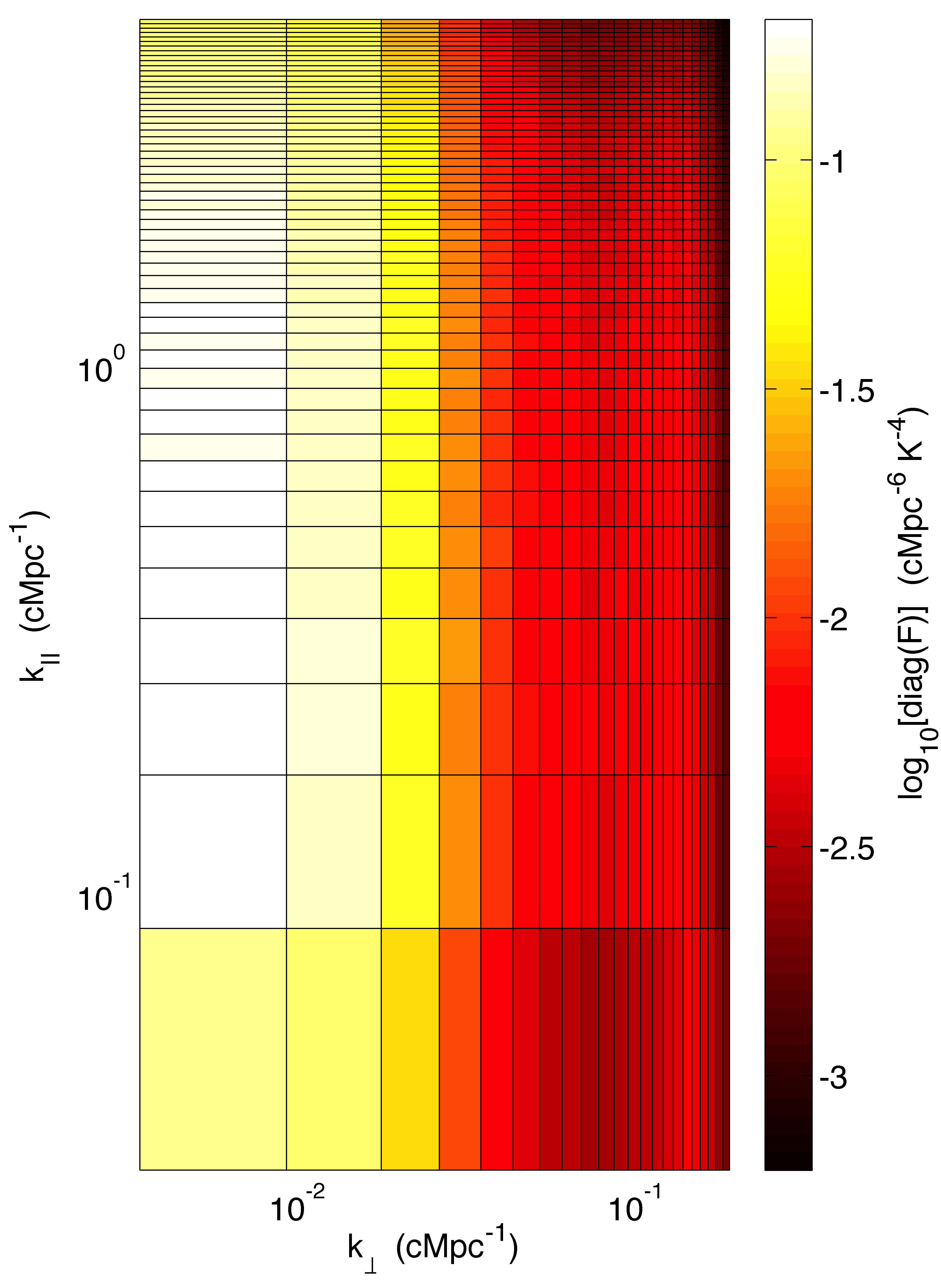}
	\caption{The diagonal of the Fisher matrix predicted for 1000 hours of observation with the MWA with 128 tiles shows the region of power spectrum space least contaminated by noise and foregrounds.  Noise, and thus array layout, dominates the shape of the region of maximum information, creating a large, vertical region at a value of $k_\perp$ corresponding to the typical separation between antennas in the compact core of the array.  The contaminating effects of the foregrounds are clearly visible at low-$k_\|$.}
	\label{MWAFisher}
\end{figure}

\begin{figure} 
	\centering 
	\includegraphics[width=.48\textwidth]{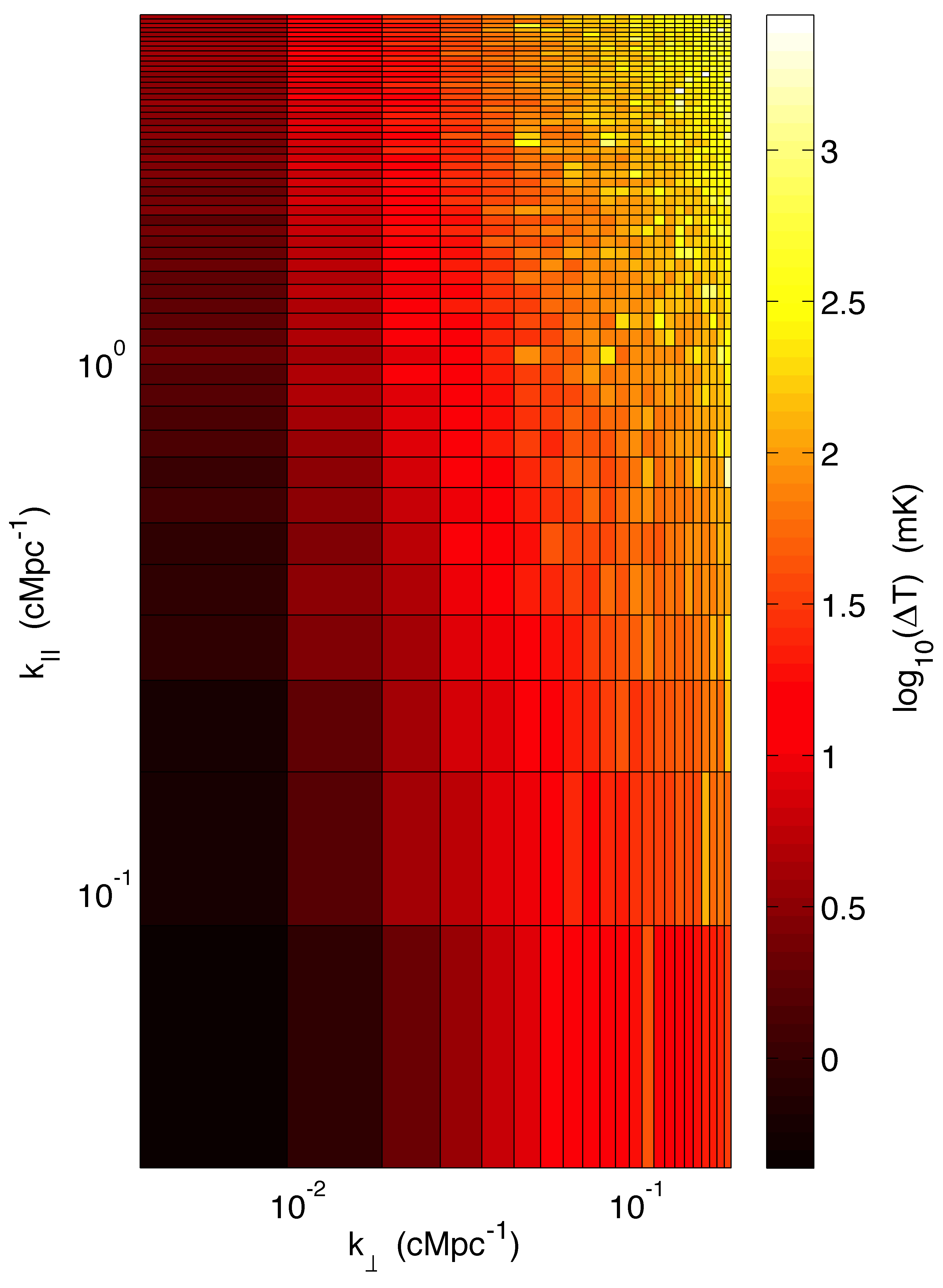}
	\caption{The expected error bars in temperature units on decorrelated estimates of the power spectrum highlight a sizable region of $k$-space where we expect to be able to use the MWA with 128 tiles to detect a fiducial 10 mK signal with a signal to noise ratio greater than 1.  Perhaps surprisingly, the smallest error bars are still on the smallest $k$ modes acessible by our method, though some of them are contaminated by large foregrounds.  This is because our conversion to temperature units includes a factor of $(k_\perp^2 k_\|)^{1/2}$, which accounts for the difference between this Figure and Figure \ref{MWAFisher}.  From the shape of the region of smallest error, we can better appreciate the extent to which noise and our array layout determines where in $k$-space we might expect to be able to detect the EoR.  The noisiness at high-$k$ is due to Monte Carlo noise and can be improved with more CPU hours.} 
	\label{MWAerror}
\end{figure}

\begin{figure*} 
	\centering 
	\includegraphics[width=.7\textwidth]{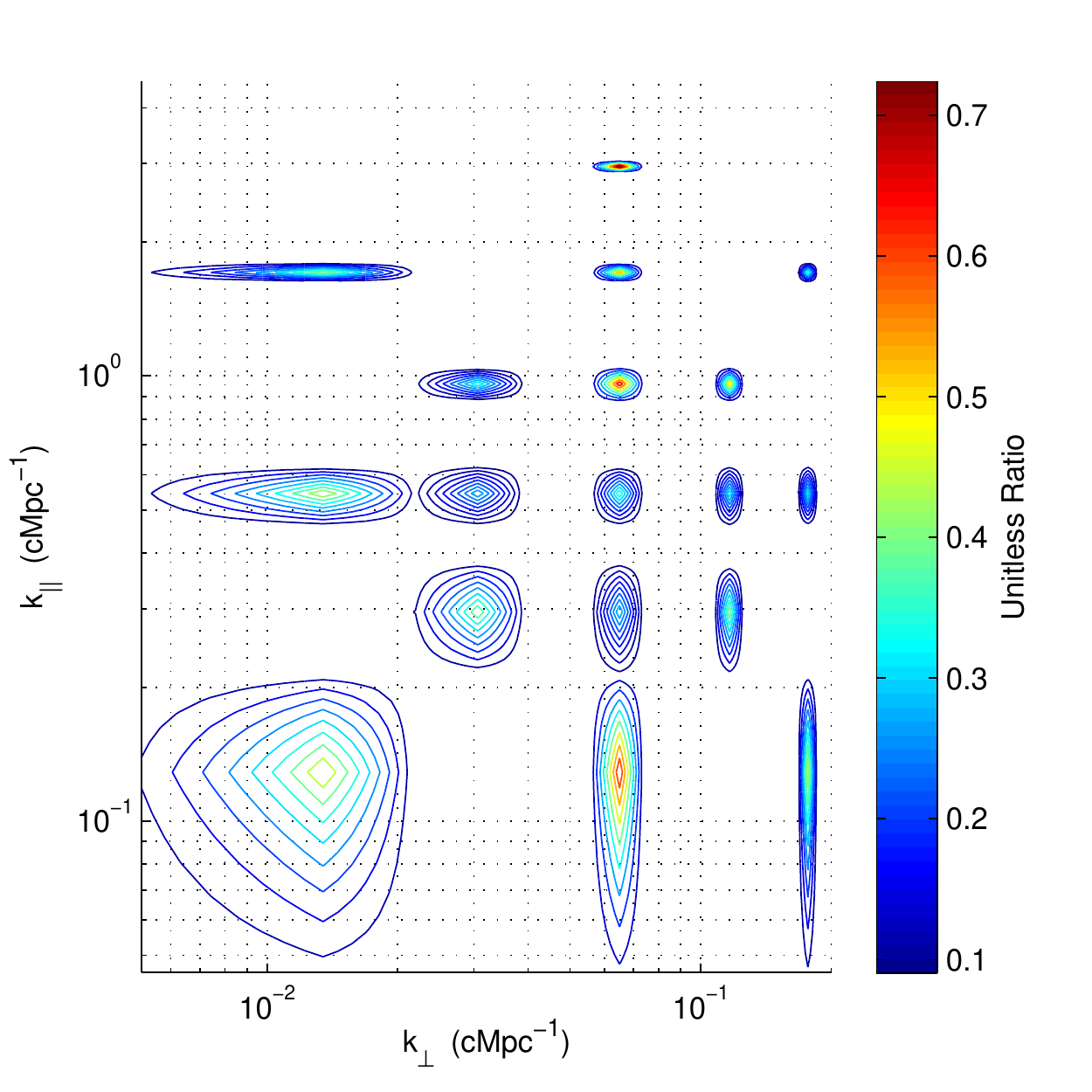}
	\caption{We can see from a sampling of window functions that our band power spectrum estimates $\widehat{p}^\alpha$ represent the weighted averages of $p^\alpha$ over a narrow range of scales, especially at higher values of $k_\perp$ and $k_\|$.  The widest window functions can be attributed to binning (with linearly binned data, low-$k$ bins look larger on logarithmic axes) and to foregrounds.  This is good news, because it will enable us to accurately make many independent measurements of the power spectrum and therefore better constrain astrophysical and cosmological parameters.}
	\label{MWAwindow}
\end{figure*}

We envision 1000 hours of integration on a field that is $9^\circ$ on each side, centered on $z=8$ with $\Delta z = 0.5$.  With a frequency resolution of 40 kHz and an angular resolution of 8 arcminutes, our data cube contains over $10^6$ voxels.  We completed over 1000 Monte Carlo iterations on our 12 core server in about one week.  We use the foreground parameters outlined above in Sections \ref{AdrianModels} and \ref{FastCov}.  In Figures \ref{MWAFisher} through \ref{MWAwindow}, we show the diagonal elements of the Fisher matrix we have calculated, the temperature power spectrum error bars, and a sampling of window functions. 

In Figure \ref{MWAFisher} we plot the diagonal elements of the Fisher matrix, which are related directly to the power spectrum errors.  Drawing from the discussion in Section \ref{CovMatBuildup}, we can see clearly the effects of the array layout (and thus noise), of foregrounds (included resolved point sources), and of pixelization.  Interestingly, until pixelization effects set in at the highest values of $k_\|$, the least contaminated region spans a large range of values of $k_\|$.  One way of probing more cosmological modes is to increase the frequency resolution of the instrument.  The number of modes accessible to the observation scales with $(\Delta \nu)^{-1}$, though the amplitude of the noise scales scales with $(\Delta \nu)^{-1/2}$. As long as the noise level is manageable and the cosmological signal is not dropping off too quickly with $k$, increasing the frequency resolution seems like a good deal.

In Figure \ref{MWAerror} we show the vertical error bars that we expect on power spectrum estimates in temperature units.  The most important fact about this plot is that there is a large region where we expect that vertical error bars will be sufficiently small that we should be able to detect a 10 mK signal with signal to noise greater than 1.  This is especially the case at fairly small values of $k$, which is surprising since these $k$ modes were supposed to be the most contaminated by foregrounds.  There are two reasons why this happens.  

First, the conversion to temperature units (Equation \ref{TemperatureUnits}) introduces a factor of $(k_\perp^2 k_\|)^{1/2}$ that raises the error bars for larger values of $k$.  Second, the strongest foreground modes overlap significantly with one of the $k=0$ modes of the discrete Fourier transform, which we exclude for our power spectrum estimate (this is just the average value of the cube in all three directions, which is irrelevant to an interferometer that is insensitive to the average).  

Another way to think about it is this: because the coherence length of the foregrounds along the line of sight is much longer than the size of any box small enough to comfortably ignore cosmological evolution, we expect that the most contaminated Fourier mode will be precisely the one we ignore.  Unlike the LT method, our method cannot easily measure modes much longer than the size of the data cube.  Along the line of sight, these modes have very wide window functions and are the most contaminated by foregrounds.  Perpendicular to the line of sight, these modes are better measured by considering much larger maps where the flat sky approximation no longer holds.  For the purposes of measuring these low-$k$ modes, the LT method can provide a useful complement to ours.  Large-scale modes from down-sampled maps can be measured by LT; smaller-scale modes from full-resolution maps can be measured by our method.  Then both can be combined to estimate the power spectrum across a large range of scales. 

And finally, in Figure \ref{MWAwindow} we show many different window functions for a selection of values of $k_\perp$ and $k_\|$ that spans the whole space.  In general, these window functions are quite narrow, meaning that each band power measurement probes only a narrow range of scales.  The widest windows we see look wide for two reasons.  First, linearly separated bins appear wider at low-$k$ when plotted logarithmically.  Second, foregrounds cause contaminated bins to leak into nearby bins, especially at low-$k_\|$ and moderate $k_\perp$.  We saw hints of this effect in Figure \ref{ComparisonToAdrian} when comparing noise-only simulations to simulations with both noise and foregrounds. 

In the vast majority of the $k_\perp$-$k_\|$ plane, the window functions seem to be dominated by the central bin and neighbors.  Except for edge cases, no window function has contributions exceeding 10\% from bins outside the central bin and its nearest neighbors.  This means that we should be happy with our choice of Fourier space binning, which was designed to have bin widths equal to those of our data cube before zero padding.  We also know that significantly finer binning would be inappropriate, so we do not have to worry about the tradeoff between fine binning of the power spectrum and the inversion of the Fisher matrix.  Therefore, with the 128-tile deployment of the MWA, we can be confident that our estimates of the power spectrum correspond to distinct modes of the true underlying power spectrum.


\section{Conclusions}

With this paper, we have presented an optimal algorithm for 21 cm power spectrum estimation which is dramatically faster than the Liu \& Tegmark (LT) method \cite{LT11}, scaling as $\BigO(N \log N)$ instead of $\BigO(N^3)$, where $N$ is the number of voxels in the 3D sky map.  By using the inverse variance weighted quadratic estimator formalism adapted to 21 cm tomography by the LT method, we preserve all accessible cosmological information in our measurement to produce the smallest possible error bars and narrow window functions.  Moreover, our method can incorporate additional information about the brightest point sources and thus further reduce our error bars at the cost of some---but by no means all---of that computational advantage. Our method is highly parallelizable and has only modest memory requirements; it never needs to store an entire $N\times N$ matrix.  

Our method achieves this computational speed-up for measuring power spectra, error bars, and window functions by eliminating the time-consuming matrix operations of the LT method.  We accomplish this using a combination of Fourier, spectral, and Monte Carlo techniques which exploit symmetries and other physical properties of our models for the noise and foregrounds.

We have demonstrated the successful simulation of error bars and window functions for the sort of massive data set we expect from the upcoming 128-tile deployment of the MWA---a data set that cannot be fully utilized using only the LT method.  Our forecast predicts that 1000 hours of MWA observation should be enough to detect the fiducial 10 mK signal across much of the $k_\|$-$k_\perp$ plane accessible to the instrument.  Moreover, we predict that the horizontal error bars on each band power estimate will be narrow, allowing each estimate to probe only a small range of scales.

Our results suggest several avenues for further research.  Of course, the most immediate application is to begin analyzing the data already being produced by interferometers like LOFAR, GMRT, MWA, and PAPER as they start accumulating the sensitivity necessary to zero in on a detection of the EoR.  The large volume of data these instruments promise to produce might make it useful to explore ways of further speeding up the Monte Carlo estimation of the Fisher matrix.  There is significant redundancy in our calculated Fisher matrix because the window function shapes vary only relatively slowly with $k$-scale.  We believe that one can reduce the number of Monte Carlo simulations needed to attain the same accuracy by adding a postprocessing step that fits the Fisher matrix to a parametrized form. This should work best in the regions of the $k_\|$-$k_\perp$ plane that are fairly uncontaminated by foregrounds, where Fisher matrix elements are expected to vary most smoothly.  It may also be possible to speed up the Monte Carlo estimation of the Fisher matrix using the trace evaluation technology of \citep{UeLiFast}.

The forecasting power of our method to see whether a particular observing campaign might reveal a particular aspect the power spectrum need not be limited to measurements of the EoR.  Our method provides an opportunity to precisely predict what kind of measurement, and what kind of instrument, might be necessary for observing 21 cm brightness temperature fluctuations during the cosmic dark ages.  Our method should prove useful for weighing a number of important design considerations: What is the optimal array configuration? What is the optimal survey volume?  What about angular resolution?  Spectral resolution?  And in what sense are these choices optimal for doing astrophysics and cosmology?  
 
To help answer such questions, our technique could be used to compare the myriad of ideas for and possible implementations of future projects like HERA and the SKA and even to help find an optimal proposal.  For example, one plan for achieving large collecting area is building a hierarchically regular array (a so-called ``Omniscope") that takes advantage of FFT correlation \citep{FFTT2} and redundant baseline calibration \citep{redundant}.  There exist many array configurations that fit into this category and it is not obvious what the optimal Omniscope might look like.

The quest to detect a statistical signal from the Epoch of Reionization is as daunting as it is exciting.  It is no easy task to find that needle in a haystack of noise and foregrounds.  However, now that we are for the first time armed with a method that can extract all the cosmological information from a massive data set without a prohibitive computational cost, we can feel confident that a sufficiently sensitive experiment can make that first detection---not just in theory, but also in practice.

\section*{Acknowledgments}
The authors wish to thank Gianni Bernardi, Michael Betancourt, Katherine Deck, Jacqueline Hewitt, Andrew Lutomirski, Michael Matejek, Miguel Morales, Ue-Li Pen, John Rutherford, Leo Stein, Christopher Williams, and Phillip Zukin for many useful discussions and advice.  This work is partially supported by the Bruno Rossi Graduate Fellowship and NSF Grant AST-1105835.  A.L. acknowledges support from the Berkeley Center for Cosmological Physics.


\appendix
\section{Toeplitz Matrices} \label{ToeplitzAppendix}

In this appendix, we briefly review how to rapidly multiply by Toeplitz matrices. We need to employ the advantages of Toeplitz matrices because the assumption that our covariance matrices are diagonal in real space or in Fourier space, as was the case in \citep{UeLiFast}, break down for covariance matrices with coherence lengths much larger than the box size.
 
A ``Toeplitz" matrix is any matrix with the same number for every entry along its main diagonal and with every other diagonal similarly constant \citep{Toeplitz}.  In general, a Toeplitz matrix is uniquely defined by the entries in its first row and its first column: if $i \ge j$ then $T_{ij} = T_{1+i-j,1}$ and if $i \le j$ then $T_{ij} = T_{1,1-i+j}$.   If the first row of a matrix is repeated with a cyclic permutation by one to the right in each successive row, then it is a special kind of Toeplitz matrix called a ``circulant" matrix.  Circulant matrices are diagonalized by the discrete Fourier transform \citep{Toeplitz}.  Given a circulant matrix $\mathbf{C}$ with first column $\mathbf{c}$, the product of $\mathbf{C}$ and some arbitrary vector $\mathbf{v}$ can be computed in $\mathcal{O}(N\log N)$ time because
\begin{equation} \label{circulant}
\mathbf{Cv} = \mathbf{F}^\dagger \mbox{ diag}(\mathbf{Fc}) \mbox{ } \mathbf{Fv},
\end{equation}
where $\mathbf{F}$ is the unitary, discrete Fourier transform matrix \citep{Toeplitz}.  Reading Equation \ref{circulant} from right to left, we see that every matrix operation necessary for this multiplication can be performed in $\mathcal{O}(N\mbox{log}N)$ time or better.  

Conveniently, any symmetric Toeplitz matrix can be embedded in a circulant matrix twice its size.  Given a symmetric Toeplitz matrix $\mathbf{T}$, we can define another symmetric Toeplitz matrix $\mathbf{S}$ with an arbitrary constant along its main diagonal.  If we specify that the rest of the first row (besides the first entry) is the reverse of the rest of the first row of $\T$ (again ignoring the first entry), the fact that the matrix is Toeplitz and symmetric completely determines the other entries.  For example,
\begin{equation}
\text{if } \mathbf{T} = \left( \begin{array}{ccc} 5 & 3 & 2 \\ 3 & 5 & 3 \\ 2 & 3 & 5 \end{array} \right), \text{ then } \mathbf{S} = \left( \begin{array}{ccc} 0 & 2 & 3 \\ 2 & 0 & 2 \\ 3 & 2 & 0 \end{array} \right).
\end{equation}
It is straightforward to verify that the matrix $\mathbf{C}$, defined as
\begin{equation}
\mathbf{C} \equiv \left( \begin{array}{cc} \mathbf{T} & \mathbf{S} \\ \mathbf{S} & \mathbf{T} \end{array} \right),
\end{equation}
is a circulant matrix. We can now can multiply $\mathbf{C}$ by a zero-padded vector so as to yield the product of the Toeplitz matrix and the original vector, $\mathbf{Tx}$, that we are looking for:
\begin{equation}
\mathbf{C} \left( \begin{array}{c} \mathbf{v} \\ \mathbf{0} \end{array} \right) = \left( \begin{array}{c} \mathbf{Tv} \\ \mathbf{Sv} \end{array} \right).
\end{equation}
Therefore, we can multiply any Toeplitz matrix by a vector in $\BigO(N \log N)$.

\section{Noise Covariance Matrix Derivation} \label{NoiseAppendix}
In this appendix, we derive the the form of $\N$, the noise covariance matrix, in Equation \ref{Nfourier} by combining the form of $P^N(\kv,\lambda)$, the noise power spectrum, in Equation \ref{NPowerSpectrum} with Equation \ref{Ndef}, which relates $\N$ to $P^N(\kv,\lambda)$.  To accomplish this, we simplify $P^N(\kv,\lambda)$ into a form that is more directly connected to our data cube.   We then approximate the integrals in Equation \ref{Ndef} by assuming that the $uv$-coverage is piecewise constant in cells corresponding to our Fourier space grid.\footnote{We also assume that measurements in nearby $uv$-cells are uncorrelated, which may not be true if the baselines are not coplanar; instead $\N$ would have to be modeled as sparse rather than diagonal in angular Fourier space.}

To simplify $P^N(\kv,\lambda)$, we first note that  because the term $B(\kv,\lambda)$ in Equation \ref{NPowerSpectrum} represents the synthesized beam and is normalized to peak at unity, we can reinterpret the factor of $(f^\text{cover})^{-1} B^{-2}(\kv,\lambda)$ as an inverted and normalized $uv$-coverage.  When $f^\text{cover} = 1$, the array has uniform coverage.  We want to replace the factor $(f^\text{cover})^{-1} B^{-2}(\kv,\lambda)$ with a quantity directly tied our choice of pixelization of the $uv$-plane and written in terms of the simplest observational specification: the total time that baselines spend observing a particular $uv$-cell, $t(\kv,\lambda)$.  We already know that the noise power is inversely proportional to that time because more time yields more independent samples.

To relate $t^{-1}(\kv,\lambda)$ to $(f^\text{cover})^{-1} B^{-2}(\kv,\lambda)$, we want to make sure that the formula yields the same answer for peak density in the case of a complete coverage.  In other words, we want to find the constant $t_\text{max}$ such that
\beq
\frac{t_\text{max}}{t(\kv,\lambda)} = (f^\text{cover})^{-1} B^{-2}(\kv,\lambda). \label{timeConversion}
\eeq
The time spent in the most observed cell is related to the size of the cell in the $uv$-plane, the density of baselines in that cell, and the total integration time of the observation, $\tau$.  The cell size is determined by the pixelization of our data cube.  We have divided each slice of our data cube of size $L_x \times L_y$ into $n_x \times n_y$ pixels.  In Fourier space, this corresponds a pixel size of
\beq
\Delta k_x = \frac{2\pi}{L_x} = \frac{2\pi}{d_M \Delta\theta_x n_x},
\eeq
where $\Delta\theta_x$ and is the angular pixelization in the $x$ direction.  An equivalent relation is true for the $y$ direction.  Since $\Delta u = \Delta k_x d_M / (2\pi)$, we have that the area in $uv$-space of each of our grid points is
\beq
\Delta v \Delta u = \frac{1}{\Delta\theta_x n_x}\frac{1}{\Delta\theta_y n_y} = \frac{1}{\Omega_\text{pix} n_x n_y}.
\eeq

The maximum density of baselines is the density of the autocorrelations,\footnote{If the use of autocorrelations (which most observations throw out, due to their unfavorable noise properties) is troubling, then it is helpful to recall that for a large and fully-filled array, the $uv$-density of the shortest baselines is approximately the same as the $uv$-density of the autocorrelations.} which is
\beq
n_\text{max} = N_\text{ant} \left( \frac{A_\text{ant}}{\lambda^2}\right)^{-1},
\eeq
where the quantity $(A_\text{ant}/\lambda^2)$ is the area in the $uv$-plane associated with a single baseline \citep{Matt3}.  We thus have that
\beq
t_\text{max} \equiv  n_\text{max} \Delta u \Delta v  \tau  = \frac{N_\text{ant} \lambda^2 \tau}{A_\text{ant} \Omega_\text{pix} n_x n_y}.
\eeq
Now we can substitute Equation \ref{timeConversion} into Equation \ref{NPowerSpectrum} to get a more useful form of $P^N(\kv,\lambda)$:
\begin{align}
P^N(\kv,\lambda) &= \frac{\lambda^4 T_\text{sys}^2 y d^2_M}{A^2_\text{ant} \Omega_\text{pix} n_x n_y}\frac{1}{t(\kv_\perp,\lambda)}.
\end{align}

In general, $t(\kv_\perp,\lambda)$ depends in a nontrivial way on the array layout.  As such, the integral expression for $\N$ in Equation \ref{Ndef} with this form of $P^N(\kv,\lambda)$ is only analytically tractable along the line of sight.  Integrating $k_z$, we get that
\begin{align}
N_{ij} =& \frac{\delta_{z_i z_j}}{\Delta z} \frac{\lambda^4 T_\text{sys}^2 y d^2_M}{A^2_\text{ant} \Omega_\text{pix} n_x n_y} \int  j_0^2(k_x \Delta x /2) j_0^2(k_y \Delta y /2) \times  \nonumber \\ &e^{i k_x (x_i-x_j) + i k_y(y_i-y_j)} \frac{1}{t(\kv_\perp,\lambda_i)} \frac{dk_x dk_y}{(2\pi)^2}. 
\end{align}
We note that $\N$ is uncorrelated between frequency channels, as we would expect.

Along the other two dimensions, we will approach the problem by approximating the integrand as piecewise constant in Fourier cells, turning the integral into a sum and the $dk$ into a $\Delta k$.  We will use the index $l$ to run over all Fourier modes perpendicular to the line of sight.  Using the fact that the line of sight voxel length $\Delta z = y \Delta \nu$ and that $L_x L_y =  \Omega_\text{pix} d_M^2 n_x n_y$, we have that
\begin{align}
N_{ij} = & \frac{\lambda^4 T_\text{sys}^2 \delta_{z_i z_j}}{A^2_\text{ant} (\Omega_\text{pix} n_x n_y)^2\Delta\nu} \sum_{l}^{\text{all x \& y}} \bigg[  j_0^2(k_{x,l} \Delta x /2)  \times  \nonumber \\ &j_0^2(k_{y,l}  \Delta y /2) e^{i k_{x,l}(x_i-x_j)} e^{i k_{y,l}(y_i-y_j)} \frac{1}{t_l(\lambda_i)}  \bigg].
\end{align}
Next, we can turn this form into one that is more clearly computationally easy to multiply by a vector by introducing another Kronecker delta:
\begin{align}
N_{ij} = & \frac{\lambda^4 T_\text{sys}^2 \delta_{z_i z_j}}{A^2_\text{ant} (\Omega_\text{pix} n_x n_y)^2\Delta\nu} \sum_{l}^{\text{all x \& y}} e^{i k_{x,l}x_i} e^{i k_{y,l}y_i} \nonumber \\ & \sum_{m}^{\text{all x \& y}} \bigg[  j_0^2(k_{x,l} \Delta x /2) j_0^2(k_{y,l} \Delta y /2) \nonumber \\ &e^{-i k_{x,m}x_j} e^{-i k_{y,m}y_j} \frac{\delta_{lm}}{t_l(\lambda_i)}  \bigg].
\end{align}
Finally, if we extend $l$ and $m$ to index over all frequency channels and all Fourier modes perpendicular to the line of sight, we can write down the noise covariance matrix as $\N = \F_\perp^\dagger \widetilde{\N} \F_\perp$ where $\F_\perp$ and $\F_\perp^\dagger$ are the discrete, unitary 2D Fourier and inverse Fourier transforms and where $\widetilde{\N}$ can be written as
\beq
\widetilde{N}_{lm} =  \frac{\lambda^4 T_\text{sys}^2 j_0^2(k_{x,l} \Delta x /2) j_0^2(k_{y,l} \Delta y /2) }{A^2_\text{ant} (\Omega_\text{pix})^2 n_x n_y\Delta\nu}   \frac{\delta_{lm}}{t_l}.
\eeq
The result, therefore, is a matrix that can be multiplied by a vector in $\BigO(N\log N)$.

\section{Construction of the Preconditioner} \label{PreconAppendix}

In this final appendix, we show how to construct the preconditioner that we use to speed up the conjugate gradient method for multiplying $\C^{-1}$ by a vector. We devise our preconditioner by looking at $\C$ piece by piece, building up pairs of matrices that make our covariances look more like the identity matrix.  We start with $\C = \N$, generalize to $\C = \U + \N$, and then finally incorporate $\R$ and $\G$ to build the full preconditioner.

\subsection{Constructing a Preconditioner for $\N$} \label{PreconNSection}
Our first task is to find a pair of preconditioning matrices that turn $\N$ into the identity:
\beq
\Pre_\N \N \Pre_\N^\dagger = \I.
\eeq
Because $\N = \F_\perp^\dagger \widetilde{\N} \F_\perp$, and because $\widetilde{\N}$ is a diagonal matrix, we define $\Pre_\N$ and $\Pre_\N^\dagger$ as follows:
\begin{align}
\Pre_\N &= \widetilde{\N}^{-1/2} \F_\perp, \nonumber \\
\Pre_\N^\dagger &= \F_\perp^\dagger \widetilde{\N}^{-1/2}. \label{PNdef}
\end{align}
Since applying $\Pre_\N$ only requires multiplying by the inverse square root of a diagonal matrix and Fourier transforming in two dimensions, the complexity of applying $\Pre_\N$ to a vector is less than $\mathcal{O}(N\log N)$.

\subsection{Constructing a Preconditioner for $\U$}\label{PreconForU}
The matrix $\U$ (Equation \ref{formOfU}) can be written as the tensor product of three Toeplitz matrices, one for each dimension, bookended by two diagonal matrices, $\D_\U$.  Furthermore, since $\D_\U$ depends only on frequency (as we saw in Section \ref{Ufast}),  its effect can be folded into $\U_z$ such that
\beq \label{Uouterproduct}
\D_\U [\U_x \otimes \U_y \otimes \U_z] \D_\U \equiv \U_x \otimes \U_y \otimes \U_z'.
\eeq 
It is generally the case that $\U_x$ and $\U_y$ are both well approximated by the identity matrix.  This reflects the fact that the spatial clustering of unresolved point sources is comparable with the angular resolution of the instrument.  This assumption turns out to be quite good for fairly compact arrays, since for an array with 1 km as its longest baseline---the sort of compact array thought to be optimal for 21 cm cosmology---we expect an angular resolution on the order of 10 arcminutes, which is comparable to the fiducial value of 7 arcminutes that LT took to describe the clustering length scale for unresolved point sources.  That value appears to be fairly reasonable given the results of \citep{BernardiForegrounds,GhoshForegrounds}.   For the purposes of devising a preconditioner only, we can therefore adopt the simplification that
\beq
\U \approx \I_x \otimes \I_y \otimes \U_z,
\eeq
where we have dropped the prime for notational simplicity.  Looking back at Figure \ref{noPrecon}, this form of $\U$ neatly explains the stair-stepping behavior of the eigenvalues: for every eigenvalue of $\U_z$, $\U$ has $n_x\times n_y$ similar eigenvalues.
 
Since only a few eigenvalues of $\U_z$ are large, it is pedagogically useful to first address a simplified version of the preconditioning problem where $\U_z$ is approximated as a rank 1 matrix by cutting off its spectral decomposition after the first eigenvalue.  We will later return to include the other relevant eigenvalues.  We therefore write $\U$ as follows:
\beq
\U \approx \I_x \otimes \I_y \otimes \lambda \ev_z \ev_z^\dagger.
\eeq
where $\ev_z$ is the normalized eigenvector of $\U$.

Let us now take a look at the action of $\Pre_\N$ and $\Pre_\N^\dagger$ on $\U + \N$:
\begin{align}
\Pre_\N& (\U + \N) \Pre_\N^\dagger \nonumber \\
&= \I + \widetilde{\N}^{-1/2} \F_\perp  (\I_x \otimes \I_y \otimes \lambda \ev_z \ev_z^\dagger)\F_\perp^\dagger \widetilde{\N}^{-1/2} \nonumber\\
 &= \I + \widetilde{\N}^{-1/2} (\I_x \otimes \I_y \otimes \lambda \ev_z \ev_z^\dagger)\widetilde{\N}^{-1/2} \nonumber \\ &\equiv \I + \Ubar.
\end{align}
Our next goal, therefore, is to come up with a new matrix $\Pre_\U$ that, when applied to $\I + \Ubar$ gives us something close to $\I$.

We now take a closer look at $\Ubar$.  Since it is a good approximation to say that $\widetilde{\N}$ only changes perpendicular to the line of sight,\footnote{Were it not for the breathing of the synthesized beam with frequency, $\widetilde{\N}$, would only change perpendicularly to the line of sight.  Since it is a small effect when considered over a modest redshift range, we can ignore it in the construction of our preconditioner.  After all, we only need to make $\Pre\C\Pre^\dagger$ close to $\I$.} we can rewrite $\Ubar$:
\begin{align}
\Ubar &\approx (\widetilde{\N}_{\perp}^{-1/2}\otimes \I_z) (\I_x \otimes \I_y \otimes \lambda {\ev}_z {\ev}_z^\dagger)(\widetilde{\N}_{\perp}^{-1/2}\otimes \I_z)  \nonumber \\ &= (\widetilde{\N}_{\perp}^{-1}) \otimes (\lambda \ev_z {\ev}_z^\dagger), 
\end{align}
where $\widetilde{\N}_{\perp}$ is still a diagonal matrix, though only in two dimensions, generated from a baseline distribution averaged over frequency slices.  We now form a pair of preconditioning matrices, $\Pre_\U$ and $\Pre_\U^\dagger$ of the form $\Pre_\U \equiv \I - \beta \Proj$ where $\Proj$ has the property that $\Proj\Ubar = \Ubar$ and that $\Ubar\Proj^{\dagger} = \Ubar$.  The matrix that fits this description is:
\begin{align}
\Proj &= \widetilde{\N}^{-1/2} (\I_x \otimes \I_y \otimes {\ev}_z {\ev}_z^\dagger)\widetilde{\N}^{1/2} \nonumber \\&\approx (\I_x \otimes \I_y \otimes {\ev}_z {\ev}_z^\dagger) = \frac{1}{\lambda}\U,
\end{align}
since $\widetilde{\N}$ only affects the $x$ and $y$ components and thus passes through the inner matrix.  This also means that $\Proj = \Proj^\dagger$ and that $\Proj = \Proj^2$.  The result for $\Pre_\U (\I + \Ubar)\Pre_\U^\dagger$ is
\begin{align}
&(\I - \beta \Proj)(\I + \Ubar)(\I - \beta \Proj^\dagger) = \nonumber \\ &\I + \Ubar - 2\beta\Ubar - 2\beta\Proj + \beta^2\Ubar + \beta^2\Proj. \label{ubarquadratic}
\end{align}

The trick now is that for each $uv$-cell, $\Ubar$ has only one eigenvalue, which we call $\lbar_l$ (again using $l$ as an index over both directions perpendicular to the line of sight):
\beq 
\lbar_{l} = \frac{\lambda}{(\widetilde{N}_\perp)_{ll}}.
\eeq
Likewise, $\Proj$ only has one eigenvalue: 1. By design, these eigenvalues correspond to the same eigenvector.  Since our goal is to have the matrix in equation \eqref{ubarquadratic} be the identity, we need only pick $\beta$ such that:
\beq
1 = 1 + \lbar_{l} - 2\beta\lbar_{l} - 2\beta + \beta^2\lbar_{l} + \beta^2. \label{BetaEquation}
\eeq
Solving the quadratic equation, we get 
\begin{align}
\Pre_\U \equiv  \I - \bigg[ \sum_{l} \left(1-\sqrt{\frac{1}{1+\lbar_{l}}}  \right) {\ev}_z {\ev}_z^\dagger& \text{ } \nonumber \\  \otimes \stackrel{\leftrightarrow}{\mathbf{\delta}}_{x,x_l} \otimes \stackrel{\leftrightarrow}{\mathbf{\delta}}_{y,y_l} \bigg]&, \label{PU}
\end{align}
where the pair of $\stackrel{\leftrightarrow}{\mathbf{\delta}}$ matrices pick out a particular $uv$-cell.  If we want to generalize to more eigenvectors of $\U_z$, we simply need to keep subtracting off sums of matrices on the right hand side of Equation \eqref{PU}:
\begin{align}
\Pre_\U \equiv \I - \sum_k \bigg[ \sum_{l} \left(1-\sqrt{\frac{1}{1+\lbar_{l,k}}}  \right) {\ev}_{z_k} {\ev}_{z_k}^\dagger \nonumber \\ \otimes \stackrel{\leftrightarrow}{\mathbf{\delta}}_{x,x_l} \otimes \stackrel{\leftrightarrow}{\mathbf{\delta}}_{y,y_l} \bigg], \label{PUmulti}
\end{align}
This works because every set of vectors corresponding to a value of $k$ is orthogonal to every other set.  Each term in the above sum acts on a different subspace of $\C$ independent of all the other terms in the sum.

If the relevant vectors ${\ev}_{z_k}$ are precomputed, applying $\Pre_\U$ can be done in $\BigO(N m(\U_z))$ where $m(\U_z)$ is defined as the number of relevant eigenvalues of $\U_z$ that need preconditioning or, equivalently, the number of ``steps" in the eigenvalues of $\U$ in Figure \ref{noPrecon} above the noise floor. We examine how $m(\U_z)$ scales with the size of the data cube in Section \ref{PreconComplexity}. Because the fall off of the eigenvalues is exponential \cite{AdrianForegrounds} we expect the scaling of $m$ to be logarithmic.

In general, we can pick some threshold $\theta \ge 1$ to compare to the largest value of $\lbar_{l,k}$ for a given $k$ and then do not precondition modes with eigenvalues smaller than $\theta$.  One might expect there to be diminishing marginal utility to preconditioning the eigenvalues nearest $1$.  We explore how to optimally cut off the spectral decomposition in Section \ref{PreconResults} by searching for a value of $\theta$ where the costs and benefits of preconditioning equalize.

\subsection{Constructing a Preconditioner for $\R$ and $\G$}\label{PreconForGamma}

We now turn our attention to the full matrix $\C$.  The fundamental challenge to preconditioning all of the matrices in $\C$ simultaneously is that the components of $\R$ and $\G$ perpendicular to the line of sight are diagonalized in completely different bases.  However, $\U$, $\G$, and $\R$ have very similar components parallel to the line of sight, due to the fact they all represent spectrally smooth radiation of astrophysical origin.

We can write down $\R$ as follows:
\begin{align}
\R = \sum_n\bigg[ &\stackrel{\leftrightarrow}{\delta}_{x,x_n}\otimes\stackrel{\leftrightarrow}{\delta}_{y,y_n}\otimes  \nonumber \\ &\left( \sum_k \lambda_{n,k} \ev_{z_{n,k}} \ev_{z_{n,k}}^\dagger \right) \bigg],
\end{align}
which can be interpreted as a set of matrices describing spectral coherence, each localized to one point source, and all of which are spatially uncorrelated.  And likewise, we can write down $\G$ as:
\beq
\G = \sum_{i,j,k} \left[ \lambda_{x_i}\lambda_{y_j}\lambda_{z_k} \ev_{x_i} \ev_{x_i}^\dagger \otimes \ev_{y_j} \ev_{y_j}^\dagger \otimes \ev_{z_k} \ev_{z_k}^\dagger \right].
\eeq
We now make two key approximations for the purposes of preconditioning.  First, we assume that all the $z_k$ eigenvectors are the same, so $\ev_{z_k} \approx \ev_{z_{n,k}}$ for all $n$, all of which are also taken to be the same as the eigenvectors that appear in the preconditioner for $\U$ in Equation \ref{PUmulti}.  Second, as in Section \ref{PreconForU}, we are only interested in acting upon the largest eigenvalues of $\R$ and $\G$.  To this end, we will ultimately only consider the largest values of $\lambda_{n,k}$ and $\lambda_{i,j,k} \equiv \lambda_{x_i}\lambda_{y_j}\lambda_{z_k}$, which will vastly reduce the computational complexity of the preconditioner.

Our strategy for overcoming the difficulty of the different bases is to simply add the two perpendicular parts of the matrices and then decompose the sum into its eigenvalues and eigenvectors.  We therefore define 
\beq
\Gam \equiv \R + \G
\eeq
(choosing the symbol $\Gam$ because it looks like $\R$ and sounds like $\G$).  Given the above approximations, we can reexpress $\Gam$ as follows:
\begin{align} \label{GammaCalc}
\Gam \approx \sum_k \left(\Gam_{\perp,k} \otimes \ev_{z_k} \ev_{z_k}^\dagger\right),
\end{align}
where we have defined each $\Gam_{\perp,k}$ as:
\begin{align}
\Gam_{\perp,k} \equiv &\left( \sum_n  \lambda_{n,k} \stackrel{\leftrightarrow}{\delta}_{x,x_n}\otimes\stackrel{\leftrightarrow}{\delta}_{y,y_n} \right) \nonumber + \\ &\left( \sum_{i,j} \lambda_{i,j,k} \ev_{x_i} \ev_{x_i}^\dagger \otimes \ev_{y_j} \ev_{y_j}^\dagger \right).
\end{align}
Due to the high spectral coherence of the foregrounds, only a few values of $k$ need to be included to precondition for $\Gam$. Considering the limit on angular box size imposed by the flat sky approximation and the limit on angular resolution imposed by the array size, this should require at most a few eigenvalue determinations of matrices no bigger than about $10^4$ entries on a side.  Moreover, those eigenvalue decompositions need only be computed once and then only partially stored for future use.  In practice, this is not a rate-limiting step, as we see in Section \ref{PreconComplexity}.

We now write down the eigenvalue decomposition of $\Gam$:
\beq
\Gam = \sum_k \left( \sum_l \lambda_{l,k} \ev_{\perp,l}\ev_{\perp,l}^\dagger \right) \ev_{z_k} \ev_{z_k}^\dagger.
\eeq
Before we attack the general case, we assume that only one value of $\lambda_{l,k}$ is worth preconditioning---we generalize to the full $\Pre_\Gam$ later.  We now know that if we have a matrix that looks like $\I + \Ubar$ we can make it look like $\I$.  So can we take $\I + \Ubar + \GammaBar$, where $\GammaBar \equiv \Pre_\N \Gam \Pre_\N^\dagger$, and turn it into $\I + \Ubar$?  Looking at $\GammaBar$,
\begin{align}
\GammaBar  =& \widetilde{\N}^{-1/2} \F_\perp \lambda_\Gam \ev \ev^\dagger \F_\perp^\dagger \widetilde{\N}^{-1/2} \nonumber \\ =& \lambda_\Gam (\widetilde{\N}_\perp^{-1/2}  \widetilde{\ev}_\perp \widetilde{\ev}_\perp^\dagger \widetilde{\N}_\perp^{-1/2}) \otimes \ev_z \ev_z^\dagger,
\end{align}
where $\lambda_\Gam$ is the sole eigenvalue we are considering and where $\widetilde{\ev}_\perp \equiv \F_\perp \ev_\perp.$

Again, we will look at a preconditioner of the $\Pre_\Gam = \I - \beta\Proj$ where:
\beq
\Proj \equiv \left(\widetilde{\N}_\perp^{-1/2} \widetilde{\ev}_\perp \widetilde{\ev}_\perp^\dagger \widetilde{\N}_\perp^{1/2}\right) \otimes \ev_z \ev_z^\dagger.
\eeq
This time, the $\widetilde{\N}^{\pm 1/2}_\perp$ matrices do not pass through the eigenvectors to cancel one another out.  We now exploit the spectral similarity of foregrounds and the fact that $\widetilde{\ev}_\perp^\dagger \widetilde{\ev}_\perp = \ev^\dagger_z \ev_z  = 1$ to obtain 
\beq 
\Pre_\Gam \Ubar \Pre_\Gam^\dagger = \Ubar + \frac{\lambda_\U}{\lambda_\Gam}(\beta^2 - 2\beta)\GammaBar.
\eeq
This is very useful because it means that if we pick $\beta$ properly, we can get the second term to cancel the $\GammaBar$ terms we expect when we calculate the full effect of $\Pre_\Gam$ and $\Pre_\N$ on $\N + \U + \Gam$.  Noting that the sole eigenvalue of $\GammaBar$ is $\overline{\lambda}_\Gam \equiv \lambda_\Gam \widetilde{\ev}_\perp^\dagger \widetilde{\N}_\perp^{-1} \widetilde{\ev}_\perp$, we also define $\overline{\lambda}_\U \equiv \lambda_\U \widetilde{\ev}_\perp^\dagger \widetilde{\N}_\perp^{-1} \widetilde{\ev}_\perp$.  Multiplying our preconditioner by our matrices, we see that the the equality of the single eigenvalues yields another quadratic equation for $\beta$: 
\begin{align}
1 + \overline{\lambda}_\U = \text{ } & 1 - 2\beta + \beta^2 + (\beta^2 - 2\beta + 1)\overline{\lambda}_\Gam \nonumber \\ &+ \overline{\lambda}_\Gam \frac{\lambda_\U}{\lambda_\Gam}(\beta^2 -2\beta).
\end{align}
Solving, we finally have our $\Pre_\Gam$ that acts on $\I+\Ubar+\GammaBar$ and yields $\I + \Ubar$:
\begin{align}
\Pre_\Gam = \I - &\left(1 - \sqrt{\frac{1+\overline{\lambda}_\U}{1+\overline{\lambda}_\U+\overline{\lambda}_\Gam}} \right) \times \nonumber \\ &\left[\left(\widetilde{\N}_\perp^{-1/2} \widetilde{\ev}_\perp \widetilde{\ev}_\perp^\dagger \widetilde{\N}_\perp^{1/2}\right) \otimes \ev_z \ev_z^\dagger \right]. \label{PreGamma}
\end{align}
Finally, generalizing to multiple eigenvalues and taking advantage of the orthonormality of the eigenvectors, we have
\begin{align}
\Pre_\Gam =\text{ }&\I - \sum_{k,m} \bigg[\left(1 - \sqrt{\frac{1+\overline{\lambda}_{\U_k}}{1+\overline{\lambda}_{\U_k}+\overline{\lambda}_{\Gam_{k,m}}}} \right) \times \nonumber \\ &\left(\left(\widetilde{\N}_\perp^{-1/2} \widetilde{\ev}_{\perp_m} \widetilde{\ev}_{\perp_m}^\dagger \widetilde{\N}_\perp^{1/2}\right) \otimes \ev_{z_k} \ev_{z_k}^\dagger \right) \bigg]. \label{PreGammaMulti}
\end{align}
The result of this somewhat complicated preconditioner is a reduction of the condition number of the matrix to be inverted by many orders of magnitude (see Figure \ref{preconditionedEigs}).

Lastly, we include Fourier transforms at the front and the back of the preconditioner, so that the result, when multiplied by a real vector, returns a real vector.  Therefore, the total preconditioner we use for $\C$ is:
\beq
\F_\perp^\dagger \Pre_\U \Pre_\Gam \Pre_\N (\R + \U + \N + \G) \Pre_\N^\dagger \Pre_\Gam^\dagger \Pre_\U^\dagger \F_\perp.
\eeq

\bibliography{21cmTomographyReferences}

\begin{thebibliography}{70}%
\makeatletter
\providecommand \@ifxundefined [1]{%
 \@ifx{#1\undefined}
}%
\providecommand \@ifnum [1]{%
 \ifnum #1\expandafter \@firstoftwo
 \else \expandafter \@secondoftwo
 \fi
}%
\providecommand \@ifx [1]{%
 \ifx #1\expandafter \@firstoftwo
 \else \expandafter \@secondoftwo
 \fi
}%
\providecommand \natexlab [1]{#1}%
\providecommand \enquote  [1]{``#1''}%
\providecommand \bibnamefont  [1]{#1}%
\providecommand \bibfnamefont [1]{#1}%
\providecommand \citenamefont [1]{#1}%
\providecommand \href@noop [0]{\@secondoftwo}%
\providecommand \href [0]{\begingroup \@sanitize@url \@href}%
\providecommand \@href[1]{\@@startlink{#1}\@@href}%
\providecommand \@@href[1]{\endgroup#1\@@endlink}%
\providecommand \@sanitize@url [0]{\catcode `\\12\catcode `\$12\catcode
  `\&12\catcode `\#12\catcode `\^12\catcode `\_12\catcode `\%12\relax}%
\providecommand \@@startlink[1]{}%
\providecommand \@@endlink[0]{}%
\providecommand \url  [0]{\begingroup\@sanitize@url \@url }%
\providecommand \@url [1]{\endgroup\@href {#1}{\urlprefix }}%
\providecommand \urlprefix  [0]{URL }%
\providecommand \Eprint [0]{\href }%
\providecommand \doibase [0]{http://dx.doi.org/}%
\providecommand \selectlanguage [0]{\@gobble}%
\providecommand \bibinfo  [0]{\@secondoftwo}%
\providecommand \bibfield  [0]{\@secondoftwo}%
\providecommand \translation [1]{[#1]}%
\providecommand \BibitemOpen [0]{}%
\providecommand \bibitemStop [0]{}%
\providecommand \bibitemNoStop [0]{.\EOS\space}%
\providecommand \EOS [0]{\spacefactor3000\relax}%
\providecommand \BibitemShut  [1]{\csname bibitem#1\endcsname}%
\let\auto@bib@innerbib\@empty
\bibitem [{\citenamefont {{Barkana}}\ and\ \citenamefont
  {{Loeb}}(2001)}]{BLreview}%
  \BibitemOpen
  \bibfield  {author} {\bibinfo {author} {\bibfnamefont {R.}~\bibnamefont
  {{Barkana}}}\ and\ \bibinfo {author} {\bibfnamefont {A.}~\bibnamefont
  {{Loeb}}},\ }\href {\doibase 10.1016/S0370-1573(01)00019-9} {\bibfield
  {journal} {\bibinfo  {journal} {\physrep}\ }\textbf {\bibinfo {volume}
  {349}},\ \bibinfo {pages} {125} (\bibinfo {year} {2001})},\ \Eprint
  {http://arxiv.org/abs/arXiv:astro-ph/0010468} {arXiv:astro-ph/0010468}
  \BibitemShut {NoStop}%
\bibitem [{\citenamefont {{Furlanetto}}\ \emph {et~al.}(2006)\citenamefont
  {{Furlanetto}}, \citenamefont {{Oh}},\ and\ \citenamefont
  {{Briggs}}}]{FurlanettoReview}%
  \BibitemOpen
  \bibfield  {author} {\bibinfo {author} {\bibfnamefont {S.~R.}\ \bibnamefont
  {{Furlanetto}}}, \bibinfo {author} {\bibfnamefont {S.~P.}\ \bibnamefont
  {{Oh}}}, \ and\ \bibinfo {author} {\bibfnamefont {F.~H.}\ \bibnamefont
  {{Briggs}}},\ }\href {\doibase 10.1016/j.physrep.2006.08.002} {\bibfield
  {journal} {\bibinfo  {journal} {\physrep}\ }\textbf {\bibinfo {volume}
  {433}},\ \bibinfo {pages} {181} (\bibinfo {year} {2006})},\ \Eprint
  {http://arxiv.org/abs/arXiv:astro-ph/0608032} {arXiv:astro-ph/0608032}
  \BibitemShut {NoStop}%
\bibitem [{\citenamefont {{Morales}}\ and\ \citenamefont
  {{Wyithe}}(2010)}]{miguelreview}%
  \BibitemOpen
  \bibfield  {author} {\bibinfo {author} {\bibfnamefont {M.~F.}\ \bibnamefont
  {{Morales}}}\ and\ \bibinfo {author} {\bibfnamefont {J.~S.~B.}\ \bibnamefont
  {{Wyithe}}},\ }\href {\doibase 10.1146/annurev-astro-081309-130936}
  {\bibfield  {journal} {\bibinfo  {journal} {\araa}\ }\textbf {\bibinfo
  {volume} {48}},\ \bibinfo {pages} {127} (\bibinfo {year} {2010})},\ \Eprint
  {http://arxiv.org/abs/0910.3010} {arXiv:0910.3010 [astro-ph.CO]} \BibitemShut
  {NoStop}%
\bibitem [{\citenamefont {{Pritchard}}\ and\ \citenamefont
  {{Loeb}}(2012)}]{PritchardLoebReview}%
  \BibitemOpen
  \bibfield  {author} {\bibinfo {author} {\bibfnamefont {J.~R.}\ \bibnamefont
  {{Pritchard}}}\ and\ \bibinfo {author} {\bibfnamefont {A.}~\bibnamefont
  {{Loeb}}},\ }\href {\doibase 10.1088/0034-4885/75/8/086901} {\bibfield
  {journal} {\bibinfo  {journal} {Reports on Progress in Physics}\ }\textbf
  {\bibinfo {volume} {75}},\ \bibinfo {pages} {086901} (\bibinfo {year}
  {2012})},\ \Eprint {http://arxiv.org/abs/1109.6012} {arXiv:1109.6012
  [astro-ph.CO]} \BibitemShut {NoStop}%
\bibitem [{\citenamefont {{McQuinn}}\ \emph {et~al.}(2006)\citenamefont
  {{McQuinn}}, \citenamefont {{Zahn}}, \citenamefont {{Zaldarriaga}},
  \citenamefont {{Hernquist}},\ and\ \citenamefont {{Furlanetto}}}]{Matt3}%
  \BibitemOpen
  \bibfield  {author} {\bibinfo {author} {\bibfnamefont {M.}~\bibnamefont
  {{McQuinn}}}, \bibinfo {author} {\bibfnamefont {O.}~\bibnamefont {{Zahn}}},
  \bibinfo {author} {\bibfnamefont {M.}~\bibnamefont {{Zaldarriaga}}}, \bibinfo
  {author} {\bibfnamefont {L.}~\bibnamefont {{Hernquist}}}, \ and\ \bibinfo
  {author} {\bibfnamefont {S.~R.}\ \bibnamefont {{Furlanetto}}},\ }\href
  {\doibase 10.1086/505167} {\bibfield  {journal} {\bibinfo  {journal} {ApJ}\
  }\textbf {\bibinfo {volume} {653}},\ \bibinfo {pages} {815} (\bibinfo {year}
  {2006})},\ \Eprint {http://arxiv.org/abs/arXiv:astro-ph/0512263}
  {arXiv:astro-ph/0512263} \BibitemShut {NoStop}%
\bibitem [{\citenamefont {{Santos}}\ and\ \citenamefont
  {{Cooray}}(2006)}]{Santos2}%
  \BibitemOpen
  \bibfield  {author} {\bibinfo {author} {\bibfnamefont {M.~G.}\ \bibnamefont
  {{Santos}}}\ and\ \bibinfo {author} {\bibfnamefont {A.}~\bibnamefont
  {{Cooray}}},\ }\href {\doibase 10.1103/PhysRevD.74.083517} {\bibfield
  {journal} {\bibinfo  {journal} {\prd}\ }\textbf {\bibinfo {volume} {74}},\
  \bibinfo {pages} {083517} (\bibinfo {year} {2006})},\ \Eprint
  {http://arxiv.org/abs/arXiv:astro-ph/0605677} {arXiv:astro-ph/0605677}
  \BibitemShut {NoStop}%
\bibitem [{\citenamefont {{Bowman}}\ \emph {et~al.}(2007)\citenamefont
  {{Bowman}}, \citenamefont {{Morales}},\ and\ \citenamefont
  {{Hewitt}}}]{juddjackiemiguel1}%
  \BibitemOpen
  \bibfield  {author} {\bibinfo {author} {\bibfnamefont {J.~D.}\ \bibnamefont
  {{Bowman}}}, \bibinfo {author} {\bibfnamefont {M.~F.}\ \bibnamefont
  {{Morales}}}, \ and\ \bibinfo {author} {\bibfnamefont {J.~N.}\ \bibnamefont
  {{Hewitt}}},\ }\href {\doibase 10.1086/516560} {\bibfield  {journal}
  {\bibinfo  {journal} {ApJ}\ }\textbf {\bibinfo {volume} {661}},\ \bibinfo
  {pages} {1} (\bibinfo {year} {2007})},\ \Eprint
  {http://arxiv.org/abs/arXiv:astro-ph/0512262} {arXiv:astro-ph/0512262}
  \BibitemShut {NoStop}%
\bibitem [{\citenamefont {{Furlanetto}}\ \emph {et~al.}(2009)\citenamefont
  {{Furlanetto}}, \citenamefont {{Lidz}}, \citenamefont {{Loeb}}, \citenamefont
  {{McQuinn}}, \citenamefont {{Pritchard}}, \citenamefont {{Shapiro}},
  \citenamefont {{Alvarez}}, \citenamefont {{Backer}}, \citenamefont
  {{Bowman}}, \citenamefont {{Burns}}, \citenamefont {{Carilli}}, \citenamefont
  {{Cen}}, \citenamefont {{Cooray}}, \citenamefont {{Gnedin}}, \citenamefont
  {{Greenhill}}, \citenamefont {{Haiman}}, \citenamefont {{Hewitt}},\ and\
  \citenamefont {et~al.}}]{Whitepaper2}%
  \BibitemOpen
  \bibfield  {author} {\bibinfo {author} {\bibfnamefont {S.~R.}\ \bibnamefont
  {{Furlanetto}}}, \bibinfo {author} {\bibfnamefont {A.}~\bibnamefont
  {{Lidz}}}, \bibinfo {author} {\bibfnamefont {A.}~\bibnamefont {{Loeb}}},
  \bibinfo {author} {\bibfnamefont {M.}~\bibnamefont {{McQuinn}}}, \bibinfo
  {author} {\bibfnamefont {J.~R.}\ \bibnamefont {{Pritchard}}}, \bibinfo
  {author} {\bibfnamefont {P.~R.}\ \bibnamefont {{Shapiro}}}, \bibinfo {author}
  {\bibfnamefont {M.~A.}\ \bibnamefont {{Alvarez}}}, \bibinfo {author}
  {\bibfnamefont {D.~C.}\ \bibnamefont {{Backer}}}, \bibinfo {author}
  {\bibfnamefont {J.~D.}\ \bibnamefont {{Bowman}}}, \bibinfo {author}
  {\bibfnamefont {J.~O.}\ \bibnamefont {{Burns}}}, \bibinfo {author}
  {\bibfnamefont {C.~L.}\ \bibnamefont {{Carilli}}}, \bibinfo {author}
  {\bibfnamefont {R.}~\bibnamefont {{Cen}}}, \bibinfo {author} {\bibfnamefont
  {A.}~\bibnamefont {{Cooray}}}, \bibinfo {author} {\bibfnamefont
  {N.}~\bibnamefont {{Gnedin}}}, \bibinfo {author} {\bibfnamefont {L.~J.}\
  \bibnamefont {{Greenhill}}}, \bibinfo {author} {\bibfnamefont
  {Z.}~\bibnamefont {{Haiman}}}, \bibinfo {author} {\bibfnamefont {J.~N.}\
  \bibnamefont {{Hewitt}}}, \ and\ \bibinfo {author} {\bibnamefont {et~al.}},\
  }in\ \href@noop {} {\emph {\bibinfo {booktitle} {astro2010: The Astronomy and
  Astrophysics Decadal Survey}}},\ \bibinfo {series} {Astronomy}, Vol.\
  \bibinfo {volume} {2010}\ (\bibinfo {year} {2009})\ pp.\ \bibinfo {pages}
  {82--+}\BibitemShut {NoStop}%
\bibitem [{\citenamefont {{Wyithe}}\ \emph {et~al.}(2008)\citenamefont
  {{Wyithe}}, \citenamefont {{Loeb}},\ and\ \citenamefont
  {{Geil}}}]{wyithe2008}%
  \BibitemOpen
  \bibfield  {author} {\bibinfo {author} {\bibfnamefont {J.~S.~B.}\
  \bibnamefont {{Wyithe}}}, \bibinfo {author} {\bibfnamefont {A.}~\bibnamefont
  {{Loeb}}}, \ and\ \bibinfo {author} {\bibfnamefont {P.~M.}\ \bibnamefont
  {{Geil}}},\ }\href {\doibase 10.1111/j.1365-2966.2007.12631.x} {\bibfield
  {journal} {\bibinfo  {journal} {MNRAS}\ }\textbf {\bibinfo {volume} {383}},\
  \bibinfo {pages} {1195} (\bibinfo {year} {2008})}\BibitemShut {NoStop}%
\bibitem [{\citenamefont {{Chang}}\ \emph {et~al.}(2008)\citenamefont
  {{Chang}}, \citenamefont {{Pen}}, \citenamefont {{Peterson}},\ and\
  \citenamefont {{McDonald}}}]{ChangDE}%
  \BibitemOpen
  \bibfield  {author} {\bibinfo {author} {\bibfnamefont {T.}~\bibnamefont
  {{Chang}}}, \bibinfo {author} {\bibfnamefont {U.}~\bibnamefont {{Pen}}},
  \bibinfo {author} {\bibfnamefont {J.~B.}\ \bibnamefont {{Peterson}}}, \ and\
  \bibinfo {author} {\bibfnamefont {P.}~\bibnamefont {{McDonald}}},\ }\href
  {\doibase 10.1103/PhysRevLett.100.091303} {\bibfield  {journal} {\bibinfo
  {journal} {Phys. Rev. Lett.}\ }\textbf {\bibinfo {volume} {100}},\ \bibinfo
  {pages} {091303} (\bibinfo {year} {2008})},\ \Eprint
  {http://arxiv.org/abs/0709.3672} {arXiv:0709.3672} \BibitemShut {NoStop}%
\bibitem [{\citenamefont {{Madau}}\ \emph {et~al.}(1997)\citenamefont
  {{Madau}}, \citenamefont {{Meiksin}},\ and\ \citenamefont {{Rees}}}]{Rees}%
  \BibitemOpen
  \bibfield  {author} {\bibinfo {author} {\bibfnamefont {P.}~\bibnamefont
  {{Madau}}}, \bibinfo {author} {\bibfnamefont {A.}~\bibnamefont {{Meiksin}}},
  \ and\ \bibinfo {author} {\bibfnamefont {M.~J.}\ \bibnamefont {{Rees}}},\
  }\href {\doibase 10.1086/303549} {\bibfield  {journal} {\bibinfo  {journal}
  {ApJ}\ }\textbf {\bibinfo {volume} {475}},\ \bibinfo {pages} {429} (\bibinfo
  {year} {1997})},\ \Eprint {http://arxiv.org/abs/arXiv:astro-ph/9608010}
  {arXiv:astro-ph/9608010} \BibitemShut {NoStop}%
\bibitem [{\citenamefont {{Tozzi}}\ \emph
  {et~al.}(2000{\natexlab{a}})\citenamefont {{Tozzi}}, \citenamefont {{Madau}},
  \citenamefont {{Meiksin}},\ and\ \citenamefont {{Rees}}}]{Tozzi2}%
  \BibitemOpen
  \bibfield  {author} {\bibinfo {author} {\bibfnamefont {P.}~\bibnamefont
  {{Tozzi}}}, \bibinfo {author} {\bibfnamefont {P.}~\bibnamefont {{Madau}}},
  \bibinfo {author} {\bibfnamefont {A.}~\bibnamefont {{Meiksin}}}, \ and\
  \bibinfo {author} {\bibfnamefont {M.~J.}\ \bibnamefont {{Rees}}},\ }\href
  {\doibase 10.1086/308196} {\bibfield  {journal} {\bibinfo  {journal} {ApJ}\
  }\textbf {\bibinfo {volume} {528}},\ \bibinfo {pages} {597} (\bibinfo {year}
  {2000}{\natexlab{a}})},\ \Eprint
  {http://arxiv.org/abs/arXiv:astro-ph/9903139} {arXiv:astro-ph/9903139}
  \BibitemShut {NoStop}%
\bibitem [{\citenamefont {{Tozzi}}\ \emph
  {et~al.}(2000{\natexlab{b}})\citenamefont {{Tozzi}}, \citenamefont {{Madau}},
  \citenamefont {{Meiksin}},\ and\ \citenamefont {{Rees}}}]{Tozzi}%
  \BibitemOpen
  \bibfield  {author} {\bibinfo {author} {\bibfnamefont {P.}~\bibnamefont
  {{Tozzi}}}, \bibinfo {author} {\bibfnamefont {P.}~\bibnamefont {{Madau}}},
  \bibinfo {author} {\bibfnamefont {A.}~\bibnamefont {{Meiksin}}}, \ and\
  \bibinfo {author} {\bibfnamefont {M.~J.}\ \bibnamefont {{Rees}}},\
  }\href@noop {} {\bibfield  {journal} {\bibinfo  {journal} {Nuclear Physics B
  Proceedings Supplements}\ }\textbf {\bibinfo {volume} {80}},\ \bibinfo
  {pages} {C509+} (\bibinfo {year} {2000}{\natexlab{b}})},\ \Eprint
  {http://arxiv.org/abs/arXiv:astro-ph/9905199} {arXiv:astro-ph/9905199}
  \BibitemShut {NoStop}%
\bibitem [{\citenamefont {{Iliev}}\ \emph {et~al.}(2002)\citenamefont
  {{Iliev}}, \citenamefont {{Shapiro}}, \citenamefont {{Ferrara}},\ and\
  \citenamefont {{Martel}}}]{Iliev}%
  \BibitemOpen
  \bibfield  {author} {\bibinfo {author} {\bibfnamefont {I.~T.}\ \bibnamefont
  {{Iliev}}}, \bibinfo {author} {\bibfnamefont {P.~R.}\ \bibnamefont
  {{Shapiro}}}, \bibinfo {author} {\bibfnamefont {A.}~\bibnamefont
  {{Ferrara}}}, \ and\ \bibinfo {author} {\bibfnamefont {H.}~\bibnamefont
  {{Martel}}},\ }\href {\doibase 10.1086/341869} {\bibfield  {journal}
  {\bibinfo  {journal} {ApJL}\ }\textbf {\bibinfo {volume} {572}},\ \bibinfo
  {pages} {L123} (\bibinfo {year} {2002})},\ \Eprint
  {http://arxiv.org/abs/arXiv:astro-ph/0202410} {arXiv:astro-ph/0202410}
  \BibitemShut {NoStop}%
\bibitem [{\citenamefont {{Furlanetto}}\ \emph
  {et~al.}(2004{\natexlab{a}})\citenamefont {{Furlanetto}}, \citenamefont
  {{Sokasian}},\ and\ \citenamefont {{Hernquist}}}]{furlanetto1}%
  \BibitemOpen
  \bibfield  {author} {\bibinfo {author} {\bibfnamefont {S.~R.}\ \bibnamefont
  {{Furlanetto}}}, \bibinfo {author} {\bibfnamefont {A.}~\bibnamefont
  {{Sokasian}}}, \ and\ \bibinfo {author} {\bibfnamefont {L.}~\bibnamefont
  {{Hernquist}}},\ }\href {\doibase 10.1111/j.1365-2966.2004.07187.x}
  {\bibfield  {journal} {\bibinfo  {journal} {MNRAS}\ }\textbf {\bibinfo
  {volume} {347}},\ \bibinfo {pages} {187} (\bibinfo {year}
  {2004}{\natexlab{a}})},\ \Eprint
  {http://arxiv.org/abs/arXiv:astro-ph/0305065} {arXiv:astro-ph/0305065}
  \BibitemShut {NoStop}%
\bibitem [{\citenamefont {{Loeb}}\ and\ \citenamefont
  {{Zaldarriaga}}(2004)}]{Loeb1}%
  \BibitemOpen
  \bibfield  {author} {\bibinfo {author} {\bibfnamefont {A.}~\bibnamefont
  {{Loeb}}}\ and\ \bibinfo {author} {\bibfnamefont {M.}~\bibnamefont
  {{Zaldarriaga}}},\ }\href {\doibase 10.1103/PhysRevLett.92.211301} {\bibfield
   {journal} {\bibinfo  {journal} {Physical Review Letters}\ }\textbf {\bibinfo
  {volume} {92}},\ \bibinfo {pages} {211301} (\bibinfo {year} {2004})},\
  \Eprint {http://arxiv.org/abs/arXiv:astro-ph/0312134}
  {arXiv:astro-ph/0312134} \BibitemShut {NoStop}%
\bibitem [{\citenamefont {{Furlanetto}}\ \emph
  {et~al.}(2004{\natexlab{b}})\citenamefont {{Furlanetto}}, \citenamefont
  {{Zaldarriaga}},\ and\ \citenamefont {{Hernquist}}}]{furlanetto2}%
  \BibitemOpen
  \bibfield  {author} {\bibinfo {author} {\bibfnamefont {S.~R.}\ \bibnamefont
  {{Furlanetto}}}, \bibinfo {author} {\bibfnamefont {M.}~\bibnamefont
  {{Zaldarriaga}}}, \ and\ \bibinfo {author} {\bibfnamefont {L.}~\bibnamefont
  {{Hernquist}}},\ }\href {\doibase 10.1086/423028} {\bibfield  {journal}
  {\bibinfo  {journal} {ApJ}\ }\textbf {\bibinfo {volume} {613}},\ \bibinfo
  {pages} {16} (\bibinfo {year} {2004}{\natexlab{b}})},\ \Eprint
  {http://arxiv.org/abs/arXiv:astro-ph/0404112} {arXiv:astro-ph/0404112}
  \BibitemShut {NoStop}%
\bibitem [{\citenamefont {{Barkana}}\ and\ \citenamefont
  {{Loeb}}(2005{\natexlab{a}})}]{Barkana1}%
  \BibitemOpen
  \bibfield  {author} {\bibinfo {author} {\bibfnamefont {R.}~\bibnamefont
  {{Barkana}}}\ and\ \bibinfo {author} {\bibfnamefont {A.}~\bibnamefont
  {{Loeb}}},\ }\href {\doibase 10.1086/429954} {\bibfield  {journal} {\bibinfo
  {journal} {ApJ}\ }\textbf {\bibinfo {volume} {626}},\ \bibinfo {pages} {1}
  (\bibinfo {year} {2005}{\natexlab{a}})},\ \Eprint
  {http://arxiv.org/abs/arXiv:astro-ph/0410129} {arXiv:astro-ph/0410129}
  \BibitemShut {NoStop}%
\bibitem [{\citenamefont {{Mack}}\ and\ \citenamefont {{Wesley}}(2008)}]{Mack}%
  \BibitemOpen
  \bibfield  {author} {\bibinfo {author} {\bibfnamefont {K.~J.}\ \bibnamefont
  {{Mack}}}\ and\ \bibinfo {author} {\bibfnamefont {D.~H.}\ \bibnamefont
  {{Wesley}}},\ }\href@noop {} {\bibfield  {journal} {\bibinfo  {journal}
  {ArXiv e-prints}\ } (\bibinfo {year} {2008})},\ \Eprint
  {http://arxiv.org/abs/0805.1531} {arXiv:0805.1531} \BibitemShut {NoStop}%
\bibitem [{\citenamefont {{Mao}}\ \emph {et~al.}(2008)\citenamefont {{Mao}},
  \citenamefont {{Tegmark}}, \citenamefont {{McQuinn}}, \citenamefont
  {{Zaldarriaga}},\ and\ \citenamefont {{Zahn}}}]{Yi}%
  \BibitemOpen
  \bibfield  {author} {\bibinfo {author} {\bibfnamefont {Y.}~\bibnamefont
  {{Mao}}}, \bibinfo {author} {\bibfnamefont {M.}~\bibnamefont {{Tegmark}}},
  \bibinfo {author} {\bibfnamefont {M.}~\bibnamefont {{McQuinn}}}, \bibinfo
  {author} {\bibfnamefont {M.}~\bibnamefont {{Zaldarriaga}}}, \ and\ \bibinfo
  {author} {\bibfnamefont {O.}~\bibnamefont {{Zahn}}},\ }\href {\doibase
  10.1103/PhysRevD.78.023529} {\bibfield  {journal} {\bibinfo  {journal}
  {\prd}\ }\textbf {\bibinfo {volume} {78}},\ \bibinfo {pages} {023529}
  (\bibinfo {year} {2008})},\ \Eprint {http://arxiv.org/abs/0802.1710}
  {arXiv:0802.1710} \BibitemShut {NoStop}%
\bibitem [{\citenamefont {{Clesse}}\ \emph {et~al.}(2012)\citenamefont
  {{Clesse}}, \citenamefont {{Lopez-Honorez}}, \citenamefont {{Ringeval}},
  \citenamefont {{Tashiro}},\ and\ \citenamefont
  {{Tytgat}}}]{ClesseBackgroundReionizationOmniscopes}%
  \BibitemOpen
  \bibfield  {author} {\bibinfo {author} {\bibfnamefont {S.}~\bibnamefont
  {{Clesse}}}, \bibinfo {author} {\bibfnamefont {L.}~\bibnamefont
  {{Lopez-Honorez}}}, \bibinfo {author} {\bibfnamefont {C.}~\bibnamefont
  {{Ringeval}}}, \bibinfo {author} {\bibfnamefont {H.}~\bibnamefont
  {{Tashiro}}}, \ and\ \bibinfo {author} {\bibfnamefont {M.~H.~G.}\
  \bibnamefont {{Tytgat}}},\ }\href {\doibase 10.1103/PhysRevD.86.123506}
  {\bibfield  {journal} {\bibinfo  {journal} {\prd}\ }\textbf {\bibinfo
  {volume} {86}},\ \bibinfo {eid} {123506} (\bibinfo {year} {2012})},\ \Eprint
  {http://arxiv.org/abs/1208.4277} {arXiv:1208.4277 [astro-ph.CO]} \BibitemShut
  {NoStop}%
\bibitem [{\citenamefont {{de Oliveira-Costa}}\ \emph
  {et~al.}(2008)\citenamefont {{de Oliveira-Costa}}, \citenamefont {{Tegmark}},
  \citenamefont {{Gaensler}}, \citenamefont {{Jonas}}, \citenamefont
  {{Landecker}},\ and\ \citenamefont {{Reich}}}]{Angelica}%
  \BibitemOpen
  \bibfield  {author} {\bibinfo {author} {\bibfnamefont {A.}~\bibnamefont {{de
  Oliveira-Costa}}}, \bibinfo {author} {\bibfnamefont {M.}~\bibnamefont
  {{Tegmark}}}, \bibinfo {author} {\bibfnamefont {B.~M.}\ \bibnamefont
  {{Gaensler}}}, \bibinfo {author} {\bibfnamefont {J.}~\bibnamefont {{Jonas}}},
  \bibinfo {author} {\bibfnamefont {T.~L.}\ \bibnamefont {{Landecker}}}, \ and\
  \bibinfo {author} {\bibfnamefont {P.}~\bibnamefont {{Reich}}},\ }\href
  {\doibase 10.1111/j.1365-2966.2008.13376.x} {\bibfield  {journal} {\bibinfo
  {journal} {MNRAS}\ }\textbf {\bibinfo {volume} {388}},\ \bibinfo {pages}
  {247} (\bibinfo {year} {2008})},\ \Eprint {http://arxiv.org/abs/0802.1525}
  {arXiv:0802.1525} \BibitemShut {NoStop}%
\bibitem [{\citenamefont {{Bernardi}}\ \emph {et~al.}(2011)\citenamefont
  {{Bernardi}}, \citenamefont {{Mitchell}}, \citenamefont {{Ord}},
  \citenamefont {{Greenhill}}, \citenamefont {{Pindor}}, \citenamefont
  {{Wayth}},\ and\ \citenamefont {{Wyithe}}}]{bernardi}%
  \BibitemOpen
  \bibfield  {author} {\bibinfo {author} {\bibfnamefont {G.}~\bibnamefont
  {{Bernardi}}}, \bibinfo {author} {\bibfnamefont {D.~A.}\ \bibnamefont
  {{Mitchell}}}, \bibinfo {author} {\bibfnamefont {S.~M.}\ \bibnamefont
  {{Ord}}}, \bibinfo {author} {\bibfnamefont {L.~J.}\ \bibnamefont
  {{Greenhill}}}, \bibinfo {author} {\bibfnamefont {B.}~\bibnamefont
  {{Pindor}}}, \bibinfo {author} {\bibfnamefont {R.~B.}\ \bibnamefont
  {{Wayth}}}, \ and\ \bibinfo {author} {\bibfnamefont {J.~S.~B.}\ \bibnamefont
  {{Wyithe}}},\ }\href {\doibase 10.1111/j.1365-2966.2010.18145.x} {\bibfield
  {journal} {\bibinfo  {journal} {\mnras}\ }\textbf {\bibinfo {volume} {413}},\
  \bibinfo {pages} {411} (\bibinfo {year} {2011})},\ \Eprint
  {http://arxiv.org/abs/1012.3719} {arXiv:1012.3719 [astro-ph.CO]} \BibitemShut
  {NoStop}%
\bibitem [{\citenamefont {{Liu}}\ and\ \citenamefont {{Tegmark}}(2011)}]{LT11}%
  \BibitemOpen
  \bibfield  {author} {\bibinfo {author} {\bibfnamefont {A.}~\bibnamefont
  {{Liu}}}\ and\ \bibinfo {author} {\bibfnamefont {M.}~\bibnamefont
  {{Tegmark}}},\ }\href {\doibase 10.1103/PhysRevD.83.103006} {\bibfield
  {journal} {\bibinfo  {journal} {Physical Review D}\ }\textbf {\bibinfo
  {volume} {83}},\ \bibinfo {eid} {103006} (\bibinfo {year} {2011})},\ \Eprint
  {http://arxiv.org/abs/1103.0281} {arXiv:1103.0281 [astro-ph.CO]} \BibitemShut
  {NoStop}%
\bibitem [{\citenamefont {{Garrett}}(2009)}]{LOFARinstrument}%
  \BibitemOpen
  \bibfield  {author} {\bibinfo {author} {\bibfnamefont {M.~A.}\ \bibnamefont
  {{Garrett}}},\ }\href@noop {} {\bibfield  {journal} {\bibinfo  {journal}
  {ArXiv e-prints}\ } (\bibinfo {year} {2009})},\ \Eprint
  {http://arxiv.org/abs/0909.3147} {arXiv:0909.3147 [astro-ph.IM]} \BibitemShut
  {NoStop}%
\bibitem [{\citenamefont {{Paciga}}\ \emph {et~al.}(2011)\citenamefont
  {{Paciga}}, \citenamefont {{Chang}}, \citenamefont {{Gupta}}, \citenamefont
  {{Nityanada}}, \citenamefont {{Odegova}}, \citenamefont {{Pen}},
  \citenamefont {{Peterson}}, \citenamefont {{Roy}},\ and\ \citenamefont
  {{Sigurdson}}}]{GMRT}%
  \BibitemOpen
  \bibfield  {author} {\bibinfo {author} {\bibfnamefont {G.}~\bibnamefont
  {{Paciga}}}, \bibinfo {author} {\bibfnamefont {T.-C.}\ \bibnamefont
  {{Chang}}}, \bibinfo {author} {\bibfnamefont {Y.}~\bibnamefont {{Gupta}}},
  \bibinfo {author} {\bibfnamefont {R.}~\bibnamefont {{Nityanada}}}, \bibinfo
  {author} {\bibfnamefont {J.}~\bibnamefont {{Odegova}}}, \bibinfo {author}
  {\bibfnamefont {U.-L.}\ \bibnamefont {{Pen}}}, \bibinfo {author}
  {\bibfnamefont {J.~B.}\ \bibnamefont {{Peterson}}}, \bibinfo {author}
  {\bibfnamefont {J.}~\bibnamefont {{Roy}}}, \ and\ \bibinfo {author}
  {\bibfnamefont {K.}~\bibnamefont {{Sigurdson}}},\ }\href {\doibase
  10.1111/j.1365-2966.2011.18208.x} {\bibfield  {journal} {\bibinfo  {journal}
  {\mnras}\ }\textbf {\bibinfo {volume} {413}},\ \bibinfo {pages} {1174}
  (\bibinfo {year} {2011})},\ \Eprint {http://arxiv.org/abs/1006.1351}
  {arXiv:1006.1351 [astro-ph.CO]} \BibitemShut {NoStop}%
\bibitem [{\citenamefont {{Tingay}}\ \emph {et~al.}(2012)\citenamefont
  {{Tingay}}, \citenamefont {{Goeke}}, \citenamefont {{Bowman}}, \citenamefont
  {{Emrich}}, \citenamefont {{Ord}}, \citenamefont {{Mitchell}}, \citenamefont
  {{Morales}}, \citenamefont {{Booler}}, \citenamefont {{Crosse}},
  \citenamefont {{Pallot}}, \citenamefont {{Wicenec}}, \citenamefont {{Arcus}},
  \citenamefont {{Barnes}}, \citenamefont {{Bernardi}}, \citenamefont
  {{Briggs}}, \citenamefont {{Burns}}, \citenamefont {{Bunton}}, \citenamefont
  {{Cappallo}}, \citenamefont {{Colegate}}, \citenamefont {{Corey}},
  \citenamefont {{Deshpande}}, \citenamefont {{deSouza}}, \citenamefont
  {{Gaensler}}, \citenamefont {{Greenhill}}, \citenamefont {{Hall}},
  \citenamefont {{Hazelton}}, \citenamefont {{Herne}}, \citenamefont
  {{Hewitt}}, \citenamefont {{Johnston-Hollitt}}, \citenamefont {{Kaplan}},
  \citenamefont {{Kasper}}, \citenamefont {{Kincaid}}, \citenamefont
  {{Koenig}}, \citenamefont {{Kratzenberg}}, \citenamefont {{Lonsdale}},
  \citenamefont {{Lynch}}, \citenamefont {{McKinley}}, \citenamefont
  {{McWhirter}}, \citenamefont {{Morgan}}, \citenamefont {{Oberoi}},
  \citenamefont {{Pathikulangara}}, \citenamefont {{Prabu}}, \citenamefont
  {{Remillard}}, \citenamefont {{Rogers}}, \citenamefont {{Roshi}},
  \citenamefont {{Salah}}, \citenamefont {{Sault}}, \citenamefont
  {{Udaya-Shankar}}, \citenamefont {{Schlagenhaufer}}, \citenamefont
  {{Srivani}}, \citenamefont {{Stevens}}, \citenamefont {{Subrahmanyan}},
  \citenamefont {{Tremblay}}, \citenamefont {{Wayth}}, \citenamefont
  {{Waterson}}, \citenamefont {{Webster}}, \citenamefont {{Whitney}},
  \citenamefont {{Williams}}, \citenamefont {{Williams}},\ and\ \citenamefont
  {{Wyithe}}}]{MWAstatus}%
  \BibitemOpen
  \bibfield  {author} {\bibinfo {author} {\bibfnamefont {S.~J.}\ \bibnamefont
  {{Tingay}}}, \bibinfo {author} {\bibfnamefont {R.}~\bibnamefont {{Goeke}}},
  \bibinfo {author} {\bibfnamefont {J.~D.}\ \bibnamefont {{Bowman}}}, \bibinfo
  {author} {\bibfnamefont {D.}~\bibnamefont {{Emrich}}}, \bibinfo {author}
  {\bibfnamefont {S.~M.}\ \bibnamefont {{Ord}}}, \bibinfo {author}
  {\bibfnamefont {D.~A.}\ \bibnamefont {{Mitchell}}}, \bibinfo {author}
  {\bibfnamefont {M.~F.}\ \bibnamefont {{Morales}}}, \bibinfo {author}
  {\bibfnamefont {T.}~\bibnamefont {{Booler}}}, \bibinfo {author}
  {\bibfnamefont {B.}~\bibnamefont {{Crosse}}}, \bibinfo {author}
  {\bibfnamefont {D.}~\bibnamefont {{Pallot}}}, \bibinfo {author}
  {\bibfnamefont {A.}~\bibnamefont {{Wicenec}}}, \bibinfo {author}
  {\bibfnamefont {W.}~\bibnamefont {{Arcus}}}, \bibinfo {author} {\bibfnamefont
  {D.}~\bibnamefont {{Barnes}}}, \bibinfo {author} {\bibfnamefont
  {G.}~\bibnamefont {{Bernardi}}}, \bibinfo {author} {\bibfnamefont
  {F.}~\bibnamefont {{Briggs}}}, \bibinfo {author} {\bibfnamefont
  {S.}~\bibnamefont {{Burns}}}, \bibinfo {author} {\bibfnamefont {J.~D.}\
  \bibnamefont {{Bunton}}}, \bibinfo {author} {\bibfnamefont {R.~J.}\
  \bibnamefont {{Cappallo}}}, \bibinfo {author} {\bibfnamefont
  {T.}~\bibnamefont {{Colegate}}}, \bibinfo {author} {\bibfnamefont {B.~E.}\
  \bibnamefont {{Corey}}}, \bibinfo {author} {\bibfnamefont {A.}~\bibnamefont
  {{Deshpande}}}, \bibinfo {author} {\bibfnamefont {L.}~\bibnamefont
  {{deSouza}}}, \bibinfo {author} {\bibfnamefont {B.~M.}\ \bibnamefont
  {{Gaensler}}}, \bibinfo {author} {\bibfnamefont {L.~J.}\ \bibnamefont
  {{Greenhill}}}, \bibinfo {author} {\bibfnamefont {J.}~\bibnamefont {{Hall}}},
  \bibinfo {author} {\bibfnamefont {B.~J.}\ \bibnamefont {{Hazelton}}},
  \bibinfo {author} {\bibfnamefont {D.}~\bibnamefont {{Herne}}}, \bibinfo
  {author} {\bibfnamefont {J.~N.}\ \bibnamefont {{Hewitt}}}, \bibinfo {author}
  {\bibfnamefont {M.}~\bibnamefont {{Johnston-Hollitt}}}, \bibinfo {author}
  {\bibfnamefont {D.~L.}\ \bibnamefont {{Kaplan}}}, \bibinfo {author}
  {\bibfnamefont {J.~C.}\ \bibnamefont {{Kasper}}}, \bibinfo {author}
  {\bibfnamefont {B.~B.}\ \bibnamefont {{Kincaid}}}, \bibinfo {author}
  {\bibfnamefont {R.}~\bibnamefont {{Koenig}}}, \bibinfo {author}
  {\bibfnamefont {E.}~\bibnamefont {{Kratzenberg}}}, \bibinfo {author}
  {\bibfnamefont {C.~J.}\ \bibnamefont {{Lonsdale}}}, \bibinfo {author}
  {\bibfnamefont {M.~J.}\ \bibnamefont {{Lynch}}}, \bibinfo {author}
  {\bibfnamefont {B.}~\bibnamefont {{McKinley}}}, \bibinfo {author}
  {\bibfnamefont {S.~R.}\ \bibnamefont {{McWhirter}}}, \bibinfo {author}
  {\bibfnamefont {E.}~\bibnamefont {{Morgan}}}, \bibinfo {author}
  {\bibfnamefont {D.}~\bibnamefont {{Oberoi}}}, \bibinfo {author}
  {\bibfnamefont {J.}~\bibnamefont {{Pathikulangara}}}, \bibinfo {author}
  {\bibfnamefont {T.}~\bibnamefont {{Prabu}}}, \bibinfo {author} {\bibfnamefont
  {R.~A.}\ \bibnamefont {{Remillard}}}, \bibinfo {author} {\bibfnamefont
  {A.~E.~E.}\ \bibnamefont {{Rogers}}}, \bibinfo {author} {\bibfnamefont
  {A.}~\bibnamefont {{Roshi}}}, \bibinfo {author} {\bibfnamefont {J.~E.}\
  \bibnamefont {{Salah}}}, \bibinfo {author} {\bibfnamefont {R.~J.}\
  \bibnamefont {{Sault}}}, \bibinfo {author} {\bibfnamefont {N.}~\bibnamefont
  {{Udaya-Shankar}}}, \bibinfo {author} {\bibfnamefont {F.}~\bibnamefont
  {{Schlagenhaufer}}}, \bibinfo {author} {\bibfnamefont {K.~S.}\ \bibnamefont
  {{Srivani}}}, \bibinfo {author} {\bibfnamefont {J.}~\bibnamefont
  {{Stevens}}}, \bibinfo {author} {\bibfnamefont {R.}~\bibnamefont
  {{Subrahmanyan}}}, \bibinfo {author} {\bibfnamefont {S.}~\bibnamefont
  {{Tremblay}}}, \bibinfo {author} {\bibfnamefont {R.~B.}\ \bibnamefont
  {{Wayth}}}, \bibinfo {author} {\bibfnamefont {M.}~\bibnamefont {{Waterson}}},
  \bibinfo {author} {\bibfnamefont {R.~L.}\ \bibnamefont {{Webster}}}, \bibinfo
  {author} {\bibfnamefont {A.~R.}\ \bibnamefont {{Whitney}}}, \bibinfo {author}
  {\bibfnamefont {A.}~\bibnamefont {{Williams}}}, \bibinfo {author}
  {\bibfnamefont {C.~L.}\ \bibnamefont {{Williams}}}, \ and\ \bibinfo {author}
  {\bibfnamefont {J.~S.~B.}\ \bibnamefont {{Wyithe}}},\ }\href@noop {}
  {\bibfield  {journal} {\bibinfo  {journal} {ArXiv e-prints}\ } (\bibinfo
  {year} {2012})},\ \Eprint {http://arxiv.org/abs/1206.6945} {arXiv:1206.6945
  [astro-ph.IM]} \BibitemShut {NoStop}%
\bibitem [{\citenamefont {{Parsons}}\ \emph {et~al.}(2010)\citenamefont
  {{Parsons}}, \citenamefont {{Backer}}, \citenamefont {{Foster}},
  \citenamefont {{Wright}}, \citenamefont {{Bradley}}, \citenamefont
  {{Gugliucci}}, \citenamefont {{Parashare}}, \citenamefont {{Benoit}},
  \citenamefont {{Aguirre}}, \citenamefont {{Jacobs}}, \citenamefont
  {{Carilli}}, \citenamefont {{Herne}}, \citenamefont {{Lynch}}, \citenamefont
  {{Manley}},\ and\ \citenamefont {{Werthimer}}}]{PAPER}%
  \BibitemOpen
  \bibfield  {author} {\bibinfo {author} {\bibfnamefont {A.~R.}\ \bibnamefont
  {{Parsons}}}, \bibinfo {author} {\bibfnamefont {D.~C.}\ \bibnamefont
  {{Backer}}}, \bibinfo {author} {\bibfnamefont {G.~S.}\ \bibnamefont
  {{Foster}}}, \bibinfo {author} {\bibfnamefont {M.~C.~H.}\ \bibnamefont
  {{Wright}}}, \bibinfo {author} {\bibfnamefont {R.~F.}\ \bibnamefont
  {{Bradley}}}, \bibinfo {author} {\bibfnamefont {N.~E.}\ \bibnamefont
  {{Gugliucci}}}, \bibinfo {author} {\bibfnamefont {C.~R.}\ \bibnamefont
  {{Parashare}}}, \bibinfo {author} {\bibfnamefont {E.~E.}\ \bibnamefont
  {{Benoit}}}, \bibinfo {author} {\bibfnamefont {J.~E.}\ \bibnamefont
  {{Aguirre}}}, \bibinfo {author} {\bibfnamefont {D.~C.}\ \bibnamefont
  {{Jacobs}}}, \bibinfo {author} {\bibfnamefont {C.~L.}\ \bibnamefont
  {{Carilli}}}, \bibinfo {author} {\bibfnamefont {D.}~\bibnamefont {{Herne}}},
  \bibinfo {author} {\bibfnamefont {M.~J.}\ \bibnamefont {{Lynch}}}, \bibinfo
  {author} {\bibfnamefont {J.~R.}\ \bibnamefont {{Manley}}}, \ and\ \bibinfo
  {author} {\bibfnamefont {D.~J.}\ \bibnamefont {{Werthimer}}},\ }\href
  {\doibase 10.1088/0004-6256/139/4/1468} {\bibfield  {journal} {\bibinfo
  {journal} {\aj}\ }\textbf {\bibinfo {volume} {139}},\ \bibinfo {pages} {1468}
  (\bibinfo {year} {2010})},\ \Eprint {http://arxiv.org/abs/0904.2334}
  {arXiv:0904.2334 [astro-ph.CO]} \BibitemShut {NoStop}%
\bibitem [{\citenamefont {{Williams}}\ \emph {et~al.}(2012)\citenamefont
  {{Williams}}, \citenamefont {{Hewitt}}, \citenamefont {{Levine}},
  \citenamefont {{de Oliveira-Costa}}, \citenamefont {{Bowman}}, \citenamefont
  {{Briggs}}, \citenamefont {{Gaensler}}, \citenamefont {{Hernquist}},
  \citenamefont {{Mitchell}}, \citenamefont {{Morales}}, \citenamefont
  {{Sethi}}, \citenamefont {{Subrahmanyan}}, \citenamefont {{Sadler}},
  \citenamefont {{Arcus}}, \citenamefont {{Barnes}}, \citenamefont
  {{Bernardi}}, \citenamefont {{Bunton}}, \citenamefont {{Cappallo}},
  \citenamefont {{Crosse}}, \citenamefont {{Corey}}, \citenamefont
  {{Deshpande}}, \citenamefont {{deSouza}}, \citenamefont {{Emrich}},
  \citenamefont {{Goeke}}, \citenamefont {{Greenhill}}, \citenamefont
  {{Hazelton}}, \citenamefont {{Herne}}, \citenamefont {{Kaplan}},
  \citenamefont {{Kasper}}, \citenamefont {{Kincaid}}, \citenamefont
  {{Koenig}}, \citenamefont {{Kratzenberg}}, \citenamefont {{Lonsdale}},
  \citenamefont {{Lynch}}, \citenamefont {{McWhirter}}, \citenamefont
  {{Morgan}}, \citenamefont {{Oberoi}}, \citenamefont {{Ord}}, \citenamefont
  {{Pathikulangara}}, \citenamefont {{Prabu}}, \citenamefont {{Remillard}},
  \citenamefont {{Rogers}}, \citenamefont {{Anish Roshi}}, \citenamefont
  {{Salah}}, \citenamefont {{Sault}}, \citenamefont {{Udaya Shankar}},
  \citenamefont {{Srivani}}, \citenamefont {{Stevens}}, \citenamefont
  {{Tingay}}, \citenamefont {{Wayth}}, \citenamefont {{Waterson}},
  \citenamefont {{Webster}}, \citenamefont {{Whitney}}, \citenamefont
  {{Williams}},\ and\ \citenamefont {{Wyithe}}}]{ChrisMWA}%
  \BibitemOpen
  \bibfield  {author} {\bibinfo {author} {\bibfnamefont {C.~L.}\ \bibnamefont
  {{Williams}}}, \bibinfo {author} {\bibfnamefont {J.~N.}\ \bibnamefont
  {{Hewitt}}}, \bibinfo {author} {\bibfnamefont {A.~M.}\ \bibnamefont
  {{Levine}}}, \bibinfo {author} {\bibfnamefont {A.}~\bibnamefont {{de
  Oliveira-Costa}}}, \bibinfo {author} {\bibfnamefont {J.~D.}\ \bibnamefont
  {{Bowman}}}, \bibinfo {author} {\bibfnamefont {F.~H.}\ \bibnamefont
  {{Briggs}}}, \bibinfo {author} {\bibfnamefont {B.~M.}\ \bibnamefont
  {{Gaensler}}}, \bibinfo {author} {\bibfnamefont {L.~L.}\ \bibnamefont
  {{Hernquist}}}, \bibinfo {author} {\bibfnamefont {D.~A.}\ \bibnamefont
  {{Mitchell}}}, \bibinfo {author} {\bibfnamefont {M.~F.}\ \bibnamefont
  {{Morales}}}, \bibinfo {author} {\bibfnamefont {S.~K.}\ \bibnamefont
  {{Sethi}}}, \bibinfo {author} {\bibfnamefont {R.}~\bibnamefont
  {{Subrahmanyan}}}, \bibinfo {author} {\bibfnamefont {E.~M.}\ \bibnamefont
  {{Sadler}}}, \bibinfo {author} {\bibfnamefont {W.}~\bibnamefont {{Arcus}}},
  \bibinfo {author} {\bibfnamefont {D.~G.}\ \bibnamefont {{Barnes}}}, \bibinfo
  {author} {\bibfnamefont {G.}~\bibnamefont {{Bernardi}}}, \bibinfo {author}
  {\bibfnamefont {J.~D.}\ \bibnamefont {{Bunton}}}, \bibinfo {author}
  {\bibfnamefont {R.~C.}\ \bibnamefont {{Cappallo}}}, \bibinfo {author}
  {\bibfnamefont {B.~W.}\ \bibnamefont {{Crosse}}}, \bibinfo {author}
  {\bibfnamefont {B.~E.}\ \bibnamefont {{Corey}}}, \bibinfo {author}
  {\bibfnamefont {A.}~\bibnamefont {{Deshpande}}}, \bibinfo {author}
  {\bibfnamefont {L.}~\bibnamefont {{deSouza}}}, \bibinfo {author}
  {\bibfnamefont {D.}~\bibnamefont {{Emrich}}}, \bibinfo {author}
  {\bibfnamefont {R.~F.}\ \bibnamefont {{Goeke}}}, \bibinfo {author}
  {\bibfnamefont {L.~J.}\ \bibnamefont {{Greenhill}}}, \bibinfo {author}
  {\bibfnamefont {B.~J.}\ \bibnamefont {{Hazelton}}}, \bibinfo {author}
  {\bibfnamefont {D.}~\bibnamefont {{Herne}}}, \bibinfo {author} {\bibfnamefont
  {D.~L.}\ \bibnamefont {{Kaplan}}}, \bibinfo {author} {\bibfnamefont {J.~C.}\
  \bibnamefont {{Kasper}}}, \bibinfo {author} {\bibfnamefont {B.~B.}\
  \bibnamefont {{Kincaid}}}, \bibinfo {author} {\bibfnamefont {R.}~\bibnamefont
  {{Koenig}}}, \bibinfo {author} {\bibfnamefont {E.}~\bibnamefont
  {{Kratzenberg}}}, \bibinfo {author} {\bibfnamefont {C.~J.}\ \bibnamefont
  {{Lonsdale}}}, \bibinfo {author} {\bibfnamefont {M.~J.}\ \bibnamefont
  {{Lynch}}}, \bibinfo {author} {\bibfnamefont {S.~R.}\ \bibnamefont
  {{McWhirter}}}, \bibinfo {author} {\bibfnamefont {E.~H.}\ \bibnamefont
  {{Morgan}}}, \bibinfo {author} {\bibfnamefont {D.}~\bibnamefont {{Oberoi}}},
  \bibinfo {author} {\bibfnamefont {S.~M.}\ \bibnamefont {{Ord}}}, \bibinfo
  {author} {\bibfnamefont {J.}~\bibnamefont {{Pathikulangara}}}, \bibinfo
  {author} {\bibfnamefont {T.}~\bibnamefont {{Prabu}}}, \bibinfo {author}
  {\bibfnamefont {R.~A.}\ \bibnamefont {{Remillard}}}, \bibinfo {author}
  {\bibfnamefont {A.~E.~E.}\ \bibnamefont {{Rogers}}}, \bibinfo {author}
  {\bibfnamefont {D.}~\bibnamefont {{Anish Roshi}}}, \bibinfo {author}
  {\bibfnamefont {J.~E.}\ \bibnamefont {{Salah}}}, \bibinfo {author}
  {\bibfnamefont {R.~J.}\ \bibnamefont {{Sault}}}, \bibinfo {author}
  {\bibfnamefont {N.}~\bibnamefont {{Udaya Shankar}}}, \bibinfo {author}
  {\bibfnamefont {K.~S.}\ \bibnamefont {{Srivani}}}, \bibinfo {author}
  {\bibfnamefont {J.~B.}\ \bibnamefont {{Stevens}}}, \bibinfo {author}
  {\bibfnamefont {S.~J.}\ \bibnamefont {{Tingay}}}, \bibinfo {author}
  {\bibfnamefont {R.~B.}\ \bibnamefont {{Wayth}}}, \bibinfo {author}
  {\bibfnamefont {M.}~\bibnamefont {{Waterson}}}, \bibinfo {author}
  {\bibfnamefont {R.~L.}\ \bibnamefont {{Webster}}}, \bibinfo {author}
  {\bibfnamefont {A.~R.}\ \bibnamefont {{Whitney}}}, \bibinfo {author}
  {\bibfnamefont {A.~J.}\ \bibnamefont {{Williams}}}, \ and\ \bibinfo {author}
  {\bibfnamefont {J.~S.~B.}\ \bibnamefont {{Wyithe}}},\ }\href {\doibase
  10.1088/0004-637X/755/1/47} {\bibfield  {journal} {\bibinfo  {journal}
  {\apj}\ }\textbf {\bibinfo {volume} {755}},\ \bibinfo {eid} {47} (\bibinfo
  {year} {2012})},\ \Eprint {http://arxiv.org/abs/1203.5790} {arXiv:1203.5790
  [astro-ph.CO]} \BibitemShut {NoStop}%
\bibitem [{\citenamefont {{Greenhill}}\ and\ \citenamefont
  {{Bernardi}}(2012)}]{HERA}%
  \BibitemOpen
  \bibfield  {author} {\bibinfo {author} {\bibfnamefont {L.~J.}\ \bibnamefont
  {{Greenhill}}}\ and\ \bibinfo {author} {\bibfnamefont {G.}~\bibnamefont
  {{Bernardi}}},\ }\href@noop {} {\bibfield  {journal} {\bibinfo  {journal}
  {ArXiv e-prints}\ } (\bibinfo {year} {2012})},\ \Eprint
  {http://arxiv.org/abs/1201.1700} {arXiv:1201.1700 [astro-ph.CO]} \BibitemShut
  {NoStop}%
\bibitem [{\citenamefont {{Tegmark}}\ and\ \citenamefont
  {{Zaldarriaga}}(2010)}]{FFTT2}%
  \BibitemOpen
  \bibfield  {author} {\bibinfo {author} {\bibfnamefont {M.}~\bibnamefont
  {{Tegmark}}}\ and\ \bibinfo {author} {\bibfnamefont {M.}~\bibnamefont
  {{Zaldarriaga}}},\ }\href {\doibase 10.1103/PhysRevD.82.103501} {\bibfield
  {journal} {\bibinfo  {journal} {\prd}\ }\textbf {\bibinfo {volume} {82}},\
  \bibinfo {eid} {103501} (\bibinfo {year} {2010})},\ \Eprint
  {http://arxiv.org/abs/0909.0001} {arXiv:0909.0001 [astro-ph.CO]} \BibitemShut
  {NoStop}%
\bibitem [{\citenamefont {{Carilli}}\ and\ \citenamefont
  {{Rawlings}}(2004)}]{SKAspecifications}%
  \BibitemOpen
  \bibfield  {author} {\bibinfo {author} {\bibfnamefont {C.~L.}\ \bibnamefont
  {{Carilli}}}\ and\ \bibinfo {author} {\bibfnamefont {S.}~\bibnamefont
  {{Rawlings}}},\ }\href {\doibase 10.1016/j.newar.2004.09.001} {\bibfield
  {journal} {\bibinfo  {journal} {\nar}\ }\textbf {\bibinfo {volume} {48}},\
  \bibinfo {pages} {979} (\bibinfo {year} {2004})},\ \Eprint
  {http://arxiv.org/abs/arXiv:astro-ph/0409274} {arXiv:astro-ph/0409274}
  \BibitemShut {NoStop}%
\bibitem [{\citenamefont {{Pen}}(2003)}]{UeLiFast}%
  \BibitemOpen
  \bibfield  {author} {\bibinfo {author} {\bibfnamefont {U.-L.}\ \bibnamefont
  {{Pen}}},\ }\href {\doibase 10.1046/j.1365-2966.2003.07118.x} {\bibfield
  {journal} {\bibinfo  {journal} {\mnras}\ }\textbf {\bibinfo {volume} {346}},\
  \bibinfo {pages} {619} (\bibinfo {year} {2003})},\ \Eprint
  {http://arxiv.org/abs/arXiv:astro-ph/0304513} {arXiv:astro-ph/0304513}
  \BibitemShut {NoStop}%
\bibitem [{\citenamefont {{Sutter}}\ \emph {et~al.}(2012)\citenamefont
  {{Sutter}}, \citenamefont {{Wandelt}},\ and\ \citenamefont
  {{Malu}}}]{GibbsPSE}%
  \BibitemOpen
  \bibfield  {author} {\bibinfo {author} {\bibfnamefont {P.~M.}\ \bibnamefont
  {{Sutter}}}, \bibinfo {author} {\bibfnamefont {B.~D.}\ \bibnamefont
  {{Wandelt}}}, \ and\ \bibinfo {author} {\bibfnamefont {S.~S.}\ \bibnamefont
  {{Malu}}},\ }\href {\doibase 10.1088/0067-0049/202/1/9} {\bibfield  {journal}
  {\bibinfo  {journal} {\apjs}\ }\textbf {\bibinfo {volume} {202}},\ \bibinfo
  {eid} {9} (\bibinfo {year} {2012})},\ \Eprint
  {http://arxiv.org/abs/1109.4640} {arXiv:1109.4640 [astro-ph.CO]} \BibitemShut
  {NoStop}%
\bibitem [{\citenamefont {{Wang}}\ \emph {et~al.}(2006)\citenamefont {{Wang}},
  \citenamefont {{Tegmark}}, \citenamefont {{Santos}},\ and\ \citenamefont
  {{Knox}}}]{xiaomin}%
  \BibitemOpen
  \bibfield  {author} {\bibinfo {author} {\bibfnamefont {X.}~\bibnamefont
  {{Wang}}}, \bibinfo {author} {\bibfnamefont {M.}~\bibnamefont {{Tegmark}}},
  \bibinfo {author} {\bibfnamefont {M.~G.}\ \bibnamefont {{Santos}}}, \ and\
  \bibinfo {author} {\bibfnamefont {L.}~\bibnamefont {{Knox}}},\ }\href
  {\doibase 10.1086/506597} {\bibfield  {journal} {\bibinfo  {journal} {ApJ}\
  }\textbf {\bibinfo {volume} {650}},\ \bibinfo {pages} {529} (\bibinfo {year}
  {2006})},\ \Eprint {http://arxiv.org/abs/arXiv:astro-ph/0501081}
  {arXiv:astro-ph/0501081} \BibitemShut {NoStop}%
\bibitem [{\citenamefont {{Gleser}}\ \emph {et~al.}(2008)\citenamefont
  {{Gleser}}, \citenamefont {{Nusser}},\ and\ \citenamefont
  {{Benson}}}]{nusserforegrounds}%
  \BibitemOpen
  \bibfield  {author} {\bibinfo {author} {\bibfnamefont {L.}~\bibnamefont
  {{Gleser}}}, \bibinfo {author} {\bibfnamefont {A.}~\bibnamefont {{Nusser}}},
  \ and\ \bibinfo {author} {\bibfnamefont {A.~J.}\ \bibnamefont {{Benson}}},\
  }\href {\doibase 10.1111/j.1365-2966.2008.13897.x} {\bibfield  {journal}
  {\bibinfo  {journal} {MNRAS}\ }\textbf {\bibinfo {volume} {391}},\ \bibinfo
  {pages} {383} (\bibinfo {year} {2008})},\ \Eprint
  {http://arxiv.org/abs/0712.0497} {arXiv:0712.0497} \BibitemShut {NoStop}%
\bibitem [{\citenamefont {{Bowman}}\ \emph {et~al.}(2009)\citenamefont
  {{Bowman}}, \citenamefont {{Morales}},\ and\ \citenamefont
  {{Hewitt}}}]{Judd08}%
  \BibitemOpen
  \bibfield  {author} {\bibinfo {author} {\bibfnamefont {J.~D.}\ \bibnamefont
  {{Bowman}}}, \bibinfo {author} {\bibfnamefont {M.~F.}\ \bibnamefont
  {{Morales}}}, \ and\ \bibinfo {author} {\bibfnamefont {J.~N.}\ \bibnamefont
  {{Hewitt}}},\ }\href {\doibase 10.1088/0004-637X/695/1/183} {\bibfield
  {journal} {\bibinfo  {journal} {ApJ}\ }\textbf {\bibinfo {volume} {695}},\
  \bibinfo {pages} {183} (\bibinfo {year} {2009})},\ \Eprint
  {http://arxiv.org/abs/0807.3956} {arXiv:0807.3956} \BibitemShut {NoStop}%
\bibitem [{\citenamefont {{Liu}}\ \emph
  {et~al.}(2009{\natexlab{a}})\citenamefont {{Liu}}, \citenamefont
  {{Tegmark}},\ and\ \citenamefont {{Zaldarriaga}}}]{paper1}%
  \BibitemOpen
  \bibfield  {author} {\bibinfo {author} {\bibfnamefont {A.}~\bibnamefont
  {{Liu}}}, \bibinfo {author} {\bibfnamefont {M.}~\bibnamefont {{Tegmark}}}, \
  and\ \bibinfo {author} {\bibfnamefont {M.}~\bibnamefont {{Zaldarriaga}}},\
  }\href {\doibase 10.1111/j.1365-2966.2009.14426.x} {\bibfield  {journal}
  {\bibinfo  {journal} {MNRAS}\ }\textbf {\bibinfo {volume} {394}},\ \bibinfo
  {pages} {1575} (\bibinfo {year} {2009}{\natexlab{a}})},\ \Eprint
  {http://arxiv.org/abs/0807.3952} {arXiv:0807.3952} \BibitemShut {NoStop}%
\bibitem [{\citenamefont {{Jeli{\'c}}}\ \emph {et~al.}(2008)\citenamefont
  {{Jeli{\'c}}}, \citenamefont {{Zaroubi}}, \citenamefont {{Labropoulos}},
  \citenamefont {{Thomas}}, \citenamefont {{Bernardi}}, \citenamefont
  {{Brentjens}}, \citenamefont {{de Bruyn}}, \citenamefont {{Ciardi}},
  \citenamefont {{Harker}}, \citenamefont {{Koopmans}}, \citenamefont
  {{Pandey}}, \citenamefont {{Schaye}},\ and\ \citenamefont
  {{Yatawatta}}}]{LOFAR}%
  \BibitemOpen
  \bibfield  {author} {\bibinfo {author} {\bibfnamefont {V.}~\bibnamefont
  {{Jeli{\'c}}}}, \bibinfo {author} {\bibfnamefont {S.}~\bibnamefont
  {{Zaroubi}}}, \bibinfo {author} {\bibfnamefont {P.}~\bibnamefont
  {{Labropoulos}}}, \bibinfo {author} {\bibfnamefont {R.~M.}\ \bibnamefont
  {{Thomas}}}, \bibinfo {author} {\bibfnamefont {G.}~\bibnamefont
  {{Bernardi}}}, \bibinfo {author} {\bibfnamefont {M.~A.}\ \bibnamefont
  {{Brentjens}}}, \bibinfo {author} {\bibfnamefont {A.~G.}\ \bibnamefont {{de
  Bruyn}}}, \bibinfo {author} {\bibfnamefont {B.}~\bibnamefont {{Ciardi}}},
  \bibinfo {author} {\bibfnamefont {G.}~\bibnamefont {{Harker}}}, \bibinfo
  {author} {\bibfnamefont {L.~V.~E.}\ \bibnamefont {{Koopmans}}}, \bibinfo
  {author} {\bibfnamefont {V.~N.}\ \bibnamefont {{Pandey}}}, \bibinfo {author}
  {\bibfnamefont {J.}~\bibnamefont {{Schaye}}}, \ and\ \bibinfo {author}
  {\bibfnamefont {S.}~\bibnamefont {{Yatawatta}}},\ }\href {\doibase
  10.1111/j.1365-2966.2008.13634.x} {\bibfield  {journal} {\bibinfo  {journal}
  {MNRAS}\ }\textbf {\bibinfo {volume} {389}},\ \bibinfo {pages} {1319}
  (\bibinfo {year} {2008})},\ \Eprint {http://arxiv.org/abs/0804.1130}
  {arXiv:0804.1130} \BibitemShut {NoStop}%
\bibitem [{\citenamefont {{Harker}}\ \emph {et~al.}(2009)\citenamefont
  {{Harker}}, \citenamefont {{Zaroubi}}, \citenamefont {{Bernardi}},
  \citenamefont {{Brentjens}}, \citenamefont {{de Bruyn}}, \citenamefont
  {{Ciardi}}, \citenamefont {{Jeli{\'c}}}, \citenamefont {{Koopmans}},
  \citenamefont {{Labropoulos}}, \citenamefont {{Mellema}}, \citenamefont
  {{Offringa}}, \citenamefont {{Pandey}}, \citenamefont {{Schaye}},
  \citenamefont {{Thomas}},\ and\ \citenamefont {{Yatawatta}}}]{Harker}%
  \BibitemOpen
  \bibfield  {author} {\bibinfo {author} {\bibfnamefont {G.}~\bibnamefont
  {{Harker}}}, \bibinfo {author} {\bibfnamefont {S.}~\bibnamefont {{Zaroubi}}},
  \bibinfo {author} {\bibfnamefont {G.}~\bibnamefont {{Bernardi}}}, \bibinfo
  {author} {\bibfnamefont {M.~A.}\ \bibnamefont {{Brentjens}}}, \bibinfo
  {author} {\bibfnamefont {A.~G.}\ \bibnamefont {{de Bruyn}}}, \bibinfo
  {author} {\bibfnamefont {B.}~\bibnamefont {{Ciardi}}}, \bibinfo {author}
  {\bibfnamefont {V.}~\bibnamefont {{Jeli{\'c}}}}, \bibinfo {author}
  {\bibfnamefont {L.~V.~E.}\ \bibnamefont {{Koopmans}}}, \bibinfo {author}
  {\bibfnamefont {P.}~\bibnamefont {{Labropoulos}}}, \bibinfo {author}
  {\bibfnamefont {G.}~\bibnamefont {{Mellema}}}, \bibinfo {author}
  {\bibfnamefont {A.}~\bibnamefont {{Offringa}}}, \bibinfo {author}
  {\bibfnamefont {V.~N.}\ \bibnamefont {{Pandey}}}, \bibinfo {author}
  {\bibfnamefont {J.}~\bibnamefont {{Schaye}}}, \bibinfo {author}
  {\bibfnamefont {R.~M.}\ \bibnamefont {{Thomas}}}, \ and\ \bibinfo {author}
  {\bibfnamefont {S.}~\bibnamefont {{Yatawatta}}},\ }\href {\doibase
  10.1111/j.1365-2966.2009.15081.x} {\bibfield  {journal} {\bibinfo  {journal}
  {\mnras}\ }\textbf {\bibinfo {volume} {397}},\ \bibinfo {pages} {1138}
  (\bibinfo {year} {2009})},\ \Eprint {http://arxiv.org/abs/0903.2760}
  {arXiv:0903.2760 [astro-ph.CO]} \BibitemShut {NoStop}%
\bibitem [{\citenamefont {{Liu}}\ \emph
  {et~al.}(2009{\natexlab{b}})\citenamefont {{Liu}}, \citenamefont {{Tegmark}},
  \citenamefont {{Bowman}}, \citenamefont {{Hewitt}},\ and\ \citenamefont
  {{Zaldarriaga}}}]{paper2}%
  \BibitemOpen
  \bibfield  {author} {\bibinfo {author} {\bibfnamefont {A.}~\bibnamefont
  {{Liu}}}, \bibinfo {author} {\bibfnamefont {M.}~\bibnamefont {{Tegmark}}},
  \bibinfo {author} {\bibfnamefont {J.}~\bibnamefont {{Bowman}}}, \bibinfo
  {author} {\bibfnamefont {J.}~\bibnamefont {{Hewitt}}}, \ and\ \bibinfo
  {author} {\bibfnamefont {M.}~\bibnamefont {{Zaldarriaga}}},\ }\href {\doibase
  10.1111/j.1365-2966.2009.15156.x} {\bibfield  {journal} {\bibinfo  {journal}
  {MNRAS}\ }\textbf {\bibinfo {volume} {398}},\ \bibinfo {pages} {401}
  (\bibinfo {year} {2009}{\natexlab{b}})},\ \Eprint
  {http://arxiv.org/abs/0903.4890} {arXiv:0903.4890} \BibitemShut {NoStop}%
\bibitem [{\citenamefont {{Harker}}\ \emph {et~al.}(2010)\citenamefont
  {{Harker}}, \citenamefont {{Zaroubi}}, \citenamefont {{Bernardi}},
  \citenamefont {{Brentjens}}, \citenamefont {{de Bruyn}}, \citenamefont
  {{Ciardi}}, \citenamefont {{Jeli{\'c}}}, \citenamefont {{Koopmans}},
  \citenamefont {{Labropoulos}}, \citenamefont {{Mellema}}, \citenamefont
  {{Offringa}}, \citenamefont {{Pandey}}, \citenamefont {{Pawlik}},
  \citenamefont {{Schaye}}, \citenamefont {{Thomas}},\ and\ \citenamefont
  {{Yatawatta}}}]{LOFAR2}%
  \BibitemOpen
  \bibfield  {author} {\bibinfo {author} {\bibfnamefont {G.}~\bibnamefont
  {{Harker}}}, \bibinfo {author} {\bibfnamefont {S.}~\bibnamefont {{Zaroubi}}},
  \bibinfo {author} {\bibfnamefont {G.}~\bibnamefont {{Bernardi}}}, \bibinfo
  {author} {\bibfnamefont {M.~A.}\ \bibnamefont {{Brentjens}}}, \bibinfo
  {author} {\bibfnamefont {A.~G.}\ \bibnamefont {{de Bruyn}}}, \bibinfo
  {author} {\bibfnamefont {B.}~\bibnamefont {{Ciardi}}}, \bibinfo {author}
  {\bibfnamefont {V.}~\bibnamefont {{Jeli{\'c}}}}, \bibinfo {author}
  {\bibfnamefont {L.~V.~E.}\ \bibnamefont {{Koopmans}}}, \bibinfo {author}
  {\bibfnamefont {P.}~\bibnamefont {{Labropoulos}}}, \bibinfo {author}
  {\bibfnamefont {G.}~\bibnamefont {{Mellema}}}, \bibinfo {author}
  {\bibfnamefont {A.}~\bibnamefont {{Offringa}}}, \bibinfo {author}
  {\bibfnamefont {V.~N.}\ \bibnamefont {{Pandey}}}, \bibinfo {author}
  {\bibfnamefont {A.~H.}\ \bibnamefont {{Pawlik}}}, \bibinfo {author}
  {\bibfnamefont {J.}~\bibnamefont {{Schaye}}}, \bibinfo {author}
  {\bibfnamefont {R.~M.}\ \bibnamefont {{Thomas}}}, \ and\ \bibinfo {author}
  {\bibfnamefont {S.}~\bibnamefont {{Yatawatta}}},\ }\href {\doibase
  10.1111/j.1365-2966.2010.16628.x} {\bibfield  {journal} {\bibinfo  {journal}
  {\mnras}\ }\textbf {\bibinfo {volume} {405}},\ \bibinfo {pages} {2492}
  (\bibinfo {year} {2010})},\ \Eprint {http://arxiv.org/abs/1003.0965}
  {arXiv:1003.0965 [astro-ph.CO]} \BibitemShut {NoStop}%
\bibitem [{\citenamefont {{Cho}}\ \emph {et~al.}(2012)\citenamefont {{Cho}},
  \citenamefont {{Lazarian}},\ and\ \citenamefont {{Timbie}}}]{ChoForegrounds}%
  \BibitemOpen
  \bibfield  {author} {\bibinfo {author} {\bibfnamefont {J.}~\bibnamefont
  {{Cho}}}, \bibinfo {author} {\bibfnamefont {A.}~\bibnamefont {{Lazarian}}}, \
  and\ \bibinfo {author} {\bibfnamefont {P.~T.}\ \bibnamefont {{Timbie}}},\
  }\href {\doibase 10.1088/0004-637X/749/2/164} {\bibfield  {journal} {\bibinfo
   {journal} {\apj}\ }\textbf {\bibinfo {volume} {749}},\ \bibinfo {eid} {164}
  (\bibinfo {year} {2012})},\ \Eprint {http://arxiv.org/abs/1203.5197}
  {arXiv:1203.5197 [astro-ph.CO]} \BibitemShut {NoStop}%
\bibitem [{\citenamefont {{McQuinn}}\ \emph {et~al.}(2007)\citenamefont
  {{McQuinn}}, \citenamefont {{Hernquist}}, \citenamefont {{Zaldarriaga}},\
  and\ \citenamefont {{Dutta}}}]{McQuinnLyman}%
  \BibitemOpen
  \bibfield  {author} {\bibinfo {author} {\bibfnamefont {M.}~\bibnamefont
  {{McQuinn}}}, \bibinfo {author} {\bibfnamefont {L.}~\bibnamefont
  {{Hernquist}}}, \bibinfo {author} {\bibfnamefont {M.}~\bibnamefont
  {{Zaldarriaga}}}, \ and\ \bibinfo {author} {\bibfnamefont {S.}~\bibnamefont
  {{Dutta}}},\ }\href {\doibase 10.1111/j.1365-2966.2007.12085.x} {\bibfield
  {journal} {\bibinfo  {journal} {\mnras}\ }\textbf {\bibinfo {volume} {381}},\
  \bibinfo {pages} {75} (\bibinfo {year} {2007})},\ \Eprint
  {http://arxiv.org/abs/0704.2239} {arXiv:0704.2239} \BibitemShut {NoStop}%
\bibitem [{\citenamefont {{Tegmark}}\ \emph {et~al.}(1998)\citenamefont
  {{Tegmark}}, \citenamefont {{Hamilton}}, \citenamefont {{Strauss}},
  \citenamefont {{Vogeley}},\ and\ \citenamefont
  {{Szalay}}}]{Maxgalaxysurvey1}%
  \BibitemOpen
  \bibfield  {author} {\bibinfo {author} {\bibfnamefont {M.}~\bibnamefont
  {{Tegmark}}}, \bibinfo {author} {\bibfnamefont {A.~J.~S.}\ \bibnamefont
  {{Hamilton}}}, \bibinfo {author} {\bibfnamefont {M.~A.}\ \bibnamefont
  {{Strauss}}}, \bibinfo {author} {\bibfnamefont {M.~S.}\ \bibnamefont
  {{Vogeley}}}, \ and\ \bibinfo {author} {\bibfnamefont {A.~S.}\ \bibnamefont
  {{Szalay}}},\ }\href@noop {} {\bibfield  {journal} {\bibinfo  {journal}
  {ApJ}\ } (\bibinfo {year} {1998})},\ \Eprint
  {http://arxiv.org/abs/arXiv:astro-ph/9708020} {arXiv:astro-ph/9708020}
  \BibitemShut {NoStop}%
\bibitem [{\citenamefont {{Barkana}}\ and\ \citenamefont
  {{Loeb}}(2005{\natexlab{b}})}]{Barkana2}%
  \BibitemOpen
  \bibfield  {author} {\bibinfo {author} {\bibfnamefont {R.}~\bibnamefont
  {{Barkana}}}\ and\ \bibinfo {author} {\bibfnamefont {A.}~\bibnamefont
  {{Loeb}}},\ }\href {\doibase 10.1086/430599} {\bibfield  {journal} {\bibinfo
  {journal} {ApJ Lett.}\ }\textbf {\bibinfo {volume} {624}},\ \bibinfo {pages}
  {L65} (\bibinfo {year} {2005}{\natexlab{b}})},\ \Eprint
  {http://arxiv.org/abs/arXiv:astro-ph/0409572} {arXiv:astro-ph/0409572}
  \BibitemShut {NoStop}%
\bibitem [{\citenamefont {{Ali}}\ \emph {et~al.}(2005)\citenamefont {{Ali}},
  \citenamefont {{Bharadwaj}},\ and\ \citenamefont {{Pandey}}}]{AliAP}%
  \BibitemOpen
  \bibfield  {author} {\bibinfo {author} {\bibfnamefont {S.~S.}\ \bibnamefont
  {{Ali}}}, \bibinfo {author} {\bibfnamefont {S.}~\bibnamefont {{Bharadwaj}}},
  \ and\ \bibinfo {author} {\bibfnamefont {B.}~\bibnamefont {{Pandey}}},\
  }\href {\doibase 10.1111/j.1365-2966.2005.09444.x} {\bibfield  {journal}
  {\bibinfo  {journal} {MNRAS}\ }\textbf {\bibinfo {volume} {363}},\ \bibinfo
  {pages} {251} (\bibinfo {year} {2005})},\ \Eprint
  {http://arxiv.org/abs/arXiv:astro-ph/0503237} {arXiv:astro-ph/0503237}
  \BibitemShut {NoStop}%
\bibitem [{\citenamefont {{Nusser}}(2005)}]{NusserAP}%
  \BibitemOpen
  \bibfield  {author} {\bibinfo {author} {\bibfnamefont {A.}~\bibnamefont
  {{Nusser}}},\ }\href {\doibase 10.1111/j.1365-2966.2005.09603.x} {\bibfield
  {journal} {\bibinfo  {journal} {MNRAS}\ }\textbf {\bibinfo {volume} {364}},\
  \bibinfo {pages} {743} (\bibinfo {year} {2005})},\ \Eprint
  {http://arxiv.org/abs/arXiv:astro-ph/0410420} {arXiv:astro-ph/0410420}
  \BibitemShut {NoStop}%
\bibitem [{\citenamefont {{Barkana}}(2006)}]{BarkanaAP}%
  \BibitemOpen
  \bibfield  {author} {\bibinfo {author} {\bibfnamefont {R.}~\bibnamefont
  {{Barkana}}},\ }\href {\doibase 10.1111/j.1365-2966.2006.10882.x} {\bibfield
  {journal} {\bibinfo  {journal} {MNRAS}\ }\textbf {\bibinfo {volume} {372}},\
  \bibinfo {pages} {259} (\bibinfo {year} {2006})},\ \Eprint
  {http://arxiv.org/abs/arXiv:astro-ph/0508341} {arXiv:astro-ph/0508341}
  \BibitemShut {NoStop}%
\bibitem [{\citenamefont {{Tegmark}}(1997)}]{Maxpowerspeclossless}%
  \BibitemOpen
  \bibfield  {author} {\bibinfo {author} {\bibfnamefont {M.}~\bibnamefont
  {{Tegmark}}},\ }\href {\doibase 10.1103/PhysRevD.55.5895} {\bibfield
  {journal} {\bibinfo  {journal} {\prd}\ }\textbf {\bibinfo {volume} {55}},\
  \bibinfo {pages} {5895} (\bibinfo {year} {1997})},\ \Eprint
  {http://arxiv.org/abs/arXiv:astro-ph/9611174} {arXiv:astro-ph/9611174}
  \BibitemShut {NoStop}%
\bibitem [{\citenamefont {{Bond}}\ \emph {et~al.}(1998)\citenamefont {{Bond}},
  \citenamefont {{Jaffe}},\ and\ \citenamefont {{Knox}}}]{BJK}%
  \BibitemOpen
  \bibfield  {author} {\bibinfo {author} {\bibfnamefont {J.~R.}\ \bibnamefont
  {{Bond}}}, \bibinfo {author} {\bibfnamefont {A.~H.}\ \bibnamefont {{Jaffe}}},
  \ and\ \bibinfo {author} {\bibfnamefont {L.}~\bibnamefont {{Knox}}},\ }\href
  {\doibase 10.1103/PhysRevD.57.2117} {\bibfield  {journal} {\bibinfo
  {journal} {\prd}\ }\textbf {\bibinfo {volume} {57}},\ \bibinfo {pages} {2117}
  (\bibinfo {year} {1998})},\ \Eprint
  {http://arxiv.org/abs/arXiv:astro-ph/9708203} {arXiv:astro-ph/9708203}
  \BibitemShut {NoStop}%
\bibitem [{\citenamefont {{Fisher}}(1935)}]{fisher}%
  \BibitemOpen
  \bibfield  {author} {\bibinfo {author} {\bibfnamefont {R.~A.}\ \bibnamefont
  {{Fisher}}},\ }\href@noop {} {\bibfield  {journal} {\bibinfo  {journal} {J.
  Roy. Stat. Soc.}\ }\textbf {\bibinfo {volume} {98}},\ \bibinfo {pages} {39}
  (\bibinfo {year} {1935})}\BibitemShut {NoStop}%
\bibitem [{\citenamefont {{Tegmark}}\ \emph {et~al.}(2002)\citenamefont
  {{Tegmark}}, \citenamefont {{Hamilton}},\ and\ \citenamefont
  {{Xu}}}]{THX2df}%
  \BibitemOpen
  \bibfield  {author} {\bibinfo {author} {\bibfnamefont {M.}~\bibnamefont
  {{Tegmark}}}, \bibinfo {author} {\bibfnamefont {A.~J.~S.}\ \bibnamefont
  {{Hamilton}}}, \ and\ \bibinfo {author} {\bibfnamefont {Y.}~\bibnamefont
  {{Xu}}},\ }\href {\doibase 10.1046/j.1365-8711.2002.05622.x} {\bibfield
  {journal} {\bibinfo  {journal} {\mnras}\ }\textbf {\bibinfo {volume} {335}},\
  \bibinfo {pages} {887} (\bibinfo {year} {2002})},\ \Eprint
  {http://arxiv.org/abs/arXiv:astro-ph/0111575} {arXiv:astro-ph/0111575}
  \BibitemShut {NoStop}%
\bibitem [{\citenamefont {{Di Matteo}}\ \emph {et~al.}(2002)\citenamefont {{Di
  Matteo}}, \citenamefont {{Perna}}, \citenamefont {{Abel}},\ and\
  \citenamefont {{Rees}}}]{dimatteo1}%
  \BibitemOpen
  \bibfield  {author} {\bibinfo {author} {\bibfnamefont {T.}~\bibnamefont {{Di
  Matteo}}}, \bibinfo {author} {\bibfnamefont {R.}~\bibnamefont {{Perna}}},
  \bibinfo {author} {\bibfnamefont {T.}~\bibnamefont {{Abel}}}, \ and\ \bibinfo
  {author} {\bibfnamefont {M.~J.}\ \bibnamefont {{Rees}}},\ }\href {\doibase
  10.1086/324293} {\bibfield  {journal} {\bibinfo  {journal} {ApJ}\ }\textbf
  {\bibinfo {volume} {564}},\ \bibinfo {pages} {576} (\bibinfo {year}
  {2002})},\ \Eprint {http://arxiv.org/abs/arXiv:astro-ph/0109241}
  {arXiv:astro-ph/0109241} \BibitemShut {NoStop}%
\bibitem [{\citenamefont {{Tegmark}}\ and\ \citenamefont
  {{Zaldarriaga}}(2009)}]{FFTT}%
  \BibitemOpen
  \bibfield  {author} {\bibinfo {author} {\bibfnamefont {M.}~\bibnamefont
  {{Tegmark}}}\ and\ \bibinfo {author} {\bibfnamefont {M.}~\bibnamefont
  {{Zaldarriaga}}},\ }\href {\doibase 10.1103/PhysRevD.79.083530} {\bibfield
  {journal} {\bibinfo  {journal} {\prd}\ }\textbf {\bibinfo {volume} {79}},\
  \bibinfo {eid} {083530} (\bibinfo {year} {2009})},\ \Eprint
  {http://arxiv.org/abs/0805.4414} {arXiv:0805.4414} \BibitemShut {NoStop}%
\bibitem [{\citenamefont {{Hogg}}(1999)}]{HoggDistance}%
  \BibitemOpen
  \bibfield  {author} {\bibinfo {author} {\bibfnamefont {D.~W.}\ \bibnamefont
  {{Hogg}}},\ }\href@noop {} {\bibfield  {journal} {\bibinfo  {journal} {ArXiv
  Astrophysics e-prints}\ } (\bibinfo {year} {1999})},\ \Eprint
  {http://arxiv.org/abs/arXiv:astro-ph/9905116} {arXiv:astro-ph/9905116}
  \BibitemShut {NoStop}%
\bibitem [{\citenamefont {{Oh}}\ \emph {et~al.}(1999)\citenamefont {{Oh}},
  \citenamefont {{Spergel}},\ and\ \citenamefont {{Hinshaw}}}]{OhCMBConjGrad}%
  \BibitemOpen
  \bibfield  {author} {\bibinfo {author} {\bibfnamefont {S.~P.}\ \bibnamefont
  {{Oh}}}, \bibinfo {author} {\bibfnamefont {D.~N.}\ \bibnamefont {{Spergel}}},
  \ and\ \bibinfo {author} {\bibfnamefont {G.}~\bibnamefont {{Hinshaw}}},\
  }\href {\doibase 10.1086/306629} {\bibfield  {journal} {\bibinfo  {journal}
  {\apj}\ }\textbf {\bibinfo {volume} {510}},\ \bibinfo {pages} {551} (\bibinfo
  {year} {1999})},\ \Eprint {http://arxiv.org/abs/arXiv:astro-ph/9805339}
  {arXiv:astro-ph/9805339} \BibitemShut {NoStop}%
\bibitem [{\citenamefont {{Jarosik}}\ \emph {et~al.}(2007)\citenamefont
  {{Jarosik}}, \citenamefont {{Barnes}}, \citenamefont {{Greason}},
  \citenamefont {{Hill}}, \citenamefont {{Nolta}}, \citenamefont {{Odegard}},
  \citenamefont {{Weiland}}, \citenamefont {{Bean}}, \citenamefont {{Bennett}},
  \citenamefont {{Dor{\'e}}}, \citenamefont {{Halpern}}, \citenamefont
  {{Hinshaw}}, \citenamefont {{Kogut}}, \citenamefont {{Komatsu}},
  \citenamefont {{Limon}}, \citenamefont {{Meyer}}, \citenamefont {{Page}},
  \citenamefont {{Spergel}}, \citenamefont {{Tucker}}, \citenamefont
  {{Wollack}},\ and\ \citenamefont {{Wright}}}]{WMAPconjugategrad}%
  \BibitemOpen
  \bibfield  {author} {\bibinfo {author} {\bibfnamefont {N.}~\bibnamefont
  {{Jarosik}}}, \bibinfo {author} {\bibfnamefont {C.}~\bibnamefont {{Barnes}}},
  \bibinfo {author} {\bibfnamefont {M.~R.}\ \bibnamefont {{Greason}}}, \bibinfo
  {author} {\bibfnamefont {R.~S.}\ \bibnamefont {{Hill}}}, \bibinfo {author}
  {\bibfnamefont {M.~R.}\ \bibnamefont {{Nolta}}}, \bibinfo {author}
  {\bibfnamefont {N.}~\bibnamefont {{Odegard}}}, \bibinfo {author}
  {\bibfnamefont {J.~L.}\ \bibnamefont {{Weiland}}}, \bibinfo {author}
  {\bibfnamefont {R.}~\bibnamefont {{Bean}}}, \bibinfo {author} {\bibfnamefont
  {C.~L.}\ \bibnamefont {{Bennett}}}, \bibinfo {author} {\bibfnamefont
  {O.}~\bibnamefont {{Dor{\'e}}}}, \bibinfo {author} {\bibfnamefont
  {M.}~\bibnamefont {{Halpern}}}, \bibinfo {author} {\bibfnamefont
  {G.}~\bibnamefont {{Hinshaw}}}, \bibinfo {author} {\bibfnamefont
  {A.}~\bibnamefont {{Kogut}}}, \bibinfo {author} {\bibfnamefont
  {E.}~\bibnamefont {{Komatsu}}}, \bibinfo {author} {\bibfnamefont
  {M.}~\bibnamefont {{Limon}}}, \bibinfo {author} {\bibfnamefont {S.~S.}\
  \bibnamefont {{Meyer}}}, \bibinfo {author} {\bibfnamefont {L.}~\bibnamefont
  {{Page}}}, \bibinfo {author} {\bibfnamefont {D.~N.}\ \bibnamefont
  {{Spergel}}}, \bibinfo {author} {\bibfnamefont {G.~S.}\ \bibnamefont
  {{Tucker}}}, \bibinfo {author} {\bibfnamefont {E.}~\bibnamefont {{Wollack}}},
  \ and\ \bibinfo {author} {\bibfnamefont {E.~L.}\ \bibnamefont {{Wright}}},\
  }\href {\doibase 10.1086/513697} {\bibfield  {journal} {\bibinfo  {journal}
  {\apjs}\ }\textbf {\bibinfo {volume} {170}},\ \bibinfo {pages} {263}
  (\bibinfo {year} {2007})},\ \Eprint
  {http://arxiv.org/abs/arXiv:astro-ph/0603452} {arXiv:astro-ph/0603452}
  \BibitemShut {NoStop}%
\bibitem [{\citenamefont {{Hestenes}}\ and\ \citenamefont
  {{Stiefel}}(1952)}]{ConjGradOriginal}%
  \BibitemOpen
  \bibfield  {author} {\bibinfo {author} {\bibfnamefont {M.~R.}\ \bibnamefont
  {{Hestenes}}}\ and\ \bibinfo {author} {\bibfnamefont {E.}~\bibnamefont
  {{Stiefel}}},\ }\href@noop {} {\bibfield  {journal} {\bibinfo  {journal}
  {Journal of Research of the National Bureau of Standards}\ }\textbf {\bibinfo
  {volume} {49}},\ \bibinfo {pages} {409} (\bibinfo {year} {1952})}\BibitemShut
  {NoStop}%
\bibitem [{\citenamefont {{Shewchuk}}(1994)}]{ConjGrad}%
  \BibitemOpen
  \bibfield  {author} {\bibinfo {author} {\bibfnamefont {J.}~\bibnamefont
  {{Shewchuk}}},\ }\href
  {http://www.cs.cmu.edu/~quake-papers/painless-conjugate-gradient.pdf}
  {\bibfield  {journal} {\bibinfo  {journal} {unpublished, available online}\ }
  (\bibinfo {year} {1994})}\BibitemShut {NoStop}%
\bibitem [{\citenamefont {{Datta}}\ \emph {et~al.}(2010)\citenamefont
  {{Datta}}, \citenamefont {{Bowman}},\ and\ \citenamefont
  {{Carilli}}}]{Dattapowerspec}%
  \BibitemOpen
  \bibfield  {author} {\bibinfo {author} {\bibfnamefont {A.}~\bibnamefont
  {{Datta}}}, \bibinfo {author} {\bibfnamefont {J.~D.}\ \bibnamefont
  {{Bowman}}}, \ and\ \bibinfo {author} {\bibfnamefont {C.~L.}\ \bibnamefont
  {{Carilli}}},\ }\href {\doibase 10.1088/0004-637X/724/1/526} {\bibfield
  {journal} {\bibinfo  {journal} {ApJ}\ }\textbf {\bibinfo {volume} {724}},\
  \bibinfo {pages} {526} (\bibinfo {year} {2010})},\ \Eprint
  {http://arxiv.org/abs/1005.4071} {arXiv:1005.4071 [astro-ph.CO]} \BibitemShut
  {NoStop}%
\bibitem [{\citenamefont {{Vedantham}}\ \emph {et~al.}(2012)\citenamefont
  {{Vedantham}}, \citenamefont {{Udaya Shankar}},\ and\ \citenamefont
  {{Subrahmanyan}}}]{VedanthamWedge}%
  \BibitemOpen
  \bibfield  {author} {\bibinfo {author} {\bibfnamefont {H.}~\bibnamefont
  {{Vedantham}}}, \bibinfo {author} {\bibfnamefont {N.}~\bibnamefont {{Udaya
  Shankar}}}, \ and\ \bibinfo {author} {\bibfnamefont {R.}~\bibnamefont
  {{Subrahmanyan}}},\ }\href {\doibase 10.1088/0004-637X/745/2/176} {\bibfield
  {journal} {\bibinfo  {journal} {\apj}\ }\textbf {\bibinfo {volume} {745}},\
  \bibinfo {eid} {176} (\bibinfo {year} {2012})},\ \Eprint
  {http://arxiv.org/abs/1106.1297} {arXiv:1106.1297 [astro-ph.IM]} \BibitemShut
  {NoStop}%
\bibitem [{\citenamefont {{Morales}}\ \emph {et~al.}(2012)\citenamefont
  {{Morales}}, \citenamefont {{Hazelton}}, \citenamefont {{Sullivan}},\ and\
  \citenamefont {{Beardsley}}}]{MoralesPSShapes}%
  \BibitemOpen
  \bibfield  {author} {\bibinfo {author} {\bibfnamefont {M.~F.}\ \bibnamefont
  {{Morales}}}, \bibinfo {author} {\bibfnamefont {B.}~\bibnamefont
  {{Hazelton}}}, \bibinfo {author} {\bibfnamefont {I.}~\bibnamefont
  {{Sullivan}}}, \ and\ \bibinfo {author} {\bibfnamefont {A.}~\bibnamefont
  {{Beardsley}}},\ }\href {\doibase 10.1088/0004-637X/752/2/137} {\bibfield
  {journal} {\bibinfo  {journal} {\apj}\ }\textbf {\bibinfo {volume} {752}},\
  \bibinfo {eid} {137} (\bibinfo {year} {2012})},\ \Eprint
  {http://arxiv.org/abs/1202.3830} {arXiv:1202.3830 [astro-ph.IM]} \BibitemShut
  {NoStop}%
\bibitem [{\citenamefont {{Trott}}\ \emph {et~al.}(2012)\citenamefont
  {{Trott}}, \citenamefont {{Wayth}},\ and\ \citenamefont
  {{Tingay}}}]{Trottwedge}%
  \BibitemOpen
  \bibfield  {author} {\bibinfo {author} {\bibfnamefont {C.~M.}\ \bibnamefont
  {{Trott}}}, \bibinfo {author} {\bibfnamefont {R.~B.}\ \bibnamefont
  {{Wayth}}}, \ and\ \bibinfo {author} {\bibfnamefont {S.~J.}\ \bibnamefont
  {{Tingay}}},\ }\href {\doibase 10.1088/0004-637X/757/1/101} {\bibfield
  {journal} {\bibinfo  {journal} {\apj}\ }\textbf {\bibinfo {volume} {757}},\
  \bibinfo {eid} {101} (\bibinfo {year} {2012})},\ \Eprint
  {http://arxiv.org/abs/1208.0646} {arXiv:1208.0646 [astro-ph.CO]} \BibitemShut
  {NoStop}%
\bibitem [{\citenamefont {{Wood}}\ and\ \citenamefont
  {{Chan}}(1994)}]{ToeplitzSimulation}%
  \BibitemOpen
  \bibfield  {author} {\bibinfo {author} {\bibfnamefont {T.~A.}\ \bibnamefont
  {{Wood}}}\ and\ \bibinfo {author} {\bibfnamefont {G.}~\bibnamefont
  {{Chan}}},\ }\href@noop {} {\bibfield  {journal} {\bibinfo  {journal}
  {Journal of Computational and Graphical Statistics}\ }\textbf {\bibinfo
  {volume} {3}},\ \bibinfo {pages} {409} (\bibinfo {year} {1994})}\BibitemShut
  {NoStop}%
\bibitem [{\citenamefont {{Liu}}\ \emph {et~al.}(2010)\citenamefont {{Liu}},
  \citenamefont {{Tegmark}}, \citenamefont {{Morrison}}, \citenamefont
  {{Lutomirski}},\ and\ \citenamefont {{Zaldarriaga}}}]{redundant}%
  \BibitemOpen
  \bibfield  {author} {\bibinfo {author} {\bibfnamefont {A.}~\bibnamefont
  {{Liu}}}, \bibinfo {author} {\bibfnamefont {M.}~\bibnamefont {{Tegmark}}},
  \bibinfo {author} {\bibfnamefont {S.}~\bibnamefont {{Morrison}}}, \bibinfo
  {author} {\bibfnamefont {A.}~\bibnamefont {{Lutomirski}}}, \ and\ \bibinfo
  {author} {\bibfnamefont {M.}~\bibnamefont {{Zaldarriaga}}},\ }\href {\doibase
  10.1111/j.1365-2966.2010.17174.x} {\bibfield  {journal} {\bibinfo  {journal}
  {\mnras}\ }\textbf {\bibinfo {volume} {408}},\ \bibinfo {pages} {1029}
  (\bibinfo {year} {2010})},\ \Eprint {http://arxiv.org/abs/1001.5268}
  {arXiv:1001.5268 [astro-ph.IM]} \BibitemShut {NoStop}%
\bibitem [{\citenamefont {{Gray}}(2006)}]{Toeplitz}%
  \BibitemOpen
  \bibfield  {author} {\bibinfo {author} {\bibfnamefont {R.~M.}\ \bibnamefont
  {{Gray}}},\ }\href@noop {} {\bibfield  {journal} {\bibinfo  {journal}
  {Foundations and Trends in Communications and Information Theory}\ }\textbf
  {\bibinfo {volume} {2}},\ \bibinfo {pages} {155} (\bibinfo {year}
  {2006})}\BibitemShut {NoStop}%
\bibitem [{\citenamefont {{Bernardi}}\ \emph {et~al.}(2009)\citenamefont
  {{Bernardi}}, \citenamefont {{de Bruyn}}, \citenamefont {{Brentjens}},
  \citenamefont {{Ciardi}}, \citenamefont {{Harker}}, \citenamefont
  {{Jeli{\'c}}}, \citenamefont {{Koopmans}}, \citenamefont {{Labropoulos}},
  \citenamefont {{Offringa}}, \citenamefont {{Pandey}}, \citenamefont
  {{Schaye}}, \citenamefont {{Thomas}}, \citenamefont {{Yatawatta}},\ and\
  \citenamefont {{Zaroubi}}}]{BernardiForegrounds}%
  \BibitemOpen
  \bibfield  {author} {\bibinfo {author} {\bibfnamefont {G.}~\bibnamefont
  {{Bernardi}}}, \bibinfo {author} {\bibfnamefont {A.~G.}\ \bibnamefont {{de
  Bruyn}}}, \bibinfo {author} {\bibfnamefont {M.~A.}\ \bibnamefont
  {{Brentjens}}}, \bibinfo {author} {\bibfnamefont {B.}~\bibnamefont
  {{Ciardi}}}, \bibinfo {author} {\bibfnamefont {G.}~\bibnamefont {{Harker}}},
  \bibinfo {author} {\bibfnamefont {V.}~\bibnamefont {{Jeli{\'c}}}}, \bibinfo
  {author} {\bibfnamefont {L.~V.~E.}\ \bibnamefont {{Koopmans}}}, \bibinfo
  {author} {\bibfnamefont {P.}~\bibnamefont {{Labropoulos}}}, \bibinfo {author}
  {\bibfnamefont {A.}~\bibnamefont {{Offringa}}}, \bibinfo {author}
  {\bibfnamefont {V.~N.}\ \bibnamefont {{Pandey}}}, \bibinfo {author}
  {\bibfnamefont {J.}~\bibnamefont {{Schaye}}}, \bibinfo {author}
  {\bibfnamefont {R.~M.}\ \bibnamefont {{Thomas}}}, \bibinfo {author}
  {\bibfnamefont {S.}~\bibnamefont {{Yatawatta}}}, \ and\ \bibinfo {author}
  {\bibfnamefont {S.}~\bibnamefont {{Zaroubi}}},\ }\href {\doibase
  10.1051/0004-6361/200911627} {\bibfield  {journal} {\bibinfo  {journal}
  {\aap}\ }\textbf {\bibinfo {volume} {500}},\ \bibinfo {pages} {965} (\bibinfo
  {year} {2009})},\ \Eprint {http://arxiv.org/abs/0904.0404} {arXiv:0904.0404
  [astro-ph.CO]} \BibitemShut {NoStop}%
\bibitem [{\citenamefont {{Ghosh}}\ \emph {et~al.}(2012)\citenamefont
  {{Ghosh}}, \citenamefont {{Prasad}}, \citenamefont {{Bharadwaj}},
  \citenamefont {{Ali}},\ and\ \citenamefont {{Chengalur}}}]{GhoshForegrounds}%
  \BibitemOpen
  \bibfield  {author} {\bibinfo {author} {\bibfnamefont {A.}~\bibnamefont
  {{Ghosh}}}, \bibinfo {author} {\bibfnamefont {J.}~\bibnamefont {{Prasad}}},
  \bibinfo {author} {\bibfnamefont {S.}~\bibnamefont {{Bharadwaj}}}, \bibinfo
  {author} {\bibfnamefont {S.~S.}\ \bibnamefont {{Ali}}}, \ and\ \bibinfo
  {author} {\bibfnamefont {J.~N.}\ \bibnamefont {{Chengalur}}},\ }\href
  {\doibase 10.1111/j.1365-2966.2012.21889.x} {\bibfield  {journal} {\bibinfo
  {journal} {\mnras}\ }\textbf {\bibinfo {volume} {426}},\ \bibinfo {pages}
  {3295} (\bibinfo {year} {2012})},\ \Eprint {http://arxiv.org/abs/1208.1617}
  {arXiv:1208.1617 [astro-ph.CO]} \BibitemShut {NoStop}%
\bibitem [{\citenamefont {{Liu}}\ and\ \citenamefont
  {{Tegmark}}(2012)}]{AdrianForegrounds}%
  \BibitemOpen
  \bibfield  {author} {\bibinfo {author} {\bibfnamefont {A.}~\bibnamefont
  {{Liu}}}\ and\ \bibinfo {author} {\bibfnamefont {M.}~\bibnamefont
  {{Tegmark}}},\ }\href {\doibase 10.1111/j.1365-2966.2011.19989.x} {\bibfield
  {journal} {\bibinfo  {journal} {\mnras}\ }\textbf {\bibinfo {volume} {419}},\
  \bibinfo {pages} {3491} (\bibinfo {year} {2012})},\ \Eprint
  {http://arxiv.org/abs/1106.0007} {arXiv:1106.0007 [astro-ph.CO]} \BibitemShut
  {NoStop}%
\end{thebibliography}%

\end{document}